\documentclass{elsart}
\usepackage{graphicx,epsf,epsfig}
\usepackage{here,citesort}
\usepackage{tabularx,rotating}

\newcommand{\jpsi}{\mbox{$J/\psi$}}
\newcommand{\psip}{\mbox{$\psi^\prime$}}
\newcommand{\ups}{\mbox{$\Upsilon$}}
\newcommand{\upss}{\mbox{$\Upsilon(1S)$}}
\newcommand{\upsss}{\mbox{$\Upsilon(2S)$}}
\newcommand{\upssss}{\mbox{$\Upsilon(3S)$}}

\newcommand{\npart}{\mbox{$N_{\mathrm{part}}$}}
\newcommand{\nbin}{\mbox{$N_{\mathrm{coll}}$}}
\newcommand{\dnchdeta}{\mbox{$dN_{\mathrm{ch}}/d\eta$}}

\newcommand{\sqrts}{\mbox{$\sqrt{s}$}}
\newcommand{\sqrtsNN}{\mbox{$\sqrt{s_{NN}}$}}
\newcommand{\ccbar}{\mbox{$c\bar{c}$}}
\newcommand{\dAu}{d+Au}

\newcommand{\AuAu}{Au+Au}
\newcommand{\CuCu}{Cu+Cu}
\newcommand{\PbPb}{Pb+Pb}
\newcommand{\SU}{S+U}

\newcommand{\pp}{$pp$}
\newcommand{\ee}{$e^+e^-$}

\newcommand{\pA}{$p+A$}
\renewcommand{\AA}{$A+A$}

\newcommand{\pT}{\mbox{$p_T$}}

\newcommand{\gev}{\mbox{$\mathrm{GeV}$}}
\newcommand{\gevcc}{\mbox{$\mathrm{GeV/}c^2$}}

\newcommand{\gevc}{\mbox{${\mathrm{GeV/}}c$}}
\newcommand{\mevc}{\mbox{${\mathrm{MeV/}}c$}}

\newcommand{\RAA}{\mbox{$R_{AA}$}}

\newcommand{\ie}{\mbox{\textit{i.e.}}}

\def\lsim{\raise0.3ex\hbox{$<$\kern-0.75em\raise-1.1ex\hbox{$\sim$}}}
\def\gsim{\raise0.3ex\hbox{$>$\kern-0.75em\raise-1.1ex\hbox{$\sim$}}} 

\begin{document}
\runauthor{Frawley, Ullrich, Vogt}
\begin{frontmatter}

\title{Heavy flavor in heavy-ion collisions at RHIC and RHIC II}

\author[FSU]{A. D. Frawley,}
\author[BNL]{T. Ullrich,}
\author[LANL,UDC,LBNL]{and R. Vogt}
\address[FSU] {Physics Department, Florida State University, Tallahassee, 
FL, USA}
\address[BNL] {Physics Department, Brookhaven National Laboratory, Upton, 
NY, USA}
\address[LANL]{Lawrence Livermore National Laboratory, 
Livermore, CA, USA}
\address[UDC] {Physics Department, University of California at Davis, Davis, 
CA, USA}
\address[LBNL]{Nuclear Science Division, Lawrence Berkeley National Laboratory,
Berkeley, CA, USA}

\begin{keyword}
Heavy Flavor, Quarkonium, Bottomonium, Quark Gluon Plasma, Relativistic Heavy-Ion 
Collisions, Relativistic Heavy-Ion Collider, RHIC
\end{keyword}

\begin{abstract}
\label{section:abstract}

In the initial years of operation, experiments at the Relativistic
Heavy Ion Collider (RHIC) have identified a new form of matter formed
in nuclei-nuclei collisions at energy densities more than 100 times
that of a cold atomic nucleus. Measurements and comparison with
relativistic hydrodynamic models indicate that the matter thermalizes
in an unexpectedly short time, has an energy density at least 15 times
larger than needed for color deconfinement, has a temperature about twice
the critical temperature predicted by lattice QCD, and appears
to exhibit collective motion with ideal hydrodynamic properties - a
"perfect liquid" that appears to flow with a near-zero viscosity to
entropy ratio - lower than any previously observed fluid and perhaps
close to a universal lower bound. However, a fundamental understanding
of the medium seen in heavy-ion collisions at RHIC does not yet exist.
The most important scientific challenge for the field in the next
decade is the \textit{quantitative} exploration of the new state of
nuclear matter. That will require new data that will, in turn, require
enhanced capabilities of the RHIC detectors and accelerator.  In this
report we discuss the scientific opportunities for an upgraded RHIC
facility - RHIC II - in conjunction with improved capabilities of the
two large RHIC detectors, PHENIX and STAR.  We focus solely on heavy
flavor probes. Their production rates are calculable using the
well-established techniques of perturbative QCD and their sizable
interactions with the hot QCD medium provide unique and sensitive
measurements of its crucial properties making them one of the key
diagnostic tools available to us.

\end{abstract}
\end{frontmatter}

\tableofcontents

\section{Introduction}

During the last 7 years, heavy{}-ion experiments at the Relativistic
Heavy-Ion Collider (RHIC) have recorded a wealth of data in \AuAu,
\dAu, \CuCu, and \pp\ collisions at a variety of energies, ranging
from $\sqrt{s_{NN}} = 19.6$ \gev\ to the highest available \AuAu\ energy, 
200 \gev.  It is at these high energies that QCD predictions of new
phenomena come into play under conditions where, over nuclear volumes,
the relevant degrees of freedom are expected to be those of quarks and
gluons rather than of hadrons, the realm of the quark{}-gluon plasma.

Measurements from the four RHIC experiments, BRAHMS, PHENIX, PHOBOS
and STAR, have revealed compelling evidence for the existence of a new
form of nuclear matter at extremely high densities and temperatures
\cite{exp_wps_BRAHMS,exp_wps_PHOBOS,exp_wps_STAR,exp_wps_PHENIX}.
Detailed analyses of these data also make it clear that this hot, dense
medium has surprising properties.

The properties of the medium are those of a strongly coupled plasma,
or sQGP, that behaves like a ``perfect liquid'' flowing with a
near{}-zero viscosity to entropy ratio \cite{miklos_larry}. The RHIC
observations have spurred significant advances in theory. However, a
fundamental understanding of the medium seen in heavy{}-ion collisions
at RHIC does not yet exist.  It requires new data that, in turn,
necessitate enhanced capabilities of the RHIC detectors and
accelerator. A detailed plan is being developed by BNL to implement
these upgrades in collaboration with the RHIC scientific community.

The main focus of this report is to outline the scientific opportunities
in the heavy flavor sector provided by upgrades of the two large RHIC
detectors\footnote{The two smaller experiments BRAHMS and 
PHOBOS were decommissioned in 2006.}, PHENIX and STAR, in conjunction 
with an upgrade of the accelerator/collider facility, referred to as 
RHIC II. The detector
upgrades will improve the acceptance, particle identification and
secondary vertex detection capabilities of PHENIX and STAR. The RHIC II
accelerator luminosity upgrade was originally proposed to be done using electron
cooling. It is now planned to use a scheme that employs stochastic cooling 
and other upgrades to achieve for \AuAu\ a factor of 5 increase in luminosity 
over present capabilities. This will provide about 70\% of the average luminosity 
expected from electron cooling, but much sooner and at a much lower cost.
The accelerator upgrades will include 
a new ion injector, EBIS, which will provide high{}-intensity beams of nuclei 
as massive as uranium.

This report is the result of the collaboration and research efforts of a
RHIC{}-wide Heavy Flavor Working Group. It
provides a comprehensive overview of the physics questions than can be
addressed by studies of open charm, open bottom and quarkonia at RHIC
II. It also includes a detailed assessment of the accelerator and
detector capabilities required to carry out these measurements with
sufficient precision to resolve many of the outstanding issues by providing 
detailed results with which to make thorough
comparisons to current and future theoretical calculations.

This report is organized as follows. After a general introduction to
heavy flavor physics, section 2 discusses the detector upgrade program
at RHIC. The projected yields of various heavy flavor measurements that
can be achieved utilizing these upgrades and the higher RHIC II luminosities
are discussed in section 3. In sections 4 and 5 we present a more
detailed discussion of the motivation for studying open heavy flavor
and quarkonia in heavy{}-ion collisions, respectively. We 
also include a review of the current theoretical and experimental
status as well as the proposed experimental program. In section 6 we
review the relationship between heavy flavor physics at RHIC II and the
LHC. We conclude in section 7.

\subsection{Motivation}

Because charm and bottom quarks are massive, they are produced almost
exclusively in the initial parton-parton interactions in heavy-ion
collisions at RHIC energies. In the absence of any nuclear effects,
the heavy flavor cross sections in \AA\ collisions at RHIC would
simply scale with the number of binary collisions. Thus departures
from binary scaling for heavy flavor production in \AA\ collisions
provide information about nuclear effects. These can be divided into
two categories: effects due to embedding the colliding partons in a
nucleus (cold matter effects) and effects due to the large energy
density in the final state. The main focus of the heavy flavor program
at RHIC is to investigate the properties of the dense matter produced
in \AA\ collisions by studying its effects on open heavy flavor and
quarkonium production. This in turn requires a detailed understanding
of cold matter effects so that they can be unfolded from the dense
matter effects.

The program thus requires detailed measurements and calculations of
\pp\ and \pA\ heavy flavor cross sections to characterize the cold
matter effects, if we are to quantify the differences between QGP and
non-QGP effects. Up-to-date benchmark calculations of the total open
heavy flavor (charm and bottom hadrons) and quarkonium ($J/\psi$ and
$\Upsilon$ families) yields and spectra are imperative.  Cold matter
effects that need to be included are nuclear shadowing, for both open
heavy flavor and quarkonium production, and nuclear absorption of
quarkonium.  Recent calculations of charm and bottom production to
Fixed-Order Next-to-Leading Logarithm (FONLL) in \pp\ collisions
have been published \cite{CNV}, along with a discussion of the
inherent theoretical uncertainties \cite{rvjoszo} and reference 
calculations of heavy quark, heavy flavor meson and their decay 
lepton spectra \cite{CNV}.  Similar
calculations have been made for quarkonium production, including
studies of shadowing and absorption effects as a function of rapidity
and centrality in \dAu\ \cite{prcda} and \AA\ \cite{rvhip} collisions
at RHIC.

A number of dense matter effects on heavy flavor production have been
predicted.  Some of these do not change the total cross section but,
instead, modify the $p_T$ spectra of heavy flavor hadrons and their
decay products.  Heavy quark energy loss
\cite{DGVW,Djordjevic:2004nq,Armesto:2003jh,Dokshitzer:2001zm,Lin:1997cn}
by collisional and radiative processes steepens the $p_T$ distribution
relative to that in \pp\ collisions.  On the other hand, random $p_T$
kicks result in transverse momentum broadening, increasing the average
$p_T$ in both cold nuclear matter~\cite{Vogt:2001nh} and in passage
through hadron bubbles in the mixed phase of a
QGP~\cite{Svetitsky:1996nj}.  If the medium surrounding the heavy
quarks after production exhibits collective motion, such as transverse
flow \cite{Greco:2003vf,Lin:2003jy}, the low $p_T$ heavy quarks
($p_T<m$) may be caught in this flow.  Strong effects of energy loss
\cite{Abelev:2006db,PHENIX_auau_electron_final_ref} on heavy flavor
decays to electrons and charm flow
\cite{PHENIX_auau_electron_final_ref} have already been seen in \AuAu\ 
collisions at RHIC.  Studying heavy flavor energy loss using single
electrons requires being able to separate electrons from $c$ and $b$
decays since the large bottom and charm quark mass difference suggests
that bottom quark energy loss is weaker than that for charm
\cite{DGVW}.  Some QGP studies require accurate baseline
determinations of the total heavy flavor cross sections to interpret
other effects.  For example, if more than one $c \overline c$ pair is
produced in an \AA\ event, uncorrelated $c$ and $\overline c$ quarks
might coalesce to form a $J/\psi$ in a QGP
\cite{BobMic,Thews:2000rj,Andronic:2003zv,Kostyuk:2003kt}.  The total
$c \overline c$ yield is needed to normalize the $J/\psi$ production
rate from this process.

Suppression of $J/\psi$ production was one of the most exciting
proposed QGP signatures at the CERN SPS~\cite{na50}.  This $J/\psi$
suppression was predicted to occur due to the shielding of the $c
\overline c$ binding potential by color screening, leading to the
breakup of the quarkonium states, first the $\chi_c$ and $\psi'$, and
finally the $J/\psi$ itself as the temperature increases
\cite{MS,KMS}.  The QGP suppression may not be so simple, as lattice
gauge theory studies of the $J/\psi$ spectral function above the
critical temperature for deconfinement, $T_c$, attest.  The $J/\psi$
may exist as a bound state for temperatures considerably larger than
$T_c$ \cite{datta}.  However, the $J/\psi$ may instead be dissociated
by hot thermal gluons in medium \cite{ks94} before it could be
suppressed by color screening.  Secondary quarkonium production from
uncorrelated $Q \overline Q$ pairs, either in the plasma phase
\cite{Thews:2000rj,Kostyuk:2003kt,pbmjs1,pbmjs2,grrapp} or in the
hadron phase \cite{krpbm,kzwz}, could counter the effects of
suppression, ultimately leading to enhanced quarkonium production.
Such secondary $J/\psi$'s would have different kinematic distributions
than those from the initial production.  Because the underlying
$c\bar{c}$ distribution falls rapidly with $p_T$, the $p_T$
distribution produced by coalescence will be softer. If the underlying
$c\bar{c}$ distribution peaks at midrapidity, the $J/\psi$ rapidity
distribution from coalescence will be narrower than that produced in
the primordial collisions.  The coalescence rapidity distribution
should be calculated with shadowing effects on the underlying
$c\bar{c}$ distribution taken into account since these can cause the
$c\bar{c}$ distribution to flatten in more central \AA\ collisions
\cite{rvhip}.  Elliptic flow is also expected to affect quarkonium as
well as open heavy flavors \cite{Greco:2003vf,Lin:2003jy}.

With higher luminosity at RHIC, the $\Upsilon$ yields could also be
measured accurately.  Since the $\Upsilon$ radius is smaller than that
of the $J/\psi$ \cite{KMS}, direct color screening in the QGP would
not occur until much higher temperatures.  The higher mass bottomonium
states, however, would likely be suppressed at RHIC, as would the
$\chi_c$ and $\psi'$ in the charmonium family.  The feed down
structure is more complicated for the $\Upsilon$ since there are three
$S$ states ($\Upsilon$, $\Upsilon'$ and $\Upsilon''$) and two sets of
$P$ states ($\chi_{b1}$ and $\chi_{b2}$) below the $B \overline B$
threshold.  The $\Upsilon$ family suppression should be measurable
over a large $p_T$ range, with QGP suppression possible on the
$\Upsilon'$ and $\Upsilon''$ up to $p_T \sim 40$~\gevc\ at the LHC
\cite{gunvogt}.  Because of the small number of $b\bar{b}$ pairs
produced at RHIC, bottomonium formation by coalescence of unrelated
pairs should be negligible.

\subsection{Overview of results from the heavy flavor program at RHIC}

Heavy flavor measurements capable of discriminating between
theoretical models need large integrated luminosity. In RHIC runs so
far, useful data sets have been acquired at 200 \gev\ for \pp, \dAu,
\CuCu\ and \AuAu\ collisions. The Run 6 \pp\ data are not fully analyzed yet,
but a sizable data set is available from Run 5. 
Recent runs for two species greatly increase the statistical reach over 
earlier runs: \AuAu\ (Run 7) and \dAu\ (Run 8). 
The data are still being analyzed, but some preliminary 
results are available from Run 7.

The \pp\ data collected to date provide an essential reference for the
heavy-ion program in the form of the underlying heavy flavor
production rates as functions of rapidity and $p_T$. Equally
essential, the data from \dAu\ collisions provide baseline information
about cold nuclear matter effects which must also contribute to heavy
flavor production in heavy-ion collisions. The analyzed \dAu\ data from Run 3 
have limited statistical precision, but they provide useful tests of models 
that include the effects of shadowing
on heavy flavor production and of $J/\psi$ absorption in cold nuclear
matter \cite{prcda}. 

The easiest way by far to measure open heavy flavor yields in heavy-ion 
collisions at RHIC is via the semileptonic decays of $D$ and $B$ mesons. 
Two very striking and unexpected results have already been seen by studying 
decay electrons from open heavy flavor in heavy-ion collisions at RHIC. The 
first is the observation that the nuclear modification factor, $R_{AA}(y,p_T) =
(d\sigma_{AA}/dp_T dy)/(\langle T_{AB} \rangle d\sigma_{pp}/dp_T dy)$,
for electrons from open heavy flavor decays shows very strong
suppression in central \AuAu\ collisions
\cite{Abelev:2006db,PHENIX_auau_electron_final_ref}, similar to that
seen for pions. The second striking result is that the elliptic flow
parameter, $v_2$, of electrons from open heavy flavor decays appears
to favor charm quark flow at low $p_T$
\cite{PHENIX_auau_electron_final_ref}.  Until recently, it had been
expected that heavy quark energy loss would be considerably smaller
than that for light quarks due to interference effects \cite{DGVW}.
Generating the necessary energy loss for charm and bottom quarks with 
realistic gluon densities in the material is a major challenge for
models \cite{DGVW,Armesto}. The relatively large $v_2$ values at low
$p_T$ imply at least some degree of charm quark equilibration with the
medium. This also implies very strong interactions of charm quarks
with the medium at lower $p_T$ \cite{Greco:2003vf,Lin:2003jy}.

A serious shortcoming of open heavy flavor measurements employing
semileptonic decays is the difficulty of separating the contributions
to the lepton spectra from charm and bottom decays.  Perturbative
calculations of the relative contributions from charm and bottom as a
function of \pT\ have large theoretical uncertainties: at midrapidity
the crossover point at which bottom decays become dominant is between
3 and 9 \gevc. Recently, however, the ratio of charm to bottom
contributions has been extracted by STAR and PHENIX from $pp$ data,
although with still large uncertainties.  The STAR measurements
\cite{star_bc_qm2008} cover an electron \pT\ of $3-9$ \gevc\ and are
based on small azimuthal angle correlations between the decay electrons and 
hadrons as well as correlation between electrons and identified $D^0$ 
mesons.  The PHENIX bottom/charm ratio  as a function
of electron \pT\ \cite{phenix_bc_qm2008} was inferred from the 
shape of the electron-hadron correlation
function. The STAR and PHENIX ratios are in good agreement and
indicate that bottom becomes the dominant contribution to the electron
spectra at $p_T \sim 3.5$ \gevc\ (see
Fig.~\ref{bb_cross_sections_all}).  A separate PHENIX measurement of
the bottom and charm cross sections was also obtained from fits to the
dielectron invariant mass spectrum, after removal of all other
contributions \cite{phenix_ppg085}. The results are in agreement with
the cross sections inferred by PHENIX by combining the correlation
measurements with electron \pT\ distributions.

PHENIX has also measured open heavy flavor yields at forward rapidity, 
$y=1.65$, using single 
muons \cite{hornback_qm2008}. The charm cross section is 
$d\sigma/dy = 0.145 \pm 1.1\% ^{+42.7\%}_{-49.8\%}$ $\mu$b. This is consistent
with the measured values at $y=0$ but the precision is currently insufficient 
to meaningfully define the rapidity 
dependence. It is likely that the displaced vertex
measurement from the planned vertex detector upgrades will be needed to
determine the rapidity distribution.  

The first high statistics charmonium results for heavy-ion collisions
at RHIC are now available \cite{PHENIX_jpsi_auau,PHENIX_jpsi_cucu}, along
with high statistics \pp\ reference data \cite{PHENIX_jpsi_pprun5}.
These include \AuAu\ and \CuCu\ results for the
$J/\psi$ $R_{AA}$ as a function of the number of participant nucleons,
$N_{\rm part}$, in the rapidity intervals $|y| < 0.35$ and $1.2< |y| <
2.2$.  A striking feature of the \AuAu\ \jpsi\ data is that, for
$N_{\rm part} > 150$, 
the suppression is considerably stronger at forward rapidity than at
midrapidity.  Comparison with existing models at
midrapidity shows that cold nuclear matter baseline calculations
\cite{rvhip} which approximately reproduce the RHIC \dAu\ $J/\psi$
rapidity distributions \cite{prcda} somewhat underpredict the
suppression observed in \CuCu\ and \AuAu\ collisions.  On the other
hand, several suppression models
\cite{Kostyuk:2003kt,capella,grandch1} which were successful in
describing $J/\psi$ suppression at the SPS are found to strongly
overpredict the suppression at RHIC. Models which incorporate strong
suppression combined with $J/\psi$ coalescence from uncorrelated
$c\bar{c}$ pairs seem to agree best with the data although the
existing models slightly underpredict the suppression. A major source
of uncertainty in all of this is due to the poor precision of the 
existing \dAu\ data, leading to poor constraints on the baseline cold
nuclear matter effects.

A recent reanalysis of the Run 3 \dAu\ \jpsi\ data 
\cite{PHENIX_jpsi_dau_new} used the higher yield 
Run 5 \pp\ data as the reference for calculating \RAA. Theory
calculations have been fit to the reanalyzed data to
explore how well the data constrain cold nuclear matter effects in heavy-ion 
collisions. They find that the constraints from the Run 3 \dAu\ data are 
insufficient for firm conclusions to be drawn about additional hot matter
effects in central heavy-ion collisions 
\cite{PHENIX_jpsi_cucu,PHENIX_jpsi_dau_new}. The newly obtained 
Run 8 \dAu\ data set, 
approximately 20 times the yield of Run 3, will provide 
much improved  statistical precision.

In the last few years, theorists have begun exploring the consequences
of $J/\psi$ coalescence on observables other than the centrality
dependence of the nuclear modification factor
\cite{BobMic,yan_zhuang_xu_2006,zhao_rapp_2007}. 
This work has led to the prediction
that $J/\psi$'s formed by coalescence of uncorrelated $c\bar{c}$ pairs
will have narrower rapidity and $p_T$ distributions due to the
presumed shape of the underlying charm quark distributions. The
coalescence contribution to $J/\psi$ production will cause many
observables to change with centrality, including the rapidity and
$p_T$ dependence of $R_{AA}$, the shape of the $p_T$ distribution
(quantified by the average $p_T^2$, $\langle p_T^2 \rangle$), and the
$J/\psi$ elliptic flow parameter, $v_2$.  Predictions
of $\langle p_T^2 \rangle$ as a function of centrality have been made
for \AuAu\ and \CuCu\ collisions with and without coalescence
\cite{BobMic,yan_zhuang_xu_2006,zhao_rapp_2007}. These predictions, 
compared to the \AuAu\ and preliminary \CuCu\ data, favor calculations 
including coalescence. The coalescence contributions
predicted for central \AuAu\ (as well as central \CuCu) collisions
\cite{BobMic} are also qualitatively expected to narrow the $J/\psi$
rapidity distributions if the underlying charm distributions are
peaked at midrapidity. The \AuAu\ data \cite{PHENIX_jpsi_auau}
show some narrowing of the rapidity distribution for the most central 
collisions although the reduction in RMS is only about $2\sigma$.
Work is still needed to quantify both the theoretical predictions and
the experimental observables. A first measurement of the $J/\psi$ 
$v_2$ may come out of the Run 7 \AuAu\ data, albeit with
large uncertainties.

The existing, but not yet analyzed \AuAu\ data from Run 7 
will quantitatively improve the measurements of many heavy flavor 
observables. The $J/\psi$ $R_{AA}$ versus centrality and $\langle p_T^2
\rangle$ as well as $R_{AA}$ and $v_2$ measurements of charm and
bottom semileptonic decays to single electrons will all improve
significantly, allowing more stringent model tests. Measurements
of other observables will be qualitatively improved. Examples are:
definitive $v_2$ measurements from semileptonic decays at intermediate
to high $p_T$ where we might hope to see the transition from charm to
bottom dominance and flow to non-flow; a possible first $J/\psi$ $v_2$
measurement; substantially improved measurements of $J/\psi$ $R_{AA}(y)$ 
to quantify the coalescence contribution; and improved measurements of
$J/\psi$ $R_{AA}(p_T)$, invaluable for understanding
coalescence and formation time effects. 

However it is clear that the RHIC heavy flavor program is now
limited by the capabilities of the accelerator and the detectors. 
The accelerator upgrades planned over the period 2009-2013 to produce a
factor of 5 greater luminosity at RHIC II, combined with the detector 
upgrades in place by that time, will be required for the
heavy flavor program at RHIC to move to the next level, as described
below.

\subsection{Overview of the proposed heavy flavor program at RHIC II}

The increase in luminosity, combined with the
increased capabilities of the upgraded PHENIX and STAR detectors, will
make it possible to add many important new probes to the heavy flavor
program at RHIC.

One of the most powerful benefits of the luminosity upgrade will be
the ability to measure yields of the excited charmonium states: the
$\psi^\prime$ and $\chi_c$. Lattice calculations predict much smaller
melting temperatures for the $\psi^\prime$ and $\chi_c$ than for the
more tightly bound $J/\psi$.  Thus these excited states should not be
able to exist in the QGP at RHIC and comparison of the
$\psi^\prime$ and $\chi_c$ yields to the $J/\psi$ yield as a function
of centrality would be a direct test of deconfinement.

Testing models in which the observed $J/\psi$ yield in heavy-ion
collisions is due to competition between gluon dissociation and
coalescence formation in the QGP requires very high luminosity. Tests
of charm coalescence models include measuring $J/\psi$ $v_2$ as a
function of $p_T$, $J/\psi$ $R_{AA}$ to much higher $p_T$ to follow
the trends of suppression as the $J/\psi$ formation time approaches
the QGP crossing time, and $J/\psi$ polarization as a function of
collision centrality. The rapidity and $p_T$ dependence of $R_{AA}$ as
functions of \sqrtsNN\ and centrality, requiring sufficient luminosity
for precision measurements at multiple energies, is not possible on a
reasonable time scale at the present RHIC luminosity.

The detailed study of bottomonium states, the $\Upsilon$ family, is only
possible at RHIC II luminosities. Like the charmonium states, the
dissociation temperatures of the bottomonium states depend on the
binding energies.  There are, however, two important differences. 
First, the bottomonium binding energies, particularly that
of the $\Upsilon(1S)$, are higher so that they should dissociate at
higher temperatures. Only the higher-lying bottomonium states are thus
likely to break up at RHIC energies. Second, the $b\bar{b}$ production
rate in central \AuAu\ collisions is only $\sim~0.05$ pairs per
collision, making coalescence production of bottomonium much less
likely. Thus bottomonium production at RHIC II will provide a very
different window on color screening effects than charmonium
production. The bottomonium yields at RHIC II should be sufficient for
measurements of $R_{AA}$ as a function of centrality in heavy-ion
collisions for the three $\Upsilon$ $S$ states. The $\Upsilon$ yields
at RHIC II and at the LHC will not be sufficient for 
$v_2$ or polarization measurements.

As mentioned earlier, measurements of semileptonic open heavy flavor
decays at RHIC have already produced strikingly different results than
expected. The strong suppression in $R_{AA}$ coupled with the large
$v_2$ suggest very large heavy quark energy loss in the medium.
The fact that these semileptonic decay spectra contain both charm and
bottom contributions remains a significant complication, in spite of recent 
successes by STAR and PHENIX at measuring the ratio of $b$ to $c$ 
quark contributions
in the semileptonic decay spectrum of \pp\ collisions. The separation of
open charm and bottom can be done in several ways. Charm can be
observed via hadronic $D^0 \rightarrow K^\pm \pi^\mp$ and $D^\pm \rightarrow
K^\pm \pi^\pm \pi^\mp$ decays, as STAR does. Precise
$R_{AA}$ and $v_2$ measurements are difficult in this channel. Since
these events cannot be triggered, they must be extracted from a
minimum bias data set that samples only a small fraction of the
available luminosity. The combinatorial background is also very large,
making statistical precision difficult. The addition of a displaced
vertex measurement in STAR will dramatically reduce the combinatorial
background but there is still no trigger for these decays.  At RHIC II
luminosity, bottom can be observed very cleanly in both PHENIX and
STAR via $B \rightarrow J/\psi X$ decays using displaced vertices,
providing good measurements of the $b \overline b$ cross section and
bottom quark $R_{AA}$. However those yields will certainly be too
small for $v_2$ measurements at RHIC II or the LHC.  Finally, the
combination of RHIC II luminosity with a displaced vertex measurement
should allow statistical separation of the charm and bottom
contributions to the semileptonic decay spectra, taking advantage of
the different $c$ and $b$ quark decay lengths. Such semileptonic decay
measurements, while less clean than the direct $D$ and $B$ decay
measurements, have the advantage of much larger yields so that
separate charm and bottom  $v_2$ measurements should be possible.

Independent measurements of open charm and bottom $R_{AA}$ and $v_2$
to high $p_T$ will be a very important capability at RHIC II. At low
$p_T$, these measurements reflect the degree of heavy quark
thermalization in the medium. At high $p_T$, they probe the energy
loss of heavy quarks in the medium, providing an independent
measurement of the initial energy density relative to the light quark
energy loss measurements. The thermalization and energy loss
mechanisms at low and high $p_T$ respectively may be quite different
due to possible resonance scattering at low $p_T$.

\subsection{Overview of the relationship of RHIC II to the LHC program}

The heavy flavor production cross sections are significantly higher at
the LHC than at RHIC since the per nucleon \PbPb\ energy at the LHC is
a factor of 27.5 higher than the maximum per nucleon \AuAu\ energy at
RHIC.  The $c \overline c$ and $b \overline b$ cross sections increase
by factors of 15 and 100 respectively \cite{Vogt:2001nh} while the
$J/\psi$ and $\Upsilon$ cross sections increase by factors of 13 and
55 respectively \cite{hfYR}. But, because of the 10 times higher average 
luminosity and three times longer heavy-ion runs, the \AuAu\ integrated 
luminosity at RHIC II will be much higher than for \PbPb\ at LHC. Therefore
the heavy flavor yields per year are expected to be similar at the two
facilities.

At $\sqrt{s} = 200$ \gev, bottom decays to leptons begin to dominate
the single electron spectrum at $p_T \sim 4$~\gevc.  As the collision
energy increases, the lepton spectra from $B$ and $D$ decays move
closer together rather than further apart \cite{Vogt:2001nh}.  Thus,
the large increase in the $b \overline b$ cross section relative to $c
\overline c$ does not make single leptons from $B$ and $D$ decays
easier to separate.  Preliminary calculations show that the $B
\rightarrow e$ decay does become larger than that of $D \rightarrow
e$, but at $p_T > 10$~\gevc.  The yields from the two lepton sources
differ by less than a factor of two up to $p_T \sim 50$~\gevc\ in the
range $|y| \leq 1$.  Thus interpretation of single lepton results on
heavy flavors will be more difficult at the LHC.  Other means of
separating charm and bottom must be found.  ALICE can reconstruct
hadronic $D^0$ decays from $p_T \sim 0$ to $p_T \sim 25$~\gevc\ 
\cite{CJ17} but, like STAR, will have to rely on minimum bias data for
these measurements because of the lack of a trigger.  While it is not
yet clear what CMS and ATLAS will do to reconstruct charm, they should
be able to make $b$-jet measurements, similar to the Tevatron.  One
way that $B$ mesons can be measured at the LHC is through their decays
to $J/\psi$, as discussed further below.  It has also been suggested
that the $B \overline B$ contribution to the dimuon continuum, the
dominant contribution above the $\Upsilon$ mass, can be used to
measure energy loss \cite{igor}.  That channel would be fairly clean
at the LHC but more difficult at RHIC.

The RHIC II upgrades and the high LHC energies make detailed studies
of $\Upsilon$ production and suppression possible.  At the LHC, higher
initial temperatures make $\Upsilon$ suppression more likely than at
RHIC II.  But the higher $b\bar{b}$ production rate ($\sim 5$ per
central \PbPb\ collision) means that, unlike RHIC, significant
coalescence contributions to $\Upsilon$ production may be expected at
the LHC. Thus measurements at the two energies complement each other.
At RHIC II, it is likely that PHENIX will be able to measure and
resolve the three $\Upsilon$ $S$ states.  STAR and ALICE will have
similar $\Upsilon$ yields but the STAR mass resolution will require
fitting to extract yields.
The CMS detector at the LHC has sufficient mass resolution to separate
all three $\Upsilon$ $S$ states.  The $\Upsilon$ states can be
measured to $p_T \sim 0$ at all LHC detectors.  Only ALICE will be
able to measure $J/\psi$ production to $p_T \sim 0$ without a special
trigger \cite{hfYR} since CMS and ATLAS require high single muon $p_T$
so that typically only $J/\psi$ with $p_T >$ several~\gevc\ are accepted.
(However, CMS is working on a higher-level trigger to measure lower
$p_T$ $J/\psi$ \cite{Olga}.)  The larger $b \overline b$ cross section
at the LHC means that $J/\psi$ production from $B \rightarrow J/\psi
X$ cannot be neglected.  These decay $J/\psi$ should be separable from
the initial production using displaced vertices \cite{hfYR}.


\section{Detector upgrade program at RHIC}

Both PHENIX and STAR have extensive upgrade programs underway that are 
extremely
important for the heavy flavor program. The upgrades that are most relevant to 
heavy flavor measurements are described here.
The impact on the heavy flavor program of these detector upgrades, in 
combination with the RHIC II luminosity increase, will be discussed in 
sections 4 and 5.\\

\subsection{PHENIX upgrades}

\begin{figure}[tbh]
  \centering
  \includegraphics[width=0.7\textwidth]{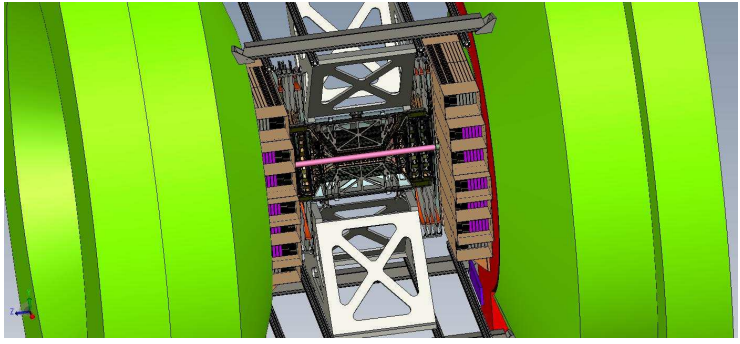}
  \caption{The central region of the PHENIX detector after the addition of the 
    barrel and endcap silicon vertex detectors and the Nose Cone Calorimeter.}
  \label{fig:phenix_ncc_fvtx_picture}
\end{figure}

\begin{figure}[tbh]
  \centering
  \includegraphics[width=0.7\textwidth]{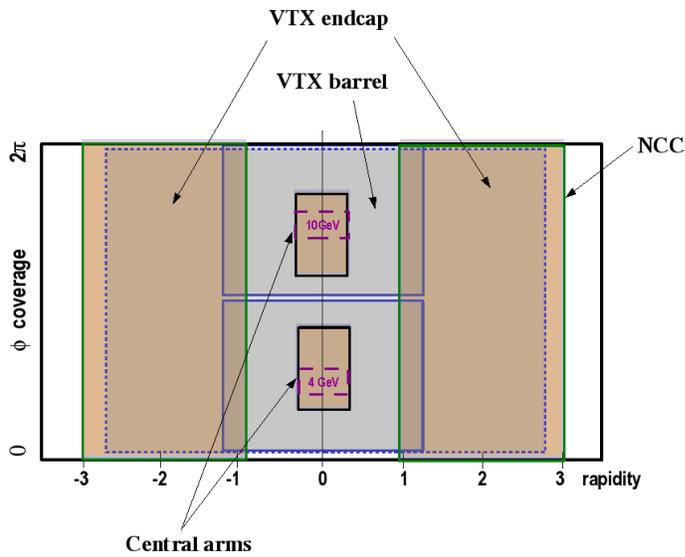}
  \caption{The pseudorapidity and azimuthal angle coverage of the PHENIX 
    barrel and endcap silicon vertex detectors and the Nose Cone Calorimeter
    (NCC) \protect\cite{phenix_ncc_ref}. 
    Two areas of the central arms that that 
    provide hadron identification to high \pT are also shown.}
  \label{fig:phenix_ncc_vtx_coverage}
\end{figure}

Several PHENIX detector upgrades that greatly enhance the heavy flavor
capability of the experiment are expected to be available in the RHIC
II time frame. The most important upgrades for the heavy flavor
program will be the barrel \cite{phenix_vtx_ref} and endcap
\cite{phenix_fvtx_ref} Silicon Vertex Detectors, the Nose Cone
Calorimeter \cite{phenix_ncc_ref} and the Muon Trigger Upgrade
\cite{phenix_muon_trigger_upgrade_ref}. The central region of the
PHENIX detector, after installation of the silicon trackers and the
Nose Cone Calorimeter, is shown in
Fig.~\ref{fig:phenix_ncc_fvtx_picture}.  The pseudorapidity and
azimuthal angle coverages of the new detectors are illustrated in
Fig.~\ref{fig:phenix_ncc_vtx_coverage}.

The Silicon Vertex Detector (SVTX) consists of a central barrel
\cite{phenix_vtx_ref} and two endcap detectors \cite{phenix_fvtx_ref},
as shown in Fig.~\ref{fig:phenix_ncc_fvtx_picture}.  The SVTX barrel
will have a displaced vertex resolution of $\sim 50$ ${\mu}$m while
the endcap resolution is $\sim 90 -
115$ ${\mu}$m.  Together, this inner tracking system provides full azimuthal
coverage over $|\eta|<2.4$.  The SVTX will tag heavy flavor decays using
displaced vertices by connecting to tracks in both
the central and muon arms, improving the quarkonium invariant mass resolution
and reducing backgrounds to heavy flavor measurements.  In the muon
arms, a loose displaced vertex cut will eliminate most muon tracks
from light hadron decays and a very tight cut, $\sim 2\sigma$ where
$\sigma$ is the resolution of the displaced vertex measurement, will
eliminate most punch-through hadrons.  The displaced vertex measurement
will greatly enhance $D^0 \rightarrow K^\pm \pi^\mp$ measurements in
the central arms, presently very difficult in PHENIX, by reducing the
contribution to the combinatorial background both from prompt tracks
(by using a tight vertex cut) and light meson decay tracks (by using a
loose cut of $\sim 1$ cm).  A loose displaced vertex cut will also
reduce high $p_T$ background tracks in the central arms due to
misidentified light hadron decays.  In addition to identifying
semileptonic heavy flavor decays, displaced vertex measurements can
help identify $J/\psi$'s from $B$ meson decays since all other
$J/\psi$'s are prompt.

The SVTX barrel is presently under construction. It will consist of
four concentric silicon layers. The two inner layers, at radii of 2.5
and 5.0 cm, consist of pixel detectors with a segmentation of 50
$\mu$m by 425 $\mu$m. The outer two layers, with radii of 10 and 14
cm, consist of 80 ${\mu}$m by 3 cm strips. The occupancy of the inner
layer will be about 4.5\% in central \AuAu\ collisions. The SVTX
barrel produces a dramatic improvement in high $p_T$ track resolution
in the central arms.  The PHENIX Drift Chamber is outside the magnetic
field so that, in the present momentum measurement, there is no
information about the initial azimuthal angle, $\phi$, of the track. 
The momentum is calculated from the difference between the $\phi$ angle of the
track after passing through the magnetic field and that
from the vertex position to the Drift Chamber.  This difference is
only $\sim$ 40\% of the total deflection.  By adding a precise
measurement of the initial $\phi$ direction, the SVTX barrel
measures the full deflection directly, decreasing the momentum resolution by a
factor of $\sim 2.5$, greatly improving the $\Upsilon$ invariant mass
resolution.  Installation of the barrel is expected starting in 2009.

The forward silicon detector endcaps will consist of four silicon
mini-strip planes.  The mini-strips have 75 $\mu$m pitch in the radial
direction and lengths in the $\phi$ direction varying from 2.8 mm to
12.1 mm, depending on the polar angle. The maximum occupancy per strip
is estimated to be less than 2.8\% in central \AuAu\ collisions. The
displaced vertex resolution of $90 - 115$ $\mu$m, depending on the
number of layers of silicon traversed by the track, should be compared
to a mean vertex displacement of 785 $\mu$m for the boosted open charm
muons. A prototype covering about 1/8 of one muon arm is presently
under construction.

The PHENIX Nose Cone Calorimeters (NCCs) \cite{phenix_ncc_ref},
tungsten-silicon calorimeters, will replace the two central arm
magnet nose cones, and will cover $0.9<|\eta|<3.5$.  The
simulated energy resolution for photons is $\sim 27$\%$/\sqrt{E}$ \gev.  
The Nose Cone Calorimeters will contain both
electromagnetic and hadronic calorimeter sections. The electromagnetic
calorimeter will contain a pre-shower detector and a shower-max
detector designed to discriminate between individual electromagnetic
showers and overlapping photons from high momentum $\pi^0$ decays. The
pre-shower and shower-max detectors are expected to resolve showers
with separations down to 2 and 4 mm, respectively. The NCCs 
should thus have good acceptance for $\chi_c
\rightarrow J/\psi + \gamma$ decays with the \jpsi\ detected in the
muon arms.

The muon trigger upgrade \cite{phenix_muon_trigger_upgrade_ref} is
required for PHENIX to be able to take complete advantage of the RHIC
II luminosity upgrade for muon arm measurements. The current muon arm
level-1 heavy vector meson triggers have sufficient rejection capability to
handle \AuAu\ collision rates of up to $\sim$ 20 kHz and \pp\ 
collision rates of up to $\sim$ 0.5 MHz.  The muon trigger upgrade
adds three layers of Resistive Plate Chamber (RPC) detectors, with two
dimensional ($\theta, \phi$) readout, in each muon arm.  These layers
follow the design of the CMS muon trigger at the LHC but the cathode
pad segmentation is optimized for PHENIX.  The front end electronics and
trigger processors will be developed within PHENIX.  The muon trigger
upgrade, with an online momentum measurement, will improve the
level-1 trigger rejection for both single muons (with a $p_T$ cut) and
muon pairs (with an invariant mass cut).  It will also
improve the high multiplicity background rejection during the final
analysis. The muon trigger upgrade is presently under construction.

\subsection{STAR upgrades}

To realize the compelling scientific opportunities in heavy flavor
physics, upgrades to the STAR detector are required to complete many
of the challenging measurements.  The collaboration has planned a
series of upgrades for the near and intermediate term to overcome the
current shortcomings and enhance its heavy flavor capabilities.
Implementation of these upgrades will also allow optimum utilization
of the increased luminosity expected from RHIC II.

\begin{figure}[tbh]
    \centering\includegraphics[width=\textwidth]{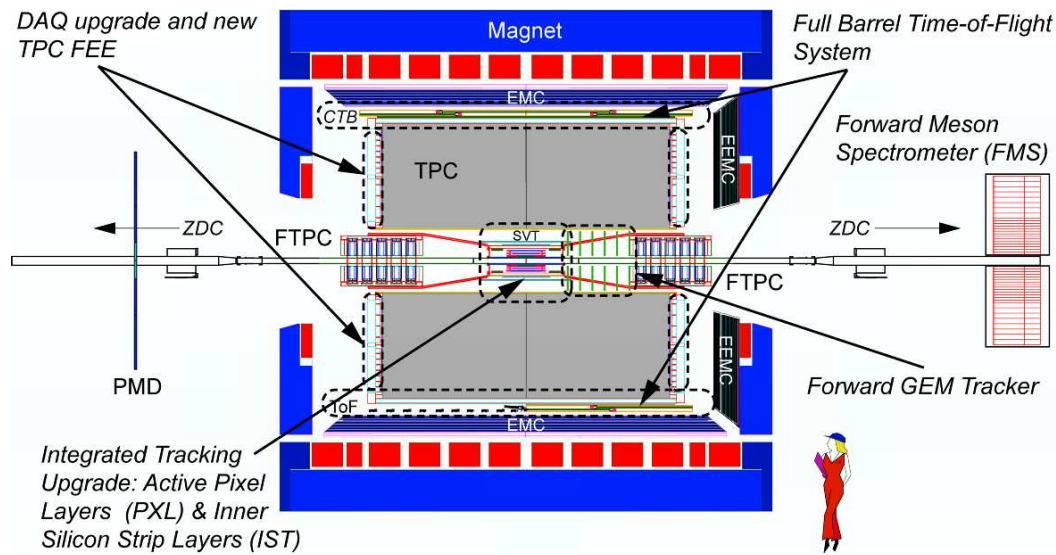}
    \caption{Layout of the STAR experiment in 2005/2006 (modified from 
      Ref.~\cite{Ackermann:2002ad}, reprinted 
      with permission from Elsevier). The locations of
      the planned upgrades are indicated by as dashed lines. See text for
      details.}
    \label{fig:star-layout}
\end{figure}

The current layout of the STAR detector is depicted in
Fig. \ref{fig:star-layout}.  The medium term upgrades to the detector
relevant for heavy flavor physics include: a full barrel Time-of-Flight
detector (ToF) replacing the current ToF patch and the Central
Trigger Barrel (CTB); new front end electronics for the large Time
Projection Chamber (TPC); an upgrade to the data acquisition system
(DAQ-1000), and a tracking upgrade including a barrel section with two
inner layers of silicon pixel sensors (PXL) and one layer of
silicon strip-pad sensors (IST), replacing the current Silicon 
Vertex Tracker (SVT).

The new ToF system covering the full outer barrel of the TPC is being
constructed and installed in STAR over the next two years.  The system
uses the multi-gap resistive plate chamber (MRPC) technology developed
at CERN and will consist of 3840 MRPC modules with 23,000 channels of
readout \cite{Llope:2005yw}. The modules will cover the TPC outer
barrel ($|\eta| <1$, $0 < \phi < 2 \pi$) and will be mounted in
120 trays which will replace the existing CTB 
scintillation counter trays and ToF patch.

The ToF triples the current momentum range over which $\pi$, $K$, and
$p$ can be identified, considerably improving charm meson and baryon 
reconstruction. When the ToF measurement
is combined with the TPC $dE/dx$ measurement, electrons can be cleanly
identified from the lowest momentum measured, $\sim$ 200 \mevc, up to
a few \gevc.  This capability complements the electromagnetic
calorimeter which works well for momenta above $\sim$ 2 \gevc.  STAR
will then be able to reconstruct soft to medium momentum electrons
with high efficiency and purity to make a
comprehensive \jpsi\ measurement. The ToF, in conjunction with the
electromagnetic calorimeter (EMC), also allows STAR to implement a
high-level trigger scheme to select $\jpsi \rightarrow $ \ee decays
in \pp\ collisions.

A series of improvements to the STAR data acquisition system over the
past several years has brought the recorded event rate capability
from the original design of 1 Hz to $50-100$ Hz.  To acquire the very large
data samples and high data rates needed for heavy flavor measurements,
a further upgrade has been initiated to achieve a minimum 1 kHz recorded
event rate which could produce data volumes that
significantly exceed the capacity for analysis and storage.
The rare-trigger data sets will especially benefit from the upgrade
since the pipelined architecture being implemented will virtually
eliminate the front end dead time, allowing STAR to make full use of 
rare-event triggers such as that designed for the \ups.

An increase in readout speed will be achieved by replacing the TPC
front end electronics (FEE), making use of circuits developed for the
ALICE experiment at CERN, in conjunction with an upgrade of the STAR
DAQ.  In addition to the increased physics capabilities from the DAQ
upgrade, the replacement of the TPC FEE, specifically the readout
boards (RDO) that collect data from the FEE boards, will make space
for a future precision tracking chamber between the TPC end planes and
the endcap calorimeter.  Replacing the TPC FEE also
assures that this system can be maintained for the next decade or
more.  The readouts of the other existing detectors, which will remain
in place for the RHIC II era, are being adapted to the new high speed DAQ
with only minor changes.

\begin{figure}[tbh]
    \centering\includegraphics[width=0.5\textwidth]{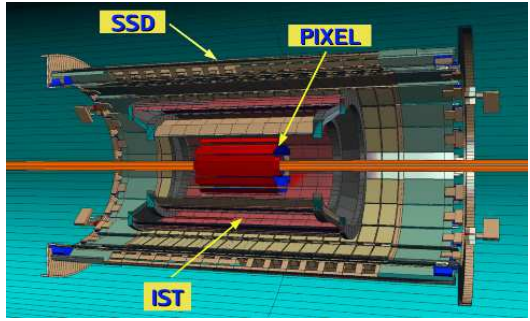}
    \caption{The proposed geometry for the STAR inner tracking
        upgrade, the Heavy Flavor Tracker.  From the inner
        to the outer radii, the 2 cm beam pipe, the two PXL layers, 
        the exoskeleton to strengthen the beam
        pipe, the one IST layers and the
        existing SSD layer are shown \cite{star_its_ref}.}
    \label{fig:star-its}
\end{figure}

In order to address heavy quark energy loss and
thermalization, it will be necessary to cleanly identify open charm.
The recent results from both STAR and PHENIX on the suppression and
flow of non-photonic electrons are intriguing.  However, without an
identified charm sample, the contributions from
semileptonic bottom decays and systematic errors on background
subtraction make a clear interpretation of these results difficult.
Measurement of the yields of various charm species will also
allow the study of charm hadrochemistry. 

Efficient topological reconstruction of open charm decays requires
a tracking ``point-back" resolution to the primary collision vertex of
$\sim 50$ $\mu$m or less.  Further, the beam pipe and innermost
detector layers must be very thin to measure the low $p_T$ particles,
which comprise the bulk of the cross section, and thus minimize the
systematic errors in extrapolating the measured yield to the total
yield.  A thin beam pipe and inner detector layers are also key
elements in efficiently vetoing photon-conversion electrons which,
along with electron identification from the TPC, ToF and
electromagnetic calorimeter, will measure the soft lepton and 
dilepton spectra.  STAR is thus developing a tracking upgrade for the central
rapidity region, the Heavy Flavor Tracker (HFT). 
The essential elements under consideration for the HFT are a new thin,
small-radius beam pipe (0.5 mm thick, 20 mm radius), two layers of
thinned (50 $\mu$m) CMOS pixel detectors at average radii of 2.5 and
8 cm (PXL) and one layer of low mass silicon strip-pad sensors (IST)
at 17 cm, see Fig.~\ref{fig:star-its}.  The existing 
layer of double-sided silicon strip sensors at a radius of 23 cm (SSD) will be
kept but upgraded to be compliant with the overall higher readout speed 
(DAQ-1000).  The two new IST layers will connect tracks from the TPC and SSD
to hits in the pixel layers.  These layers will replace the existing
three layers of silicon drift detector (SVT).  It will be necessary to
replace the SVT since, when RHIC II becomes operational, the SVT will
be over 10 years old with a readout too slow to be compatible with the
upgraded DAQ.  Its infrastructure (cables
and cooling) also adds undesirable mass in the region $1<|\eta| <2$.


\section{Projected RHIC yields}

In this section we present some estimates of the quality of the heavy flavor 
measurements that can be achieved at RHIC with increased luminosities and the 
upgraded detectors, reflected in the heavy flavor yields.

\begin{table}[tbh]
\begin{center}
\begin{tabular}{|ccc||c|c|c|c|} 
\hline
{} & {} & {} & Up to 2008 & 2009 & 2011 & 2013   \\
Species & Energy & Units & Obtained & Projected & Projected & Projected \\ 
\hline 
\AuAu & 200 & ${\mu}$b$^{-1}$ & 380 & 610 & 1450 & 1820    \\ 
\pp & 200 & pb$^{-1}$ & 7.0 & 14.6 & 31.1 & 40    \\ \hline
\pp & 500 & pb$^{-1}$ & $-$ &  36.5 &  78 & 100   \\  \hline
\end{tabular}
\end{center}
\caption{The anticipated weekly luminosity delivered by RHIC. When estimating rates, 
this delivered luminosity has to be reduced by a factor that accounts for detector up 
time and collision vertex cuts imposed by the detectors. The 
RHIC projected luminosities are maximum values taken from the RHIC Collider 
Accelerator Division projections for 2009-2013 as of February 23, 2008.
They reflect the progressive implementation of stochastic cooling and 
other accelerator upgrades and improvements. 
The  numbers in the ``obtained'' column are the best weekly luminosities 
observed through Run 8. }
\label{rhic1_rhic2_lumi}
\end{table}

Table~\ref{rhic1_rhic2_lumi} is a summary of the weekly-integrated, 
delivered luminosity 
estimates for RHIC spanning the period when the luminosity 
upgrade is implemented. 
The year 2009 should be taken to mean RHIC Run 9.
The weekly-luminosity expectations are based 
on RHIC Collider Accelerator Division guidance as of February 23, 2008.
The root-mean square (RMS) of the collision diamond 
is assumed to be 20 cm beginning in 2009.
We assume that 80\% of the RHIC beam is in the central bucket (and 
thus usable by experiments) prior to 2012, when the addition of a 56 MHz RF 
cavity will bring this up to nearly 100\%.

The PHENIX SVTX barrel is expected to be in place starting in 2010. The STAR
HFT is expected to be in place starting in 2012. In both cases the usable  
collision vertex range will be limited to $\pm 10$ cm by the acceptance of
the silicon upgrade detectors. Because the RHIC diamond RMS length is
$\pm 20$ cm this will result in a reduction in effective luminosity for the
experiments. This is taken into account when calculating yields.

The quarkonium cross sections are taken from Ref.~\cite{Gavai:1994in}
with an assumed $\psi^\prime$ to $J/\psi$ ratio of 0.14.  The
charmonia cross sections are reduced by a factor of 0.43 in \AuAu\ 
interactions, approximately accounting for the suppression measured by
PHENIX.  No $\Upsilon$ suppression is assumed.

\subsection{Projected PHENIX yields}

Table~\ref{rhic1_rhic2_yields_phenix} summarizes the projected PHENIX
yields per year for critical heavy flavor signals for the period 2009-2013 for 
12 week physics runs, and also includes the yields
observed in RHIC runs to date. The estimated yields are based on a number of
criteria.  The detector acceptances are from PHENIX simulations.  The
minimum bias trigger efficiency for hard processes is assumed to be
0.75 for \pp\ and 0.92 for \AuAu\ interactions.  Where appropriate, an
additional, realistic, level-1 trigger efficiency of 0.8 is used.

Realistic lepton pair reconstruction efficiencies of 0.8 in \pp\ and
0.4 in \AuAu\ collisions are used.  An additional efficiency factor of
0.4 is assumed for a 1 mm displaced vertex cut to identify $B
\rightarrow J/\psi X$ decays.  The PHENIX vertex detector is assumed
to be in place after 2010, requiring a collision vertex cut of $\pm 10$
cm.

\begin{table}[tbh]
\begin{tabular}{|c|cc||c|c|c|c|} \hline
Species & Signal & $|\eta|$ & To Date & 2009 & 2011 & 2013 \\ \hline
\pp & $J/\psi \rightarrow e^+e^-$ & $<0.35$                & $\sim$ 1,500 
& 30,000 & 29,000 & 46,000 \\
    & $J/\psi \rightarrow \mu^+\mu^-$ & $1.2-2.4$          & $\sim$ 8,000 
& 256,000 & 249,000 & 393,000 \\
    & $\psi^\prime \rightarrow e^+e^-$ & $<0.35$            & $-$    
& 540 & 530 & 830 \\
    & $\psi^\prime \rightarrow \mu^+\mu^-$ & $1.2-2.4$      & $-$    
& 4,600 & 4,500 & 7,100  \\
    & $\chi_c \rightarrow e^+e^-\gamma$ & $ <0.35$         & $-$    
& 2,000 & 1,900 & 3,000 \\
    & $\chi_c \rightarrow \mu^+\mu^-\gamma$ & $1.2-2.4$    & $-$    
&  & 37,000 & 116,000 \\
    & $\Upsilon \rightarrow e^+e^-$ & $<0.35$              & $-$ 
& 115 & 110 & 180 \\
    & $\Upsilon \rightarrow \mu^+\mu^-$ & $1.2-2.4$        &  $\sim$ 27 
& 290 & 280 & 440 \\
    & $B \rightarrow J/\psi X \rightarrow e^+e^-$ & $<0.35$  & $-$    
&  & 155 & 240  \\
    & $B \rightarrow J/\psi X \rightarrow \mu^+\mu^-$ & $1.2-2.4$ & $-$    
&  & 1,500 & 2,400 \\ \hline
\AuAu & $J/\psi \rightarrow e^+e^-$ & $<0.35$              & $\sim$ 800 
& 13,500 & 14,600 & 22,400 \\
    & $J/\psi \rightarrow \mu^+\mu^-$ & $1.2-2.4$          & $\sim$ 7,000 
& 119,000 & 129,000 & 198,000 \\
    & $\psi^\prime \rightarrow e^+e^-$ & $<0.35$            & $-$    
& 240 & 260 & 400 \\
    & $\psi^\prime \rightarrow \mu^+\mu^-$ & $1.2-2.4$      & $-$    
& 2,150 & 2,300 & 3,600 \\
    & $\chi_c \rightarrow e^+e^-\gamma$ & $ <0.35$         & $-$    
& 890 & 960 & 1,500 \\
    & $\chi_c \rightarrow \mu^+\mu^-\gamma$ & $1.2-2.4$    & $-$    
&  & 19,000 & 59,000 \\
    & $\Upsilon \rightarrow e^+e^-$ & $<0.35$              & $-$ 
& 120 & 130 & 200 \\
    & $\Upsilon \rightarrow \mu^+\mu^-$ & $1.2-2.4$       & $-$ 
& 310 & 340 & 520 \\
    & $B \rightarrow J/\psi X \rightarrow e^+e^-$ & $<0.35$ & $-$ 
&  & 190 & 290 \\
    & $B \rightarrow J/\psi X \rightarrow \mu^+\mu^-$ & $1.2-2.4$ & $-$    
&  & 1,900 & 2,900 \\ \hline
\end{tabular}
\caption{
The projected yields of several heavy flavor signals in PHENIX for 12 
week physics runs at RHIC. The yields are shown for both \pp\ and \AuAu\ 
collisions at \sqrtsNN = 200 \gev. The approximate yields 
obtained at RHIC to date are also shown. These reflect the fact that RHIC had
not yet achieved the full luminosity development for \AuAu\ by Run 4, or 
for \pp\  by Run 5. The Run 6 \pp\ and Run 7 \AuAu\ data have not yet been 
fully analyzed. The projected RHIC values for 2011 and 2013 assume that the 
PHENIX SVTX detector is in place, limiting the usable 
collision vertex range to $\pm 10$ cm. The collision diamond RMS will 
be $\pm 20$ cm.  The reduction in usable luminosity 
is compensated by improvements in signal to background ratio for most 
measurements. The $\chi_c \rightarrow \mu^+\mu^-\gamma$ measurement 
requires the NCC.  It is assumed that one NCC arm is installed for 2011 
and both are installed for 2013.}
\label{rhic1_rhic2_yields_phenix}
\end{table}

\begin{table}[tbh]
\begin{center}
\begin{tabular}{|c|cc||c|c|c|c|} \hline
Species & Signal & $|\eta|$ & To Date & 2009 & 2011 & 2013 \\ \hline
\pp & $J/\psi \rightarrow e^+e^-$  & $<1.0$ & $\sim 400$ & 490,000 & 1,100,000 & 750,000  \\
    & $\psi^\prime \rightarrow e^+e^-$   &   & $-$  & 10,400 & 24,000 & 16,000  \\
    & $\Upsilon \rightarrow e^+e^-$     &   & $\sim 170$ & 4,100 & 9,300 & 6,300  \\
    & $B \rightarrow J/\psi X \rightarrow e^+e^-$ &   & $-$ &  &  & 370 \\  \hline
\AuAu & $J/\psi \rightarrow e^+e^-$~(min. bias) &  $<1.0$  & $\sim 350 $ & 25,000 & 25,000 & 25,000 \\
      & $J/\psi_{p_T > 5.5\ \mathrm{GeV}/c} \rightarrow e^+e^-  $ &  $ {} $  & $-$  & 1,800 & 4,400 & 3,100 \\
       & ${\psi^\prime} \rightarrow e^+e^-$ (min.bias) &             & $-$ & 550 & 550 & 550 \\
      & $\Upsilon \rightarrow e^+e^-$ &        & $-$    & 5,800 & 13,800 & 9,700  \\
      & $B \rightarrow J/\psi X \rightarrow e^+e^-$ &   & $-$  &  &  & 540  \\ 
      & $D^0 \rightarrow K^\pm \pi^\mp$ (min. bias) & & 400,000 & 9,000,000 & 9,000,000 & 9,000,000 \\ 
      & {}               & & ($\frac{S}{B} \sim \frac{1}{600}$) & ($\frac{S}{B} \sim \frac{1}{600}$) 
& ($\frac{S}{B} \sim \frac{1}{600}$) & ($\frac{S}{B} \sim \frac{1}{10}$)
\\ \hline
\end{tabular}
\end{center}
\caption{The projected yields of several heavy flavor signals in STAR for a 
    physics run of 12 weeks at RHIC at \sqrtsNN =
    200 \gev.  The approximate yields obtained to date are
    also shown.  The projected values assume that the
    STAR Heavy Flavor Tracker, the full ToF barrel, and the DAQ
    upgrades are in place for 2013. From that time on the inner tracking limits the usable
    collision vertex range to $\pm 10$ cm. 
    The collision diamond RMS will be $\pm 20$ cm.
    Projections for signals marked
    (min. bias) are based on minimum bias events recorded with a rate of 500 Hz 
    and a 33\%
    combined experiment and machine duty cycle. Dedicated triggers cannot be 
    efficiently deployed for these measurements.
    The corresponding signal-to-background ratios ($S/B$) are also given
    for the $D^0$ measurements.}
\label{rhic1_rhic2_yields_star}
\end{table}

\begin{table}[hbt]
\begin{center}
\begin{tabular}{|c|cc||c|c|c|} \hline
Experiment & Signal & $|\eta|$ &  2009 & 2011 & 2013 \\ \hline
PHENIX & $J/\psi \rightarrow e^+e^-$ & $<0.35$ & 166,000 & 161,000 & 254,000 \\
    & $J/\psi \rightarrow \mu^+\mu^-$ & $1.2-2.4$ & 1,500,000 & 1,450,000 & 2,300,000 \\
    & $\psi^\prime \rightarrow e^+e^-$ & $<0.35$  & 3,000 & 2,900 & 4,600 \\
    & $\psi^\prime \rightarrow \mu^+\mu^-$ & $1.2-2.4$ & 27,000 & 26,000 & 41,000 \\
    & $\chi_c \rightarrow e^+e^-\gamma$ & $ <0.35$ & 28,000 & 27,000 & 43,000 \\
    & $\chi_c \rightarrow \mu^+\mu^-\gamma$ & $1.2-2.4$ &  & 550,000 & 1,700,000 \\
    & $\Upsilon \rightarrow e^+e^-$ & $<0.35$ & 820 & 800 & 1,250 \\
    & $\Upsilon \rightarrow \mu^+\mu^-$ & $1.2-2.4$ & 2,100 & 2,000 & 3,200 \\
    & $B \rightarrow J/\psi X \rightarrow e^+e^-$ & $<0.35$ &  & 2,000 & 3,200  \\
    & $B \rightarrow J/\psi X \rightarrow \mu^+\mu^-$ & $1.2-2.4$ &  & 20,000 & 32,000 \\ \hline
STAR & $J/\psi \rightarrow e^+e^-$ & $<1.0$ & 1,200,000 & 2,800,000 & 1,900,000 \\
    & $\psi^\prime \rightarrow e^+e^-$ & & 25,000 & 57,000 & 39,000  \\
    & $\Upsilon \rightarrow e^+e^-$ &  & 31,000 & 70,000 & 47,000  \\
    & $B \rightarrow J/\psi X \rightarrow e^+e^-$ & &  &  & 4,100  \\  \hline
\end{tabular}
\end{center}
\caption{
Projected heavy flavor yields in PHENIX and STAR for a 12 week $\sqrt{s} = 
500$ \gev\ \pp\ run at RHIC. The projected PHENIX yields for 2011 and 2013 
assume that the PHENIX SVTX detector is in place, limiting the usable 
collision vertex range to $\pm 10$ cm. The STAR HFT is assumed to be in 
place from 2012, and it also will limit the usable 
collision vertex range to $\pm 10$ cm. The collision diamond RMS 
will be $\pm 20$ cm.  The $\chi_c \rightarrow \mu^+\mu^-\gamma$ measurement 
requires the NCC.  It is assumed that one NCC arm is installed for 2011 
and both are installed for 2013.}
\label{rhic1_rhic2_yields_500gev}
\end{table}

\begin{table}[hbtp]
\begin{center}
\begin{tabular}{|c|cc|cc|cc|} \hline
Signal & ALICE & $|\eta|$ & CMS &  $|\eta|$ & ATLAS &  $|\eta|$  
\\ \hline
$J/\psi \rightarrow \mu^+\mu^-$ &  677,000 & $2.5-4$  & 184,000 & $< 2.4$ & 
$8,000- 100,000$ & $< 2.5$ \\
$J/\psi \rightarrow e^+e^-$ &  121,100 & $< 0.9$  &  &  &  &  \\
$\psi^\prime \rightarrow \mu^+\mu^-$ &  18,900 & $2.5-4$  &  &  & 
$1,400-1,800$  & $<2.5$   \\
$\psi^\prime \rightarrow e^+e^-$ &   &  &  &  & &  \\
$\Upsilon \rightarrow \mu^+\mu^-$ &  9,600 & $2.5-4$  & 37,700 & $< 2.4$ 
& 15,000~(21,200) & $< 2.0$~($< 2.5$) \\
$\Upsilon \rightarrow e^+e^-$ &  1,800 & $< 0.9$  & &  &  &  \\
$D^0 \rightarrow K^\pm \pi^\mp$ &  13,000 & $<0.9$  &  & &  &  \\ \hline
\end{tabular}
\end{center}
\caption{The estimated LHC heavy flavor yields for a
$10^6$ s \PbPb\ run at \sqrtsNN\ = 5.5 TeV
with 500 $\mu$b$^{-1}$ integrated 
luminosity (one year), reported by the LHC experiments.
As for the RHIC tables, 
the $\Upsilon$ rates include all three states. The ALICE 
yields~\cite{ALICE_yield_estimates} assume binary collision scaling and 
include shadowing. The ALICE dielectron yield estimates 
are for the 10\% most central events while the $D^0$ yields are for 
$10^7$ central event triggers.  The CMS yields~\cite{CMS_yield_estimates} 
assume \dnchdeta\ = 2500 and include shadowing with no quarkonia suppression. 
For ATLAS~\cite{ATLAS_yield_estimates} the range of the $J/\psi$ 
yield corresponds to different assumed muon \pT\ trigger thresholds. Good 
separation of $\Upsilon$ 
and $\Upsilon'$ states is expected in ATLAS for  $|\eta| < 2.0$.}
\label{LHC_yields_all}
\end{table}

\subsection{Projected STAR yields}

Table~\ref{rhic1_rhic2_yields_star} shows a summary of the STAR projected
yields per year for various critical heavy flavor signals for RHIC and RHIC II.
The detector acceptances and efficiencies are from STAR simulations,
or, where available, derived from existing measurements. The
assumptions of luminosity, interaction vertices, cross sections and
suppression effects are identical to those used for the PHENIX projections
discussed above.

The upgrade of the inner tracking system is central to the future STAR 
heavy flavor program. It will provide sufficient tracking
resolution to determine the heavy
flavor meson decay vertices, providing the  means to virtually unambiguously
identify these probes.  However, the complexity and volume of the data from
this subsystem precludes the use of dedicated triggers that are based
on the displaced vertex identification. Certain
measurements, such as $D^0 \rightarrow K^\pm \pi^\mp$, can only be
made by analyzing large quantities of minimum bias data.  The
substantial increase in readout speed from the DAQ upgrade (DAQ1000)
can record events with a rate of up to 1 kHz, {\it i.e.}, 10 times
the current rate. In the projections discussed here, we
assume a more conservative rate of 500 Hz. In order to minimize tape
costs, STAR plans to filter these events offline in computing farms
either before or after transfer to the RHIC Computing Facility (RCF).
The $D^0$ yields reflected in Table~\ref{rhic1_rhic2_yields_star} are
therefore limited not by luminosity but by \textit{bandwidth}.

The same bandwidth limitation holds for the $J/\psi$ trigger in \AA\ 
collision. The STAR $J/\psi$ trigger deployed in \pp\ events is based on the
coincident signal above threshold of two calorimeter towers in
conjunction with signals in the referring time-of-flight (ToF)
segments to veto photon showers. The large multiplicity
in \AA\ events, however, renders this scheme impossible. The $J/\psi$ \AA\
yields are therefore bandwidth limited while those in \pp\ 
depend directly on the delivered luminosity. However, detailed studies 
from recent runs indicate that a slightly modified $J/\psi$ trigger 
can be effectively deployed for $J/\psi$'s with $\pT > 5 - 6\,\gevc$. 
Projections for this specific trigger are shown in 
Table~\ref{rhic1_rhic2_yields_star}.

The bottom measurement, $B \rightarrow J/\psi X \rightarrow
e^+e^-$, is based on a high-\pT\ $J/\psi$ trigger, thereby avoiding
the issues inherent to the low-\pT\ trigger.  The projections shown in
Table~\ref{rhic1_rhic2_yields_star} are based on preliminary results
from high-\pT\ $J/\psi$ measurements in $pp$, folding in the expected
efficiencies of the inner tracking system. The \pT-dependent ratio of
$J/\psi$ from $B$ decays to directly produced $J/\psi$ was
derived from calculations described later.  No bottom quark energy
loss was assumed for the \AuAu\ yields.

The decay electrons from $\Upsilon \rightarrow e^+e^-$ are at
considerably higher \pT\ than those of the $J/\psi$, making the
calorimeter-based $\Upsilon$ trigger highly efficient ($\epsilon > 90\%$). The
combination of an efficient trigger and large acceptance makes the study
of the bottomonium in \pp\ and \AA\ the main point of the STAR
quarkonium program and complements PHENIX's strong charmonium
program.

\subsection{Yields at higher energies}

Table~\ref{rhic1_rhic2_yields_500gev} summarizes the expected 
PHENIX and STAR heavy flavor yields per year in $pp$ collisions
at $\sqrts = 500$ \gev. Although not 
directly comparable with heavy-ion yields from collisions at \sqrtsNN\ = 
200 \gev, the order of magnitude larger heavy flavor yields at 500 \gev\ 
should help in the understanding of the reaction
mechanism in \pp\ collisions.

Table~\ref{LHC_yields_all} contains a summary of the projected yields per year 
from 
the LHC detector collaborations for various critical heavy flavor signals in 
$10^6$ s of data taking at \sqrtsNN\ = 5.5 TeV \PbPb\ run, the standard 
planning number for a one year run 
\cite{ALICE_yield_estimates,CMS_yield_estimates,ATLAS_yield_estimates}. 
The ALICE muon spectrometer covers only one side of forward rapidity. 

Comparison of 
Tables~\ref{rhic1_rhic2_yields_phenix},~\ref{rhic1_rhic2_yields_star} 
and ~\ref{LHC_yields_all} reveal that the projected heavy flavor yields 
for one year of running are similar at the LHC and at RHIC II. The much 
larger heavy flavor cross sections at
the higher LHC energy are largely compensated at RHIC II by integrated 
luminosities that result from three times longer runs and a factor of 10 
higher luminosity. 


\section{Open heavy flavor}

In this section we present a more detailed discussion of the
theoretical motivation for studying open heavy flavor in heavy-ion
collisions, of the present experimental and theoretical status, and of
the proposed experimental program of open heavy flavor measurements at
RHIC II.

As described in the introduction, dense matter effects in nuclear
collisions may change the kinematic distributions and the total cross
sections of open heavy flavor production.  Effects such as energy loss
and flow can significantly modify the heavy flavor $p_T$ distributions
but do not, in fact, change the total yields.  In a finite acceptance
detector, however, the measured yields may appear to be enhanced or
suppressed, depending on the acceptance.  Energy loss steepens the
slope of the heavy flavor $p_T$ distribution because the heavy quark
$p_T$ is reduced.  If the momentum is reduced sufficiently for the
quarks to be stopped within the medium, the heavy quarks can take the
same velocity as the surrounding medium and `go with the flow'.  The
present RHIC results on $R_{AA}$ and $v_2$ for heavy flavor decays to
leptons show that these effects are indeed important for charm quarks and 
even for bottom quarks.
However, higher $p_T$ measurements and reconstructed charm hadrons are
needed to solidify and quantify the results.  In addition, clean
separation of leptons from charm and bottom decays is necessary to
provide a sensitive measurement of the electron $R_{AA}$ from bottom decays.

Effects that may modify the total heavy flavor yields are the initial
parton distributions in the nuclei and secondary charm production in the 
medium.  The parton distribution functions needed for perturbative QCD 
calculations of heavy flavor production are modified in the nucleus, as
was observed in nuclear deep-inelastic scattering \cite{Arneodo}.  
At very small momentum
fractions, $x$, the gluon fields may be treated as classical color fields.
The modifications of the parton distributions in nuclei relative to free
protons would affect the total yields.  The effect is expected to be small
at midrapidity and moderate $p_T$ at RHIC but is likely to be more important
at large rapidity where lower $x$ values are probed.  Although thermal charm
production from the medium is likely to be small at RHIC energies, it could
moderately enhance the total yields.

Since the $J/\psi$ yields may be enhanced in nuclear collisions by
coalescence of uncorrelated $c$ and $\overline c$ quarks in the medium,
it is important for charmonium production in heavy-ion collisions to be 
properly normalized.  The ratio of $J/\psi$ to open charm production in \pp\
collisions is not a strong function of energy.  Thus the total charm yield
sets the scale against which $J/\psi$ suppression relative to enhancement
can be quantified, as discussed in more detail in Section~5.

\subsection{Open Heavy Flavor Theory}

\subsubsection{Theoretical Baseline Results}
\label{sec:theoryHFbaseline}
We now discuss the most recent theoretical baseline calculations of the 
transverse momentum distributions of
charm and bottom quarks, the charm and bottom hadron distributions
resulting from fragmentation and,
finally, the electrons produced in semileptonic decays of the hadrons
\cite{CNV}.
 
The theoretical prediction of the electron spectrum includes
three  main components: the $p_T$ and rapidity
distributions of the heavy quark $Q$ in \pp\ collisions at $\sqrt{s} =
200$~\gev, calculated in perturbative QCD; fragmentation of the
heavy quarks into heavy hadrons, $H$, described by
phenomenological input extracted from $e^+e^-$ data; and the decay of
$H$ into electrons according to spectra available from other
measurements. This
cross section is schematically written as
\begin{eqnarray}
\frac{E d^3\sigma(e)}{dp^3} &=& \frac{E_Q d^3\sigma(Q)}{dp^3_Q} \otimes 
D(Q\to H) \otimes f(H \to e) 
\end{eqnarray}
where the symbol $\otimes$ denotes a generic convolution. The electron decay
spectrum, $f(H \to e)$, includes the branching ratios.

\begin{figure}[tbh]
    \centering\includegraphics[width=0.95\textwidth]{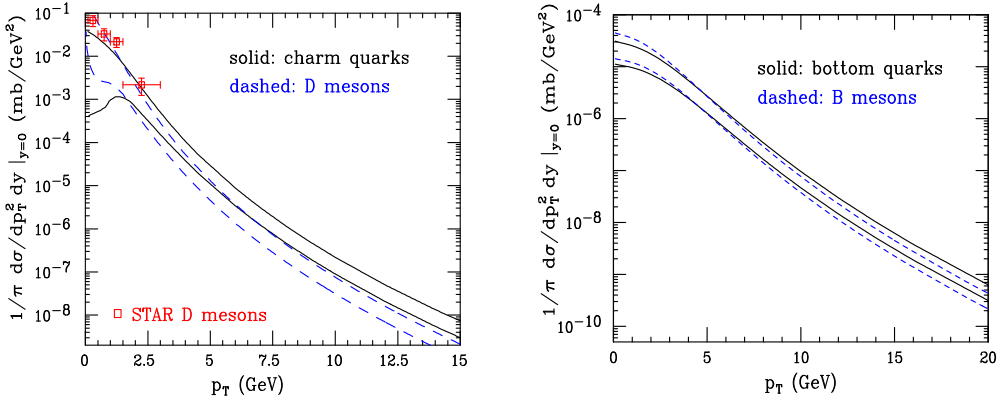}
    \caption{Left-hand side: The theoretical uncertainty bands for $c$
        quark and $D$ meson $p_T$ distributions in \pp\ collisions at
        $\sqrt{s} = 200$~\gev, using BR($c \to D$) = 1. The
        final~\protect\cite{Adams:2004fc} STAR \dAu\ data (scaled
        to \pp\ using \nbin\ = 7.5) are also shown.  Right-hand side:
        The same for $b$ quarks and $B$ mesons. Modified from 
        Ref.~\protect\cite{rvqm05}, reprinted with permission from Elsevier.}
    \label{qQ}
\end{figure}
 
The distribution $E d^3\sigma(Q)/dp^3_Q$ is evaluated at Fixed-Order
plus Next-to-Leading-Log (FONLL) level~\cite{Cacciari:1998it}.  In
addition to including the full fixed-order NLO
result~\cite{Nason:1987xz,Beenakker:1990ma}, the FONLL calculation
also resums~\cite{Cacciari:1993mq} large perturbative terms
proportional to $\alpha_s^n\log^k(p_T/m_Q)$ to all orders with
next-to-leading logarithmic (NLL) accuracy ({\it i.e.} $k=n,\,n-1$) where
$m$ is the heavy quark mass. The perturbative parameters are the heavy
quark mass, $m_Q$, and the value of the strong coupling, $\alpha_s$.
Since the FONLL calculation treats the heavy quark as an active light
flavor at $p_T >> m$, the number of light flavors used to calculate
$\alpha_s$ includes the heavy quark, i.e. $n_{\rm lf} + 1$ where, for
charm, $n_{\rm lf} = 3$ ($u$, $d$ and $s$).  The same number of
flavors, $n_{\rm lf} + 1$, is also used in the fixed-order scheme
where the quark mass is finite.  The central heavy quark masses are
$m_c = 1.5$~\gevcc\ and $m_b = 4.75$~\gevcc. The masses are varied in
the range $1.3< m_c <1.7$~\gevcc\ for charm and $4.5<m_b< 5$ \gevcc\ 
for bottom to estimate the mass uncertainties. The five-flavor QCD
scale is the CTEQ6M value, $\Lambda^{(5)} = 0.226$ \gev.  The
perturbative calculation also depends on the factorization ($\mu_F$)
and renormalization ($\mu_R$) scales.  The scale sensitivity, a
measure of the perturbative uncertainty, is calculated using
$\mu_{R,F}^2 = \mu_0^2 = p_T^2 + m_Q^2$ as the central value while
varying $\mu_F$ and $\mu_R$ independently within a `fiducial' region
defined by $\mu_{R,F} = \xi_{R,F}\mu_0$ with $0.5 \le \xi_{R,F} \le 2$
and $0.5 \le \xi_R/\xi_F \le 2$ so that $\{(\xi_R,\xi_F)\}$ = \{(1,1),
(2,2), (0.5,0.5), (1,0.5), (2,1), (0.5,1), (1,2)\}.  The envelope
containing the resulting curves defines the uncertainty.  The mass and
scale uncertainties are added in quadrature so that
\begin{eqnarray}
 \frac{d\sigma_{\rm max}}{dp_T} & = &
\frac{d\sigma_C}{dp_T}
+ \sqrt{\Bigg(\frac{d\sigma_{\mu ,{\rm max}}}{dp_T} -
\frac{d\sigma_C}{dp_T}\Bigg)^2
+ \Bigg(\frac{d\sigma_{m, {\rm max}}}{dp_T} -
\frac{d\sigma_C}{dp_T}\Bigg)^2} \label{fonllmax} \\
\frac{d\sigma_{\rm min}}{dp_T} & = & \frac{d\sigma_C}{dp_T}
- \sqrt{\Bigg(\frac{d\sigma_{\mu ,{\rm min}}}{dp_T}
- \frac{d\sigma_C}{dp_T}\Bigg)^2
+ \Bigg(\frac{d\sigma_{m, {\rm min}}}{dp_T}
- \frac{d\sigma_C}{dp_T}\Bigg)^2} \label{fonllmin} \,\, .
\end{eqnarray}
where $C$ is the distribution for the central value,
$\mu$, max ($\mu$, min) is the maximum (minimum) cross section
obtained by choosing the central value with the scale factors 
in our seven fiducial sets,
and $m$, max ($m$, min) is the maximum (minimum) cross section obtained
with $\xi_R=\xi_F=1$ and the lower and upper limits on the quark mass 
respectively.

This range of inputs leads to a FONLL total $c\bar c$ cross section
(with $n_{\rm lf} + 1$ active flavors) in \pp\ 
collisions of $\sigma_{c\bar c}^{\rm FONLL} =
256^{+400}_{-146}$~$\mu$b at $\sqrt{s} = 200$~\gev.  The theoretical
uncertainty is evaluated as described above. The corresponding NLO
prediction is $244^{+381}_{-134}$~$\mu$b.  The predictions in
Ref.~\cite{Vogt:2001nh}, using $m_c = 1.2$~\gevcc\ and $\mu_R = \mu_F
= 2\mu_0$ gives $\sigma_{c\bar c}^{\rm NLO} = 427$~$\mu$b, within the
uncertainties.  Since the FONLL and NLO calculations tend to coincide
at small $p_T$, which dominates the total cross section, the two
results are very similar.  Thus the two calculations are equivalent at
the total cross section level, within the large perturbative
uncertainties.  The total cross section for bottom production is
$\sigma_{b\bar b}^{\rm FONLL} = 1.87^{+0.99}_{-0.67}$~$\mu$b.
 
When the total $c \overline c$ cross section is calculated with the
next-to-leading order scaling functions of Ref.~\cite{Nason:1987xz}, it does
not depend on any kinematic variables, only on the quark mass, $m$,
and the renormalization and factorization scales with central value
$\mu_{R,F} = \mu_0 = m$.  The heavy quark is always considered massive
in the calculation of the total cross section and is thus not an
active flavor in the production calculation.  Therefore, the number of
light quark flavors, $n_{\rm lf}$, does not include the heavy quark.
With $n_{\rm lf}$ light flavors and a fixed scale, the charm and
bottom NLO total cross sections at $\sqrt{s} = 200$ GeV are $\sigma_{c
    \overline c}^{\rm NLO_{n_{\rm lf}}} = 301^{+1000}_{-210} \, \,
\mu{\rm b}$ and $\sigma_{b \overline b}^{\rm NLO_{n_{\rm lf}}} =
2.06^{+1.25}_{-0.81} \, \, \mu{\rm b}$, respectively \cite{rvjoszo}.  While the
central values are only about 25\% and 10\% higher respectively than
the FONLL results due to a fixed rather than running scale in
$\alpha_s$, the uncertainty is considerably larger, especially for
charm.  The larger upper limit on the uncertainty is primarily due to
the behavior of the CTEQ6M gluon density at factorization scales below
the minimum scale of the parton density and the larger value of
$\alpha_s$ for $n_{\rm lf}$ light flavors \cite{rvjoszo}.  Other parton 
densities, with lower initial scales and smaller $\Lambda_{\rm QCD}$ give
a reduced upper limit.  When both
the NLO and FONLL calculations use the same number of light flavors,
the results for the total cross section and its uncertainty are in
agreement \cite{Matteopriv}.

The fragmentation functions, $D(c\to D)$ and $D(b\to B)$, where $D$
and $B$ indicate a generic admixture of charm and bottom hadrons, are
consistently extracted from $e^+e^-$ data in the FONLL
context~\cite{Cacciari:2002pa}.

The measured spectra for primary $B\to e$ and $D \to e$ decays are
assumed to be equal for all bottom and charm hadrons, respectively.
The contribution of electrons from secondary $B$ decays, $B\to D\to
e$, was obtained by convoluting the $D\to e$ spectrum with a
parton-model prediction of $b\to c$ decay.  The resulting electron
spectrum is very soft, giving a negligible contribution to the total.
The decay spectra are normalized using the branching ratios for bottom
and charm hadron mixtures \cite{Eidelman:2004wy}: BR$(B\to e) = 10.86
\pm 0.35$\%, BR$(D\to e) = 10.3 \pm 1.2$\%, and BR$(B\to D\to e) = 9.6
\pm 0.6$\%.

\begin{figure}[tbh]
    \centering\includegraphics[width=0.95\textwidth]{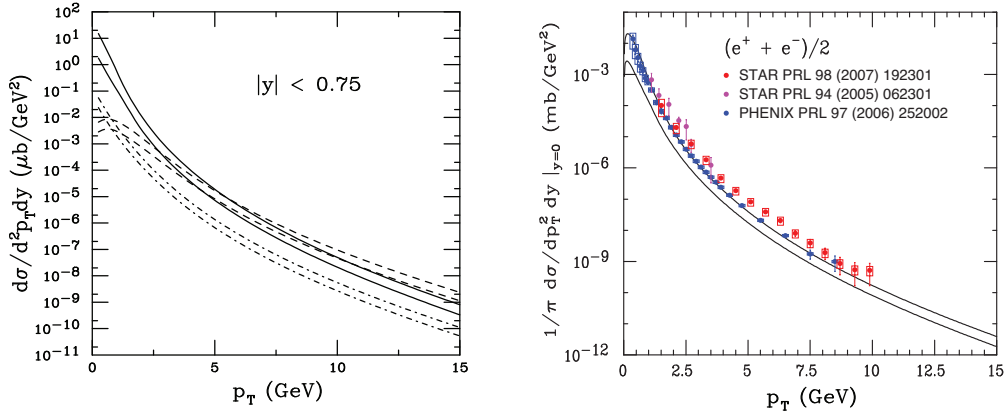}
    \caption{Left-hand side: The theoretical uncertainty bands for $D
        \to e$ (solid), $B \to e$ (dashed) and $B\to D \to e$
        (dot-dashed) as a function of $p_T$ in $\sqrt{s} = 200$ \gev\
        \pp\ collisions for $|y| < 0.75$.  Right-hand side: The final
        electron uncertainty band in \pp\ collisions is compared to
        the PHENIX~\protect\cite{Adare:2006hc} and 
        STAR~\protect\cite{Adams:2004fc,Abelev:2006db} data.  Modified from
        Ref.~\protect\cite{rvqm05}, reprinted with permission from Elsevier.}
    \label{electrons}
\end{figure}

The left-hand side of Fig.~\ref{qQ} shows the theoretical uncertainty
bands for $c$ quarks and $D$ mesons, obtained by summing the mass and
scale uncertainties in quadrature.  The band is broader at low $p_T$
due to the large value of $\alpha_s$ and the behavior of the CTEQ6M
parton densities at low scales as well as the increased sensitivity of
the cross section to the charm quark mass.  The rather hard
fragmentation function causes the $D$ meson and $c$ quark bands to
separate only at $p_T > 9$~\gevc.  The right-hand side of
Fig.~\ref{qQ} shows the same results for $b$ quarks and $B$ mesons.
The harder $b\to B$ fragmentation function causes the two bands to
partially overlap until $p_T \simeq 20$~\gevc.
 
Figure~\ref{electrons} shows the individual uncertainty bands for the
$D\to e$, $B \to e$ and $B\to D \to e$ decays to electrons on the
left-hand side and compares the RHIC data to the total band on the
right-hand side.  The upper and lower limits of the total band are
obtained by summing the upper and lower limits of each component.  The
secondary $B\to D\to e$ spectrum is extremely soft, only exceeding the
primary $B\to e$ decays at $p_T<1$ \gevc.  It is always negligible
with respect to the total yield. While, for the central parameter
sets, the $B \to e$ decays begin to dominate the $D \to e$ decays at
$p_T \simeq 4$~\gevc, a comparison of the bands shows that the
crossover may occur over a rather broad range of electron $p_T$,
assuming that the two bands are uncorrelated.


\subsection{Models of Heavy Quark Energy Loss}

While the heavy quarks are in the medium, they can undergo energy loss
by two means: elastic collisions with light partons in the system
(collisional) and gluon bremsstrahlung (radiative).  We will briefly
review some of the predicted results for $-dE/dx$ of heavy quarks for
both collisional and radiative loss.  We then show the predicted
effect on the charm and bottom contributions to single electrons at
RHIC \cite{DGVW}.

The collisional energy loss of heavy quarks through processes such as
$Qg \rightarrow Qg$ and $Qq \rightarrow Qq$ depends logarithmically on
the extremes of the heavy quark momentum, $-dE/dx \propto \ln(q_{\rm max}/q_{\rm
    min})$.  Treatments of the collisional loss vary with the values
assumed or calculated for the cutoffs.  These cutoffs are sensitive to
the energy of the heavy quark and the temperature and strong coupling
constant in the medium.  Thus the quoted value of the energy loss is
usually for a certain energy and temperature.  The calculation was
first done by Bjorken \cite{Bjorken:1982tu} who found $-dE/dx \approx
0.2$ \gev/fm for a 20 \gevc\ quark at $T = 250$ MeV.  Further work
refined the calculations of the cutoffs
\cite{Thoma:1990fm,Thomas:1991ea,Mrowczynski:da}, with similar
results.  Braaten and Thoma calculated the collisional loss in the
limits $E \ll m_Q^2/T$ and $E \gg m_Q^2/T$ in the hard thermal loop
approximation, removing the cutoff ambiguities.  They obtained $-dE/dx
\approx 0.3$ \gev/fm for a 20 \gevc\ charm quark and 0.15 \gev/fm for
a 20 \gev\ bottom quark at $T = 250$ MeV \cite{Braaten:1991we}.

Other models of heavy quark energy loss were presented in the context
of $J/\psi$ suppression: Could a produced $c \overline c$ pair stay
together in the medium long enough to form a $J/\psi$?  Svetitsky
\cite{Svetitsky:gq} calculated the effects of diffusion and drag on
the $c \overline c$ pair in the Boltzmann approach and found a strong
effect.  The drag\footnote{The drag coefficient $A(p^2)$ is related to
the energy loss per unit length by $A(p^2) = (-dE/dx)/p^2$.}
stopped the $c \overline c$ pair after traveling about 1 fm but
Brownian diffusion drove the quarks apart quickly.  The diffusion
effect increased at later times.  Svetitsky essentially predicted that
the heavy quarks would be stopped and then go with the flow.  His
later calculations of $D$ meson breakup and rehadronization
\cite{Svetitsky:1996nj} while moving through plasma droplets reached a
similar conclusion.  Koike and Matsui calculated energy loss of a
color dipole moving through a plasma using kinetic theory and found
$-dE/dx \sim 0.4-1.0$ \gev/fm for a 10 \gevc\ $Q \overline Q$
\cite{Koike:xs}.  The collisional loss was thus predicted to be small,
less than 1 \gev/fm for reasonable assumptions of the temperature.
The loss increases with energy and temperature.  Using the hard
thermal loop approach, Mustafa {\it et al.} found $-dE/dx \approx 1-2$
\gev/fm for a 20 \gevc\ quark at $T = 500$ MeV \cite{Mustafa:1997pm}.

The first application of radiative loss to heavy quarks was perhaps by
Mustafa {\it et al.}  \cite{Mustafa:1997pm}.  They included the
effects of only a single scattering/gluon emission, $Q q \rightarrow
Qqg$ or $Q g \rightarrow Qgg$.  In this case, the loss grows as the
square of the momentum logarithm, $\ln^2(q_{\rm max}/q_{\rm min})$,
one power more than the collisional loss, but is of the same order in
the strong coupling constant \cite{Braaten:1991we}.  Thus the
radiative loss is guaranteed to be larger than the collisional in this
approximation.  The heavy quark mass enters these expressions only in
the definition of $q_{\rm max}$ so that the mass dependence of the
energy loss is rather weak.  They found, for a 20 \gevc\ quark at $T =
500$ MeV, $-dE/dx \approx 12$ \gev/fm for charm and 10 \gev/fm for
bottom.

These large values suggested that energy loss could be quite important
for heavy quarks.  If true, there would be a strong effect on the $Q
\overline Q$ contribution to the dilepton continuum.  Shuryak
\cite{Shuryak:1996gc} was the first to consider this possibility for
\AA\ collisions.  He assumed that low mass $Q \overline Q$ pairs would
be stopped in the medium, substantially suppressing the dilepton
contribution from these decays.  Heavy quarks are piled up at low
$p_T$ and midrapidity if stopped completely.  However, stopped heavy
quarks should at least expand with the medium rather than coming to
rest, as discussed by Svetitsky \cite{Svetitsky:gq}.  Lin {\it et al.}
then calculated the effects of energy loss at RHIC, including thermal
fluctuations, for constant $-dE/dx = 0.5-2$ \gev/fm \cite{Lin:1997cn}.
These results showed that heavy quark contributions to the dilepton
continuum would be reduced but not completely suppressed.  In any
case, the energy loss does not affect the total cross section.

Dokshitzer and Kharzeev pointed out that soft gluon radiation from
heavy quarks is suppressed at angles less than $\theta_0 = m_Q/E$
\cite{Dokshitzer:2001zm}.  Thus bremsstrahlung is suppressed for heavy
quarks relative to light quarks by the factor $(1 +
\theta_0^2/\theta^2)^{-2}$, the `dead cone' phenomenon.  The radiative
energy loss of heavy quarks could then be quite small.  However,
Armesto {\it et al.} \cite{Armesto:2003jh} later showed that
medium-induced gluon radiation could `fill the dead cone', leading to
non-negligible energy loss for heavy flavors.  They also found that
the energy loss would be larger for charm than bottom quarks.

So far the RHIC heavy-ion measurements are not for heavy flavored
hadrons but for the electrons from their semileptonic decays.  If the
effects of energy loss are substantially different for charm and
bottom quarks, then the results in Fig.~\ref{electrons} which show
that, at high $p_T$, the single electron spectrum is dominated by $b$
decays, would suggest that, if charm quarks lose more energy than
bottom quarks, $b$-quark dominance of the electron spectra would begin
at smaller values of electron $p_T$ in \AA\ collisions.  This would,
in turn, limit the electron suppression factor, $R_{AA}$, at moderate
$p_T$ since the large bottom contribution would make $R_{AA}$ larger
than expected if the spectrum arose primarily from charm quark decays.
 
\begin{figure}[t] 
\vspace*{5.5cm}

\includegraphics{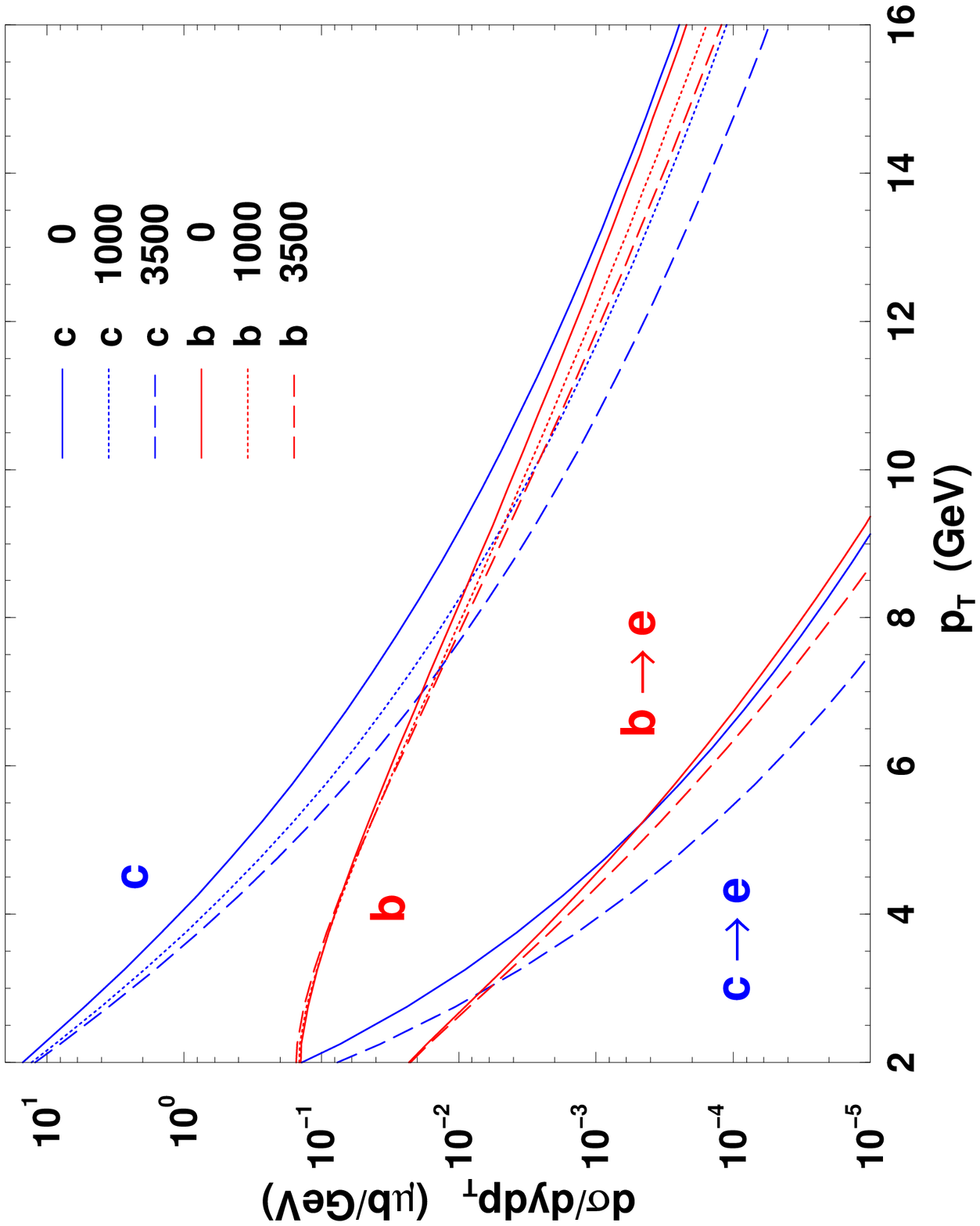} 
  \includegraphics{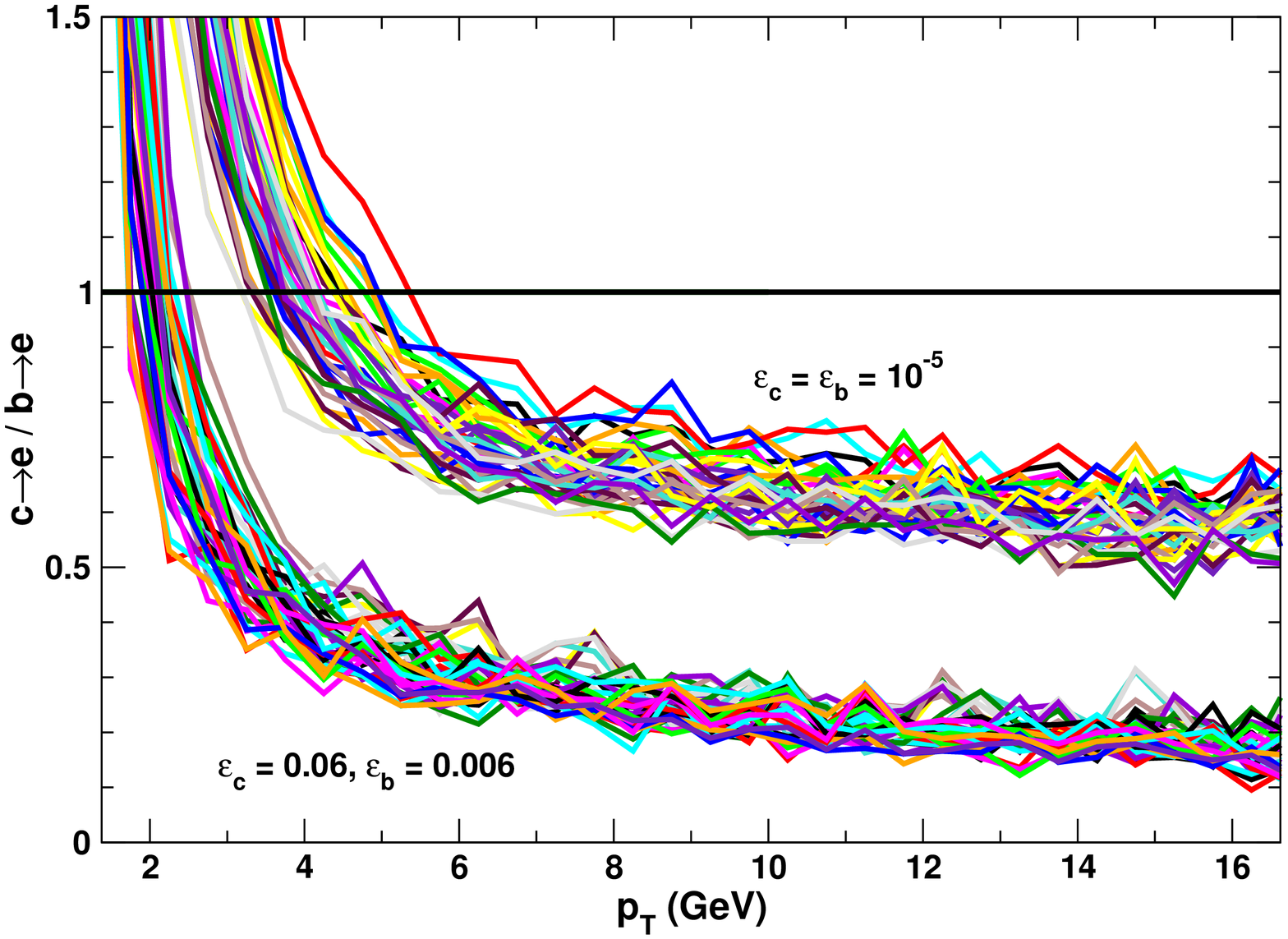}
\vspace*{0.7cm}
 \begin{tabular}{cc}\end{tabular}
  \caption{\label{fig:Pt_dist} Left-hand side:  The differential cross section
    (per nucleon pair) of charm and bottom quarks calculated to NLO in QCD
    {\protect~\cite{CNV}} compared to single electron distributions
    calculated with the fragmentation and decay scheme of
    Ref.~{\protect \cite{CNV}}. The solid, dotted and long dashed
    curves show the
    effects of heavy quark energy loss with initial gluon rapidity densities of
    $dN_g/dy=0,1000,\; {\rm{and}}\; 3500$, respectively.
    Right-hand side: The ratio of charm to bottom decays to
    electrons obtained by varying the quark masses and scale factors.
    The effect of changing the Peterson function
    parameters from $\epsilon_c = 0.06$, $\epsilon_b = 0.006$ (lower band) to
    $\epsilon_c = \epsilon_b = 10^{-5}$ (upper band) is also illustrated
    for correlated $b$ and $c$ scales.
    From Ref.~\protect\cite{DGVW}, reprinted with permission from Elsevier.}
\end{figure}

\begin{figure}[bht] 
 \begin{tabular}{cc}
  \includegraphics[width=0.4\textwidth,angle=-90]{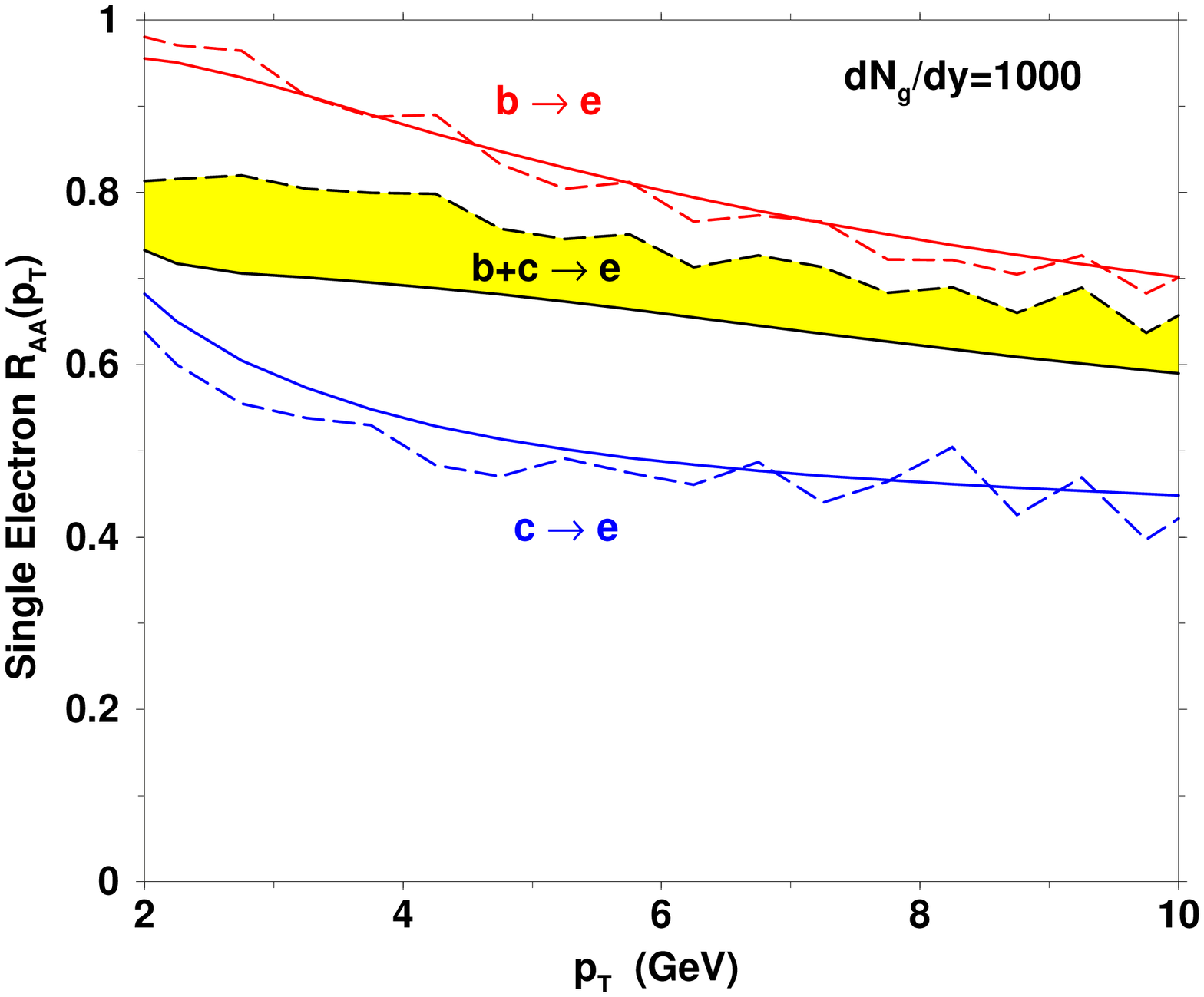} &
  \includegraphics[width=0.4\textwidth,angle=-90]{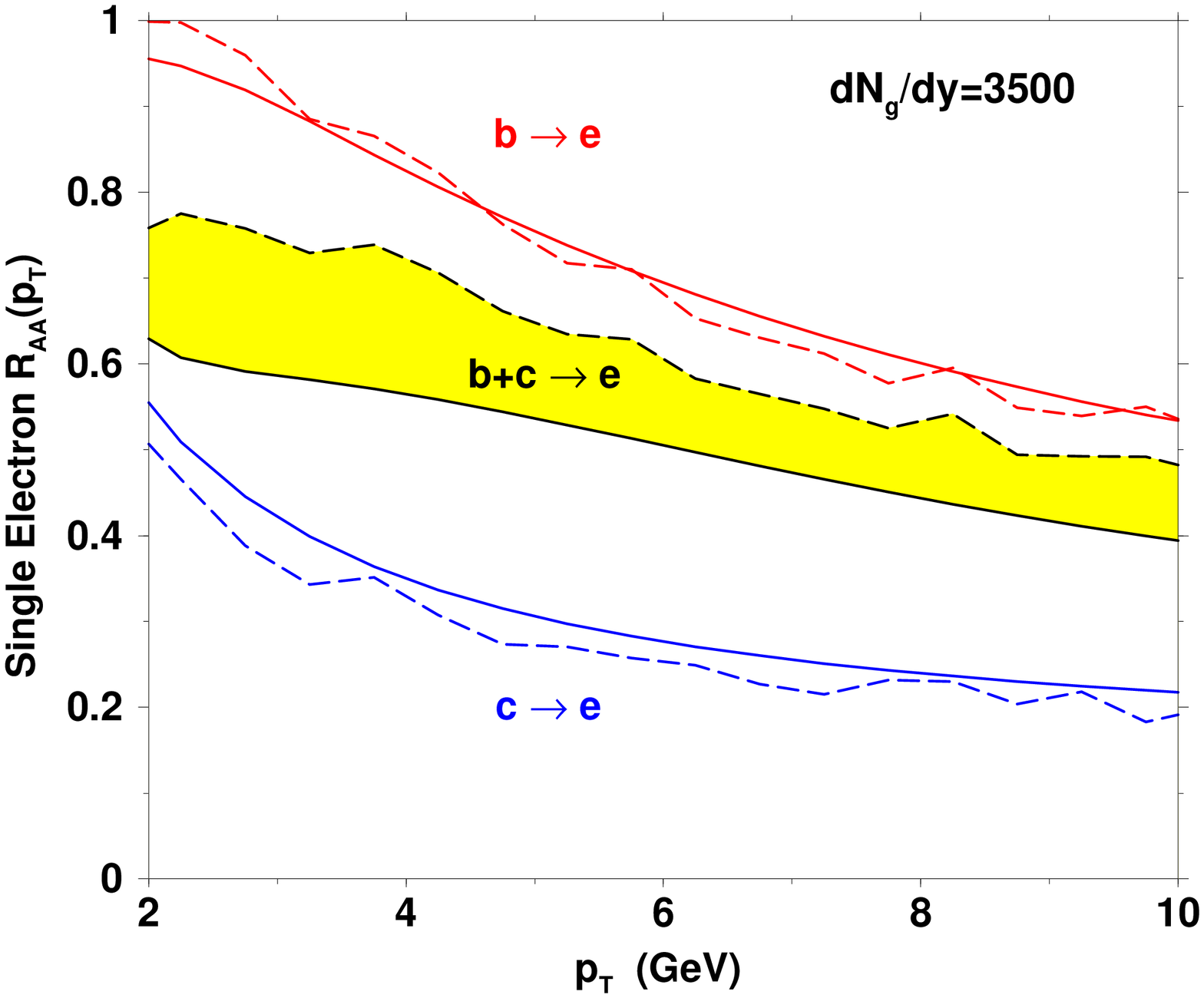}
\end{tabular}
    \caption{\label{fig:eRAA} Single electron attenuation pattern for
      $dN_g/dy=1000$, left,  and $dN_g/dy=3500$, right. The solid curves
      employ the fragmentation scheme and lepton decay parameterizations of
      Ref.~{\protect \cite{CNV}} while the dashed curves use the
      {Peterson function with $\epsilon_c = 0.06$ and $\epsilon_b = 0.006$} 
      and the decay to leptons employed by the PYTHIA Monte Carlo.
      Even for the extreme case on  the right, the less quenched $b$ 
      quarks dilute $R_{AA}$ so much that
      the modification of the combined electron yield from both $c$ and $b$
      decays does not fall below $\sim 0.5-0.6$ near $p_T\sim 5$ \gevc. From 
      Ref.~\protect\cite{DGVW}, reprinted with permission from Elsevier.}
\end{figure}

The left-hand side of Fig.~\ref{fig:Pt_dist} is an example calculation 
comparing the $c$ and $b$
distributions at midrapidity, as well as their contributions to single
electrons.  Electrons from bottom decays dominate the single
electron spectra at $p_T\sim 5$~\gevc\ for all gluon rapidity
densities. This conclusion is further supported by the right-hand side
of Fig.~\ref{fig:Pt_dist} where the ratio of charm relative to bottom
decays to electrons is shown.  The crossover region here is rather
narrow because $\mu_F$ and $\mu_R$ for $c$ and $b$ are are assumed 
to be correlated while they are uncorrelated in Fig.~\ref{electrons}.  
In all cases, the bottom
contribution to single electrons is large and cannot be neglected in
the computation of single electron suppression, shown in
Fig.~\ref{fig:eRAA}.  Since bottom energy loss is greatly reduced
relative to charm \cite{DGVW}, the possible effect on the electron
spectrum is reduced, leading to $R_{AA}(p_T<6~\gevc;~e)>0.5 \pm 0.1$.
A calculation by Armesto {\it et al.} \cite{Armesto}, with a somewhat
different model of energy loss, showed similar results to those in
Fig.~\ref{fig:eRAA}.

Djordjevic and Heinz found that finite-size effects on radiative energy loss 
in a dynamical QCD medium induce a nonlinear path length dependence of
the energy loss \cite{magda_uli}.  The light quark and charm quark loss is
similar at RHIC while nonlinearities only become large for bottom quarks in
LHC conditions.
Recent calculations by Wicks {\it et al.} \cite{WHDG} 
have revisited the importance of
elastic energy loss and have shown that this component may make a
larger contribution to the suppression factor than previously
expected.

A calculation by Adil and Vitev \cite{adil_vitev} of collisional
dissociation of heavy flavor hadrons suggests that the $D$ and $B$ meson
attenuation in a quark-gluon plasma are similar.  The $B$ dissociation rate 
is much faster than the $D$ dissociation rate due to its shorter formation
time so that it can undergo more fragmentation/dissociation cycles in the
plasma than the $D$, resulting in a similar $R_{AA}$ for $D$ and $B$ mesons
and relatively good agreement with the measured nonphotonic electron $R_{AA}$.
The heavy quark flow has not been calculated in this approach. 

Moore and Teaney \cite{mortea} and Rapp {\it et
al.}~\cite{rapp_langevin_QM05,vHGR} calculated $R_{AA}$ and
the non-photonic electron elliptic flow, $v_2$, in a Langevin model of
the time evolution of heavy quarks in the medium.  Both these groups
emphasize that elastic (collisional) energy loss should be important
at low $p_T$ relative to radiative loss since the boost for heavy
flavor hadrons in the medium should not be large.  Both also find a
strong correlation between $R_{AA}$ and $v_2$.  Although the
approaches differ somewhat, the trends are similar in the two
calculations.

Moore and Teaney \cite{mortea} calculate the diffusion and drag
coefficients for charm quarks in perturbative QCD.  The diffusion
coefficient is proportional to the inverse square of the strong
coupling constant, $\alpha_s$, {\it e.g.} $D (2\pi T) \propto
\alpha_s^{-2}$.  They present the effects of a range of values for $D
(2\pi T)$ on $R_{AA}$ and $v_2$, finding the largest effects at high
$p_T$ for small $D (2\pi T)$, corresponding to large $\alpha_s$ or
strong coupling in the plasma.

Rapp {\it et al.}~\cite{rapp_langevin_QM05,vHGR} calculated the
diffusion and drag coefficients assuming that resonant $D$ and $B$
states in the QGP elastically scatter in the medium.  Resonance
scattering reduces the thermalization times for heavy flavors relative
to those calculated with perturbative QCD matrix elements for fixed
$\alpha_s = 0.4$.  The effect is larger for charm than for the more
massive bottom quarks.  Including these states thus reduces the
electron $R_{AA}$ at high $p_T$ relative to the results in
Ref.~\cite{DGVW} while increasing the electron $v_2$ to $\sim 10$\% at
$p_T \sim 2$~\gevc, in relative agreement with the data.

Thus, given sufficiently strong coupling and/or resonant states, both
$R_{AA}$ and $v_2$ can be described within transport approaches using
elastic scattering.  More and better data is necessary to distinguish
the two approaches.

\subsection{RHIC open heavy flavor measurements to date}

Open heavy flavor production cross sections can be measured by
reconstructing the invariant mass of the heavy quark hadron from its
hadronic decay products or by detecting leptons from semileptonic
decays of those hadrons. While both PHENIX and STAR can measure heavy
flavor cross sections by either technique, PHENIX has some advantages
for semileptonic decay measurements and STAR has advantages for the
hadronic decay measurements.

In both cases the signal to background ratio can be greatly improved
if a precise measurement of the decay vertex position is available,
since hadrons containing $c$ or $b$ quarks typically travel several
hundred microns from the collision point before decaying. Both PHENIX
and STAR have plans to add secondary vertex detectors capable of the
necessary precision. The PHENIX SVTX detectors will be installed starting
in 2010, the STAR HFT in 2012. In addition to reducing the background rates 
for open heavy
flavor decays to leptons and hadrons, the secondary vertex detectors
make a clean bottom cross section measurement possible
using displaced vertex decays to $J/\psi$, given sufficient
luminosity.

Open heavy flavor cross section measurements based on semileptonic
decays of charm and bottom mesons are feasible because a small lepton
signal can be identified in a very large hadron background. The
background lepton sources are both small and well enough understood
that they can be subtracted to get the open heavy flavor signal.
However, the loss of information about the decaying heavy meson due to
the recoil kinematics is a disadvantage of semileptonic decay
measurements.  Thus charm and bottom decays cannot easily be
distinguished. Open charm measurements using hadronic decay products
have two advantages: the $D$ meson kinematic properties are
reconstructed and separation of charm from bottom is far easier
because only a small fraction of $D$ mesons arise from bottom decays
\cite{CNV}. A disadvantage of hadronic decay measurements is the huge
combinatorial background present in heavy-ion collisions.

PHENIX has measured open heavy flavor yields via semileptonic decays
to electrons at midrapidity ($|\eta| < 0.35$) using the Ring Imaging
Cerenkov detector and electromagnetic calorimeter for electron
identification.  At forward and backward rapidity ($1.2 < |\eta| <
2.2$) the two muon spectrometers are used. PHENIX results are
available for \pp\ at midrapidity \cite{Adare:2006hc} and forward
rapidity \cite{PHENIX_charm_pp_mu} as well as for \dAu\ 
\cite{PHENIX_charm_dau_e} and \AuAu\ at midrapidity
\cite{PHENIX_auau_electron_final_ref}. No open charm cross sections from
hadronic decays have yet been reported by PHENIX since the small
central arm acceptance is a disadvantage for such measurements.

STAR has measured open heavy flavor yields at midrapidity ($|\eta| <
1.0$) via semileptonic decays using either a combination of the time
projection chamber (TPC) and time of flight (ToF) for electron
identification or a combination of the TPC and the electromagnetic
calorimeter.  The backgrounds that must be subtracted are much larger
than they are for PHENIX because of the larger photon conversion rates
in STAR and the lack of a hadron blind electron identifier. This is
compensated somewhat by the larger acceptance. STAR electron results
are available for \pp, \dAu\ and \AuAu\ collisions
\cite{Abelev:2006db}. STAR has also measured open charm yields in the
range $|\eta| < 1.0$ through hadronic $D$ meson decays for \dAu\ 
\cite{Adams:2004fc}, \AuAu\ \cite{Zhong:2007iq}, and \CuCu\ 
\cite{Baumgart:2007eu} collisions.

Because the charm cross section is much larger than the bottom cross
section at RHIC and dominates the semileptonic decay spectrum for $p_T
< 2.5$~\gevc, the integrated non-photonic lepton cross section is
usually assumed to be equal to the charm cross section.

\subsubsection{Baseline measurements}

Before any conclusions can be drawn about the hot, dense final state from the 
results for heavy-ion collisions, some 
baseline information is required. Data from \pp\ collisions are needed to 
establish the underlying cross sections and kinematic distributions for 
open heavy flavor and \pA\ data are needed to isolate 
effects due to gluon saturation and the intrinsic $k_T$ distributions
in the colliding nuclei.

\begin{figure}[tbh]
  \centering
  \includegraphics[width=0.5\textwidth]{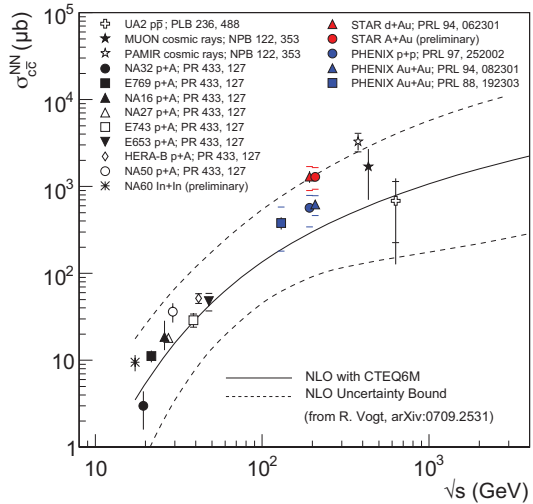}
  \caption{Comparison of total cross section measurements. The STAR
    and PHENIX results are given as cross section per binary
    collisions. Vertical lines reflect the statistical errors, 
    horizontal bars indicate the systematic uncertainties (where available).
    The NLO calculations and the depicted uncertainty bands are described 
    in the text.}
  \label{fig:charm_sigma_vs_energy}
  \vspace{0.5cm}
\end{figure}

Both PHENIX and STAR have measured charm production cross sections 
at midrapidity.
These measurements have been extrapolated to all rapidities to yield
total cross sections. These total cross sections are compared to
results at other energies and to pQCD calculations in
Fig.~\ref{fig:charm_sigma_vs_energy}.  The results for \dAu\  and \AuAu\ 
collisions are divided by the number of binary collisions, \nbin, to
directly compare with the \pp\ results.  
The STAR data are from
combined fits to hadronic and semileptonic decay data.  The PHENIX
data are from semileptonic decay measurements only.

The STAR values are somewhat higher than those for PHENIX. The total 
charm cross sections differ by about 1.5 times the combined standard 
deviations of the two measurements, obtained by adding all statistical 
and systematic uncertainties in quadrature.  

The NLO calculations and their uncertainties are those of the fixed-order
calculation with three light flavors for charm, as described in 
Ref.~\cite{rvjoszo}.  While the larger uncertainty of the fixed-order
calculation shown here can accommodate the larger STAR cross section, it is
not clear which estimate of the uncertainty is best.  Obviously the large
theoretical uncertainty means that there is little predictive power in the
total charm cross section.  The FONLL estimate of the uncertainty in the
$p_T$ distribution is more reliable for finite $p_T$ \cite{rvjoszo}.

\begin{figure}[tbh]
  \centering
  \includegraphics[width=0.45\textwidth]{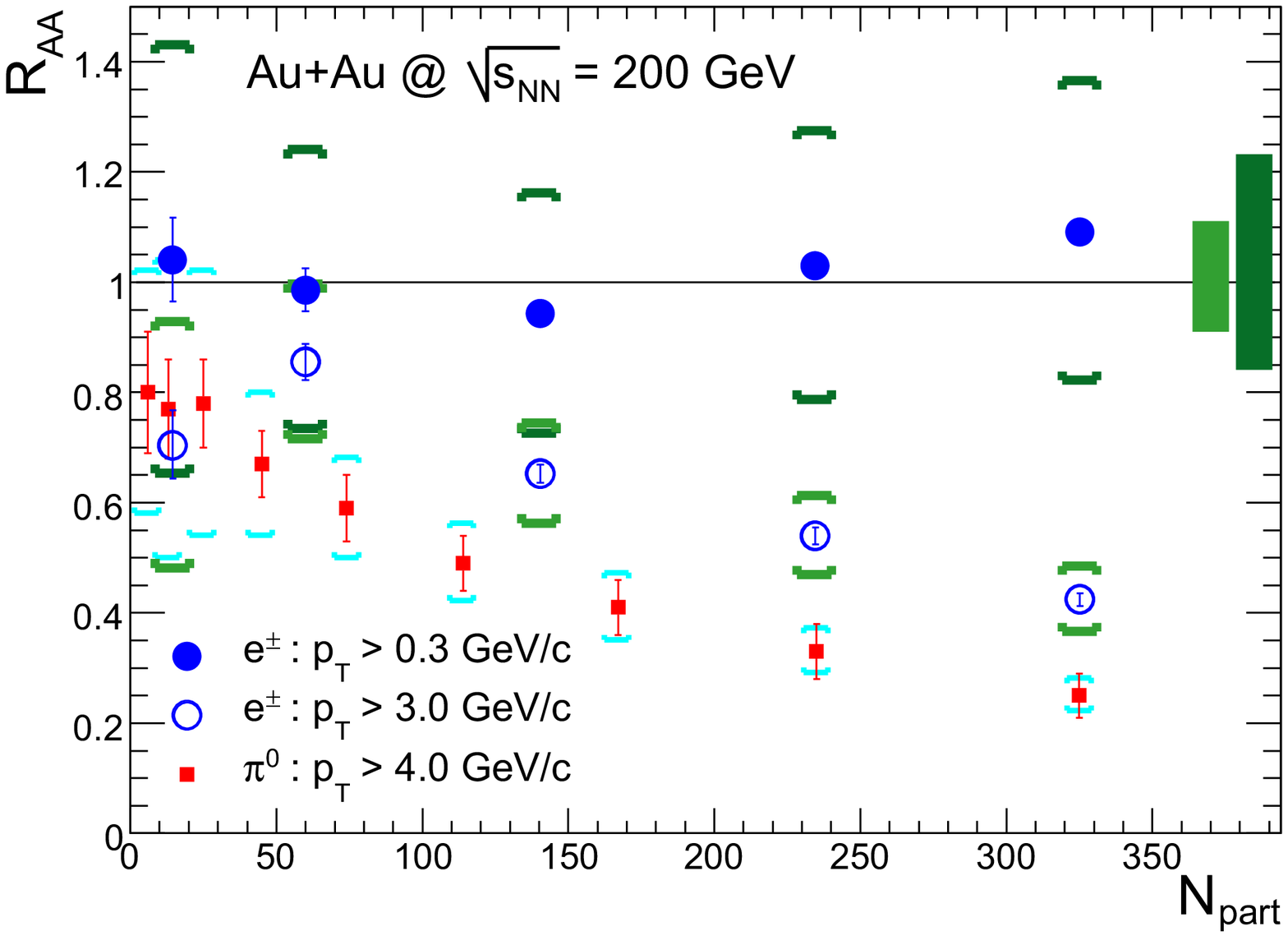}
  \includegraphics[width=0.47\textwidth]{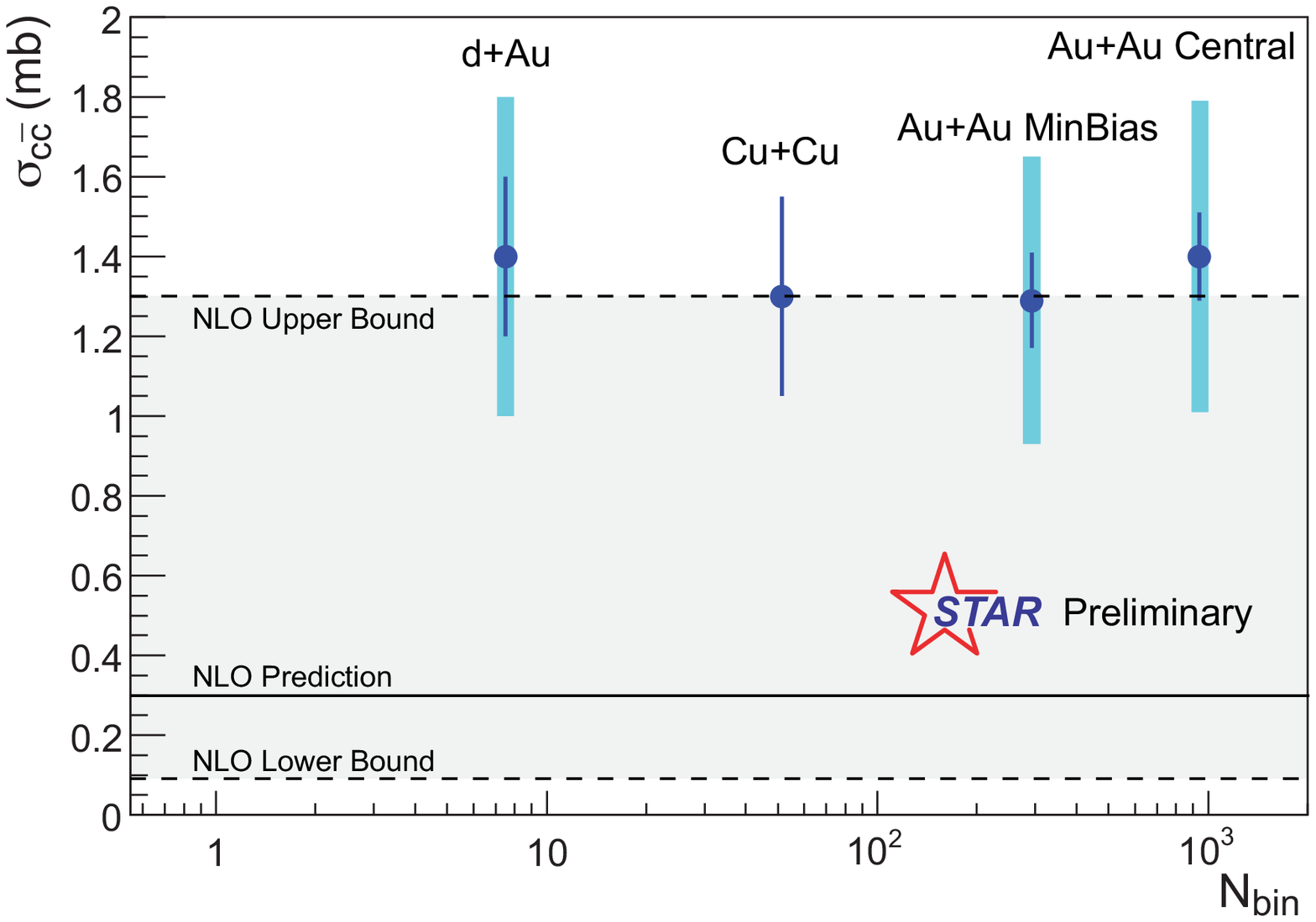}
  \caption{Left: PHENIX measurements of the midrapidity
    invariant charm yields scaled by \nbin\ as a function of 
    \npart\ \protect\cite{PHENIX_auau_electron_final_ref}. 
    More than half of the heavy 
    flavor decay electrons included in the integral are above 0.3 \gevc. 
    Right: STAR measurements of the charm cross section derived 
    from identified $D^0$ mesons in \dAu, \CuCu, and \AuAu\ collisions
    as a function of \nbin.} 
  \label{fig:charm_sigma_vs_ncoll}
  \vspace{0.5cm}
\end{figure}

Since charm and bottom quarks are expected to be produced only in the
initial nucleon-nucleon interactions, their yield should scale as  
the number of binary collisions, \nbin. 
The left-hand side of Fig.~\ref{fig:charm_sigma_vs_ncoll} 
shows the PHENIX measurement  
of the charm invariant yield in \AuAu\ collisions, scaled by \nbin,
at midrapidity as a function of  \npart\ 
\cite{PHENIX_auau_electron_final_ref}. 
The PHENIX data integrated over $p_T > 0.3$ \gevc\ are consistent with no  
\npart\ dependence, as expected. The effect of the final state medium can be 
seen when the distributions are integrated above 3 \gevc.  
These $p_T$-dependent effects are discussed in the following section.
The comparison of charm production measured by STAR in \pp, \dAu, \AuAu, 
and \CuCu\ indicates that the total charm cross section scales with 
$N_{\rm coll}$, as expected for hard processes, see the right-hand side of 
Fig.~\ref{fig:charm_sigma_vs_ncoll}.  The \dAu\ and \AuAu\ cross sections 
were further constrained by using the low-\pT\ muons and medium-\pT\ electrons 
from semileptonic charm decays.  The measured cross section for each system 
and centrality was divided by the appropriate value of \nbin\ to reflect the
\pp\ equivalent cross section \cite{Baumgart:2007eu}. The NLO lines are 
the bounds on the total $c \overline c$ cross section from 
Fig.~\ref{fig:charm_sigma_vs_energy}.

\begin{figure}[tbh]
\vspace{0.5cm}
 \centering
 \includegraphics[width=0.95\textwidth]{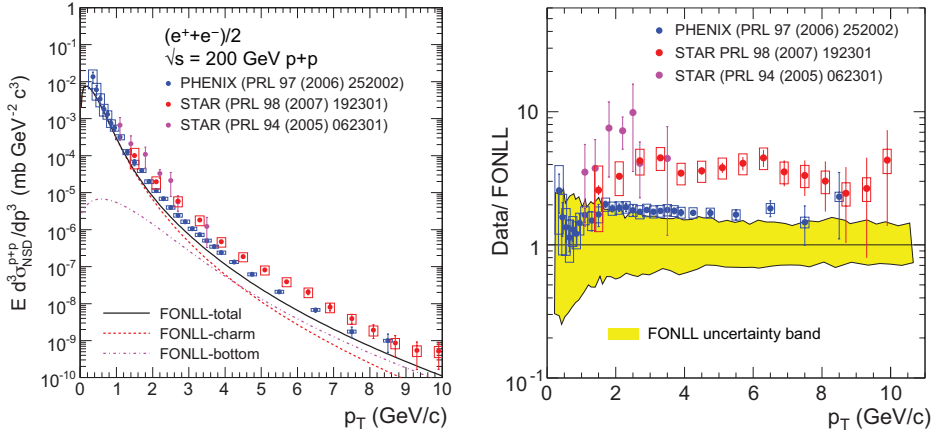}
 \caption{Left: Compilation of PHENIX \protect\cite{Adare:2006hc} and STAR 
   measurements \protect\cite{Adams:2004fc,Abelev:2006db} 
   of the $p_T$ dependence of the semileptonic 
   decay open heavy flavor cross section from 200 \gev\ \pp\ collisions,   
   compared with FONLL calculations \protect\cite{Adare:2006hc}. 
   Right: The
   ratio of the data to the FONLL calculation. The band depicts the 
   theoretical uncertainty of the calculation. Note, that they are
    smaller than those from the calculation of the total cross-section
   shown in Fig.~\ref{fig:charm_sigma_vs_energy} and 
   \ref{fig:charm_sigma_vs_ncoll} (right) as detailed earlier 
   in this section. From  Ref.~\protect\cite{AlexQM}.}
 \label{fig:phenix_star_pp_vs_fonnl}
 \vspace{0.5cm}
\end{figure}


The left-hand side of Fig.~\ref{fig:phenix_star_pp_vs_fonnl} shows the
measured $p_T$ dependence of semileptonic open heavy flavor decays to
electrons from 200 \gev\ \pp\ collisions by PHENIX \cite{Adare:2006hc}
and STAR \cite{Adams:2004fc,Abelev:2006db}.  On the right-hand side,
the ratio of the measured cross sections to the FONLL calculation
\cite{CNV} is shown. The FONLL calculation reproduces the $p_T$ slope
of the data in both cases. The magnitude of the FONLL cross section is
1.5 times smaller than that of the PHENIX data, just at the limit of
the theoretical uncertainty band.  However, the FONLL cross section is
a factor of $\sim 4$ smaller than the STAR data, well outside the
theoretical uncertainties.  The discrepancy of a factor of 
$\sim 2$ between the PHENIX and STAR cross sections in
Fig.~\ref{fig:phenix_star_pp_vs_fonnl} is not yet explained.

\begin{figure}[tbh]
\vspace{0.5cm}
 \centering
 \includegraphics[width=0.48\textwidth]{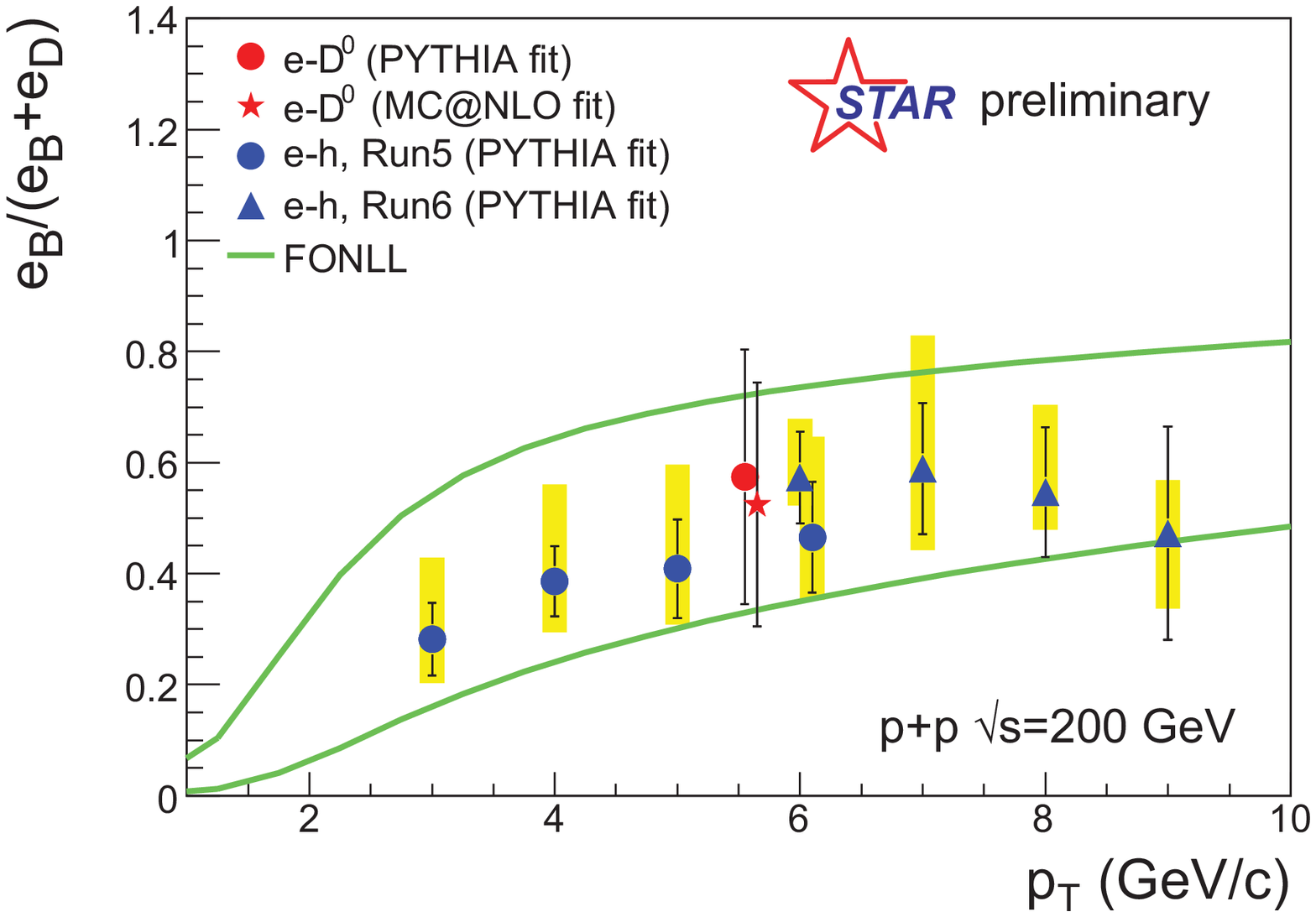}
 \includegraphics[width=0.50\textwidth]{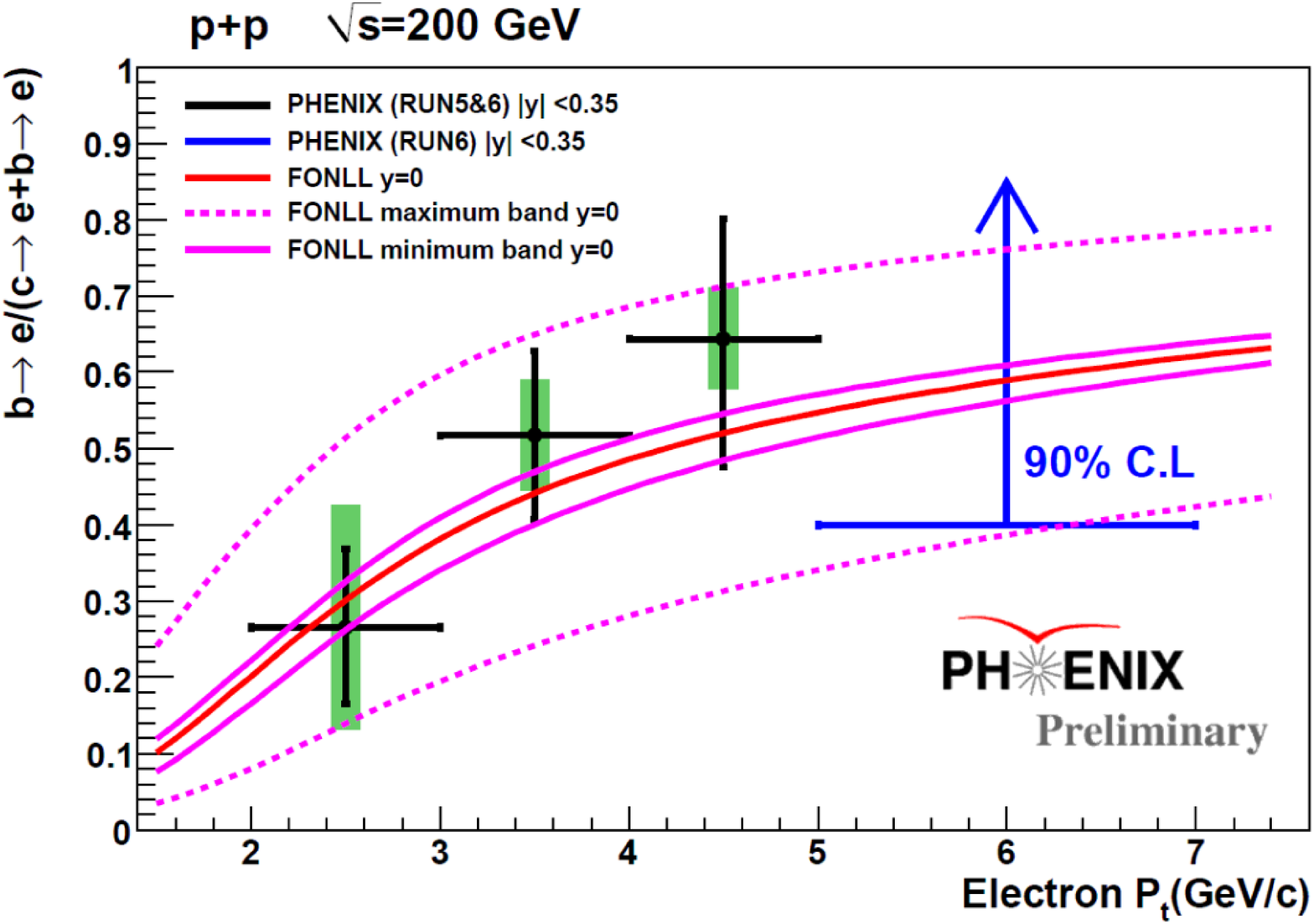}
 \caption{Measurements of the $p_T$ dependence of $b/(c+b)$ determined 
   from semileptonic heavy flavor decays in 200 \gev\ \pp\ collisions by STAR
   \protect\cite{star_bc_qm2008} (left) and PHENIX 
   \protect\cite{phenix_bc_qm2008} (right).}
 \label{fig:phenix_star_bc_ratio}
 \vspace{0.5cm}
\end{figure}

A limitation of open heavy flavor measurements using semileptonic
decays is the difficulty of separating charm and bottom decay
contributions.  Recently, the ratio of charm to bottom contributions
has been extracted as a function of decay electron \pT\ from $pp$ data
by STAR and PHENIX.  STAR measurements were made by fitting measured
small angle electron-hadron correlation functions to a combination of
the $D$ and $B$ correlation shapes given by their decay kinematics
simulated through PYTHIA.  Additional \pT\ integrated measurements were
made using like-sign electron-kaon correlations, exploiting the
difference in the correlation shape between $B$ and $D$ decays. A new
recent measurement is based on the small angle correlation of
semileptonic decay electrons with identified $D^0$ mesons yielding
consistent results with the former method \cite{star_bc_qm2008}.  The
PHENIX bottom/charm ratio \cite{phenix_bc_qm2008} was inferred by
fitting the shape of the electron-kaon reconstructed mass spectrum
with a combination of the charm and bottom decay line shapes,
taken from PYTHIA simulations.  The measured STAR and PHENIX $b/(c+b)$
ratios are compared in Fig.~\ref{fig:phenix_star_bc_ratio}. They are
in good agreement and indicate that bottom decays become dominant at
$p_T \sim 3.5$ \gevc.

A separate measurement of the bottom and charm cross sections was 
also obtained by PHENIX from fits to the dielectron invariant mass spectrum, 
after removal of all non-heavy flavor contributions \cite{phenix_ppg085}. 
The results are in  agreement with the cross sections inferred by PHENIX 
from combining the measured $b/(c+b)$ ratios with semileptonic
decay electron \pT\ distributions. The bottom cross sections determined 
by PHENIX are summarized in Table~\ref{bb_cross_sections_all}.

\begin{table}[tbh]
\begin{tabular}{|c|c|c|} 
\hline
Method & $d\sigma_{b\overline{b}}/dy$ ($\mu$b) & $\sigma_{b\overline{b}}$ 
($\mu$b)  \\ 
\hline 
$e h$ mass shape & $1.34 \pm 0.38 ^{+0.74}_{-0.64}({\rm sys})$ 
& $4.61 \pm 1.31 ^{+2.57}_{-2.22}({\rm sys})$  \\
$e^{+}e^{-}$ mass spectrum & $-$ & $3.9 \pm 2.5 ^{+3}_{-2}({\rm sys})$ \\
\hline
\end{tabular}
\caption{Summary of PHENIX \cite{phenix_bc_qm2008,phenix_ppg085} measurements 
of the open bottom cross section in 200 \gev\ \pp\ collisions. }
\label{bb_cross_sections_all}
\end{table}

Unlike the $J/\psi$ measurements discussed in the next section, the
current \dAu\ and \pp\ open heavy flavor results are not precise
enough for any conclusions to be drawn about either shadowing or $k_T$
broadening.  Obtaining more precise open heavy flavor baseline results
is an important priority for the RHIC program over the next few years.

\subsubsection{Heavy-ion measurements}

Both PHENIX and STAR have released striking results on suppression of
single electrons from open heavy flavor decays in central \AuAu\ 
collisions.  PHENIX also has results for the electron $v_2$ from open
heavy flavor decays.

\begin{figure}[tbh]
\vspace{0.5cm}
 \centering
  \includegraphics[width=0.6\textwidth]{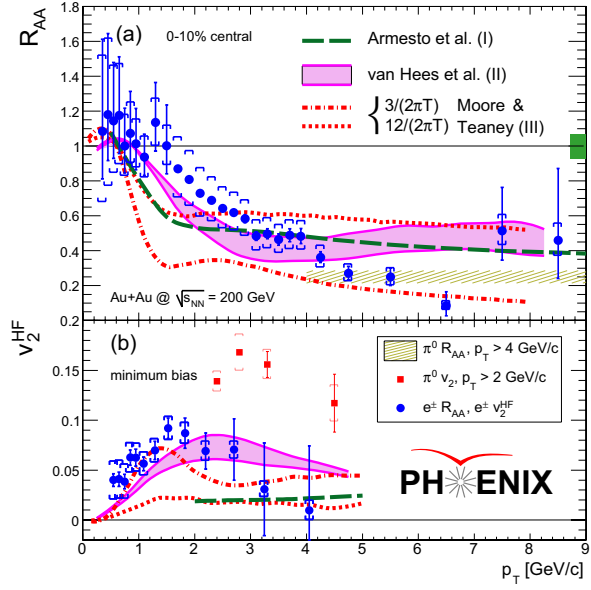}
  \caption{PHENIX measurement of the non-photonic 
    electron $R_{AA}$ (upper panel) and $v_2$ (lower panel) 
    in central \AuAu\ collisions \protect\cite{PHENIX_auau_electron_final_ref}.
    The theory curves are discussed in the text. }
  \label{fig:phenix_auau_final_electron_data}
  \vspace{0.5cm}
\end{figure}

\begin{figure}[tbh]
    \vspace{0.5cm} \centering
    \includegraphics[width=0.5\textwidth]{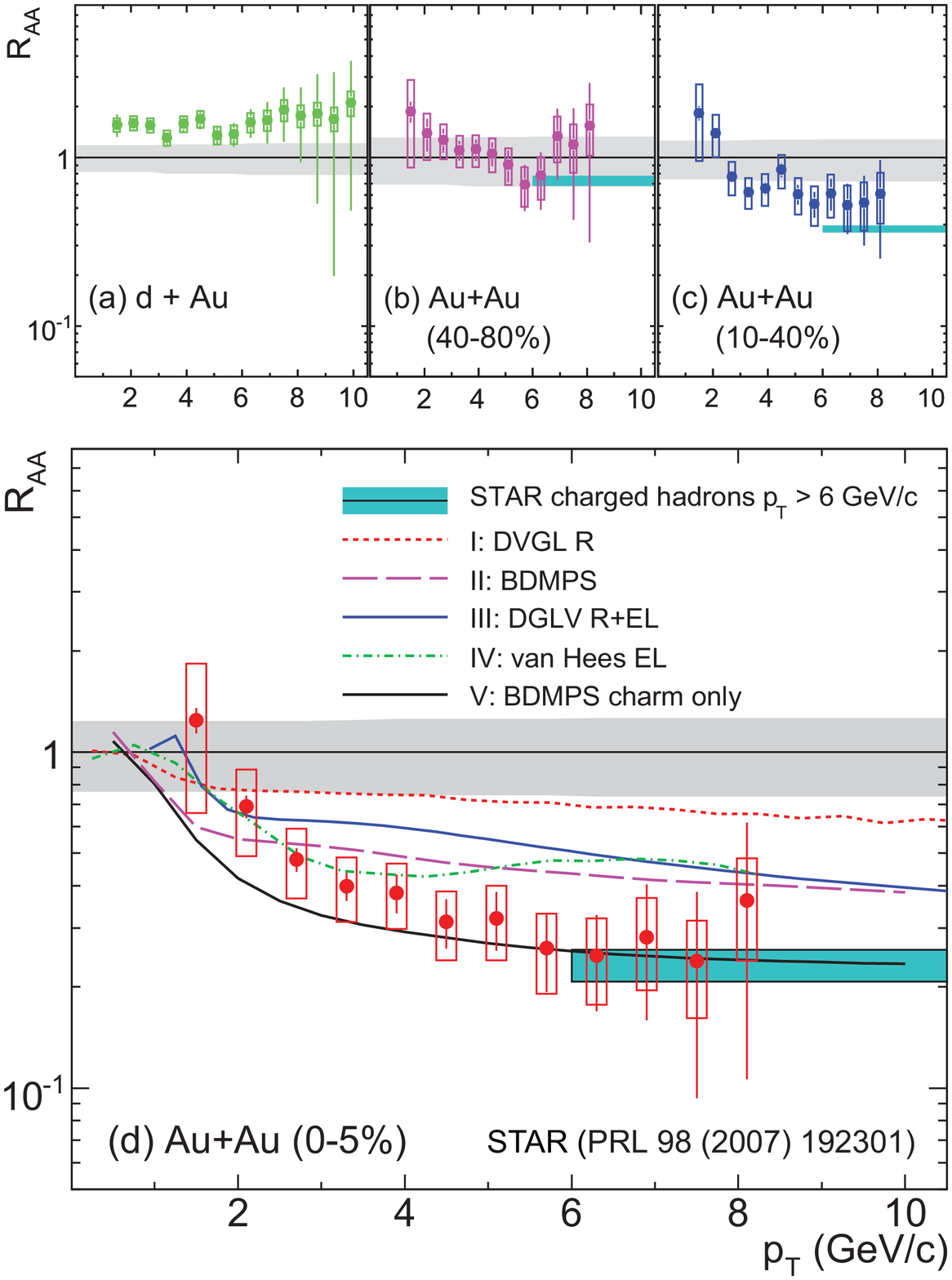}
    \caption{Measurement of the nuclear modification factor of
        non-photonic electrons by STAR for \dAu\ and \AuAu\ collisions
        at different centralities \protect\cite{Abelev:2006db}. The theory
        curves are discussed in the text.}
    \label{fig:star_auau_final_electron_raa}
    \vspace{0.5cm}
\end{figure}

The nuclear modification factors and elliptic flow parameters for
electrons from semileptonic decays of open heavy flavor in \AuAu\ 
central collisions from PHENIX \cite{PHENIX_auau_electron_final_ref}
are shown in Fig.~\ref{fig:phenix_auau_final_electron_data}.
Figure~\ref{fig:star_auau_final_electron_raa} \cite{Abelev:2006db}
shows the STAR $R_{AA}$ measured in \AuAu\ and \dAu\ collisions.
The \AuAu\ $R_{AA}$ data from the two experiments are in reasonable
agreement.  Both show very strong suppression in central collisions at
high $p_T$.   When comparing the non-photonic electron
$R_{AA}$ data to theory, recall that while the electron data contain
contributions from both charm and bottom decays, the bottom
contribution is expected to dominate for $p_T$ above $\sim 4$~\gevc\ 
\cite{CNV}.  Since energy loss is predicted to be weaker for bottom, the
similarity of the electron $R_{AA}$, $R_{AA} \sim 0.2-0.3$, to light-quark 
hadrons \cite{exp_wps_PHENIX} is surprising.  

In Fig.~\ref{fig:phenix_auau_final_electron_data}, the PHENIX $R_{AA}$
and $v_2$ data are compared with models that calculate both quantities
simultaneously. Curve I is a perturbative QCD calculation with
radiative energy loss \cite{Armesto}. It describes the $R_{AA}$ data
using a large transport coefficient, $\hat{q} = 14$ GeV$^2$/fm, that
also works well for light hadron suppression. However, in this model
$v_2$ arises only from the path length dependence of energy loss and
clearly underpredicts the $v_2$ data. Band II is a Langevin-based
heavy quark transport calculation \cite{vHGR}, including elastic
scattering mediated by resonance excitation, also compared to the STAR
data in Fig.~\ref{fig:star_auau_final_electron_raa}. The best
simultaneous description of the $R_{AA}$ and $v_2$ data is achieved
with a small heavy quark relaxation time. The curve labeled III is
also from a transport calculation \cite{mortea} where the diffusion
and drag coefficients are calculated in perturbative QCD.  The
diffusion coefficients required by the data are small in both cases,
implying a ratio of viscosity to entropy small enough to be at or near
the conjectured quantum bound \cite{PHENIX_auau_electron_final_ref},
consistent with estimates of elliptic flow and fluctuation analyses
for light-quark hadrons \cite{lacey_2006,gavin_2006}.

\begin{figure}[tbh]
\vspace{0.5cm}
 \centering
  \includegraphics[width=0.6\textwidth]{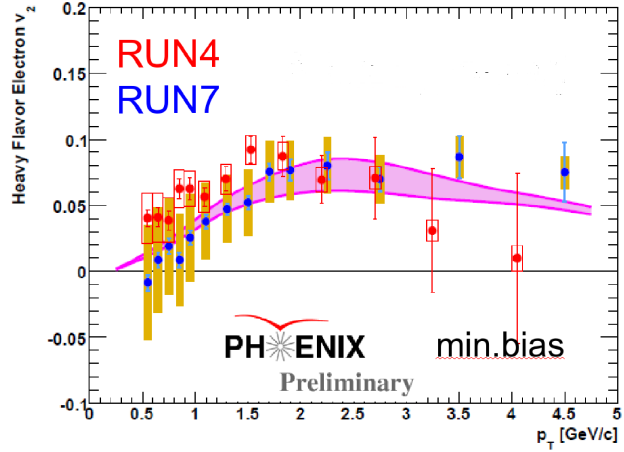}
  \caption{PHENIX preliminary measurement of the non-photonic 
    electron $v_2$ in minimum bias \AuAu\ collisions 
    \protect\cite{phenix_bc_qm2008}, 
    providing good statistical significance to 5 \gevc.
    The band is the van Hees (II) result shown in 
    Fig.~\ref{fig:phenix_auau_final_electron_data}. }
  \label{fig:phenix_run7_heavy_v2}
  \vspace{0.5cm}
\end{figure}

The STAR $R_{AA}$ data are compared to model calculations in
Fig.~\ref{fig:star_auau_final_electron_raa}.  The theory curves are
from calculations discussed previously.  The curves labeled I-IV are
calculations of the electron spectra from both $D$ and $B$ decays
incorporating final-state $c$ and $b$ quark energy loss.  Curve I
\cite{DGVW} uses only DLGV radiative energy loss with $dN_g/dy =
1000$, while Curve II \cite{Armesto} employs BDMPS radiative energy
loss (Curve I in Fig.~\ref{fig:phenix_auau_final_electron_data}).
Both calculations predict much less suppression than observed.  Curves
III are from a DLGV-based calculation including both collisional and
radiative energy loss \cite{WHDG}.  While the addition of collisional
energy loss substantially increases the suppression over the
calculation with DGLV radiative loss alone, shown in Curve I, it still
underpredicts the suppression seen in the data.  Curve IV is a
Langevin-based calculation \cite{vHGR}, the center of Band II in
Fig.~\ref{fig:phenix_auau_final_electron_data}. Resonance effects
reduce the $R_{AA}$ for charm-decay electrons to 0.2, but have a
smaller effect on bottom quark energy loss, causing $R_{AA}$ to
increase at higher $p_T$.  Curve V is the same as Curve II but with
only $D$ decays included.

There is considerable interest in the behavior of the non-photonic
electron $v_2$ for $p_T > 2$~\gevc\ where, as discussed earlier, 
the bottom contribution is expected to become important. New preliminary 
PHENIX semileptonic decay $v_2$ results were shown at Quark Matter 
2008. These show that the $v_2$ remains high up to 4.5 \gevc. Together with 
the observations from PHENIX and STAR that the bottom cross section starts 
to dominate the semileptonic decay spectra at about 3.5 \gevc\ in \pp\ 
collisions, the existing \RAA\ data and the new preliminary $v_2$ data 
strongly suggest that bottom quarks experience large energy loss and 
significant flow at RHIC.

In spite of the recent $pp$ measurements of the ratio of bottom to charm  
with \pT\ in the semileptonic decay spectra, the lack of electron 
separation between charm and bottom decays remains a
serious limitation. Sensitive measurements of bottom quark \RAA\ and 
$v_2$, in particular, in heavy-ion reactions will require 
clean separation of bottom and charm contributions. 
The ways in which this will be addressed in the future RHIC program
are discussed in the next section.

\subsection{Proposed open heavy flavor experimental program at RHIC II}

Here we focus on the new open heavy flavor physics that becomes
available with the the RHIC detector upgrades and the RHIC II
luminosity upgrade.

The measurement of charm mesons through their \textit{hadronic} decay
channels relies on the efficient reconstruction of the secondary decay
vertices. With the Heavy Flavor Tracker (HFT) upgrade the STAR
experiment will be able to measure $D$ mesons down to low \pT.
However, to-date no high-level trigger scheme exist that would allow
to trigger on displaced vertices in heavy-ion collisions.
Consequently, the direct reconstruction of $D$ mesons does require
large minimum bias data sets and the subsequent offline reconstruction
of these events. These studies do not depend on large beam luminosity.
The situation is different for measurements of open heavy flavor
through their \textit{semileptonic} decays where the high-\pT\ 
lepton can be used to tag candidate events. Triggers for high-\pT\ 
leptons are already deployed in PHENIX and STAR and will be mandatory
at RHIC-II luminosities.

\begin{figure}[tbh]
  \includegraphics[width=0.49\textwidth]{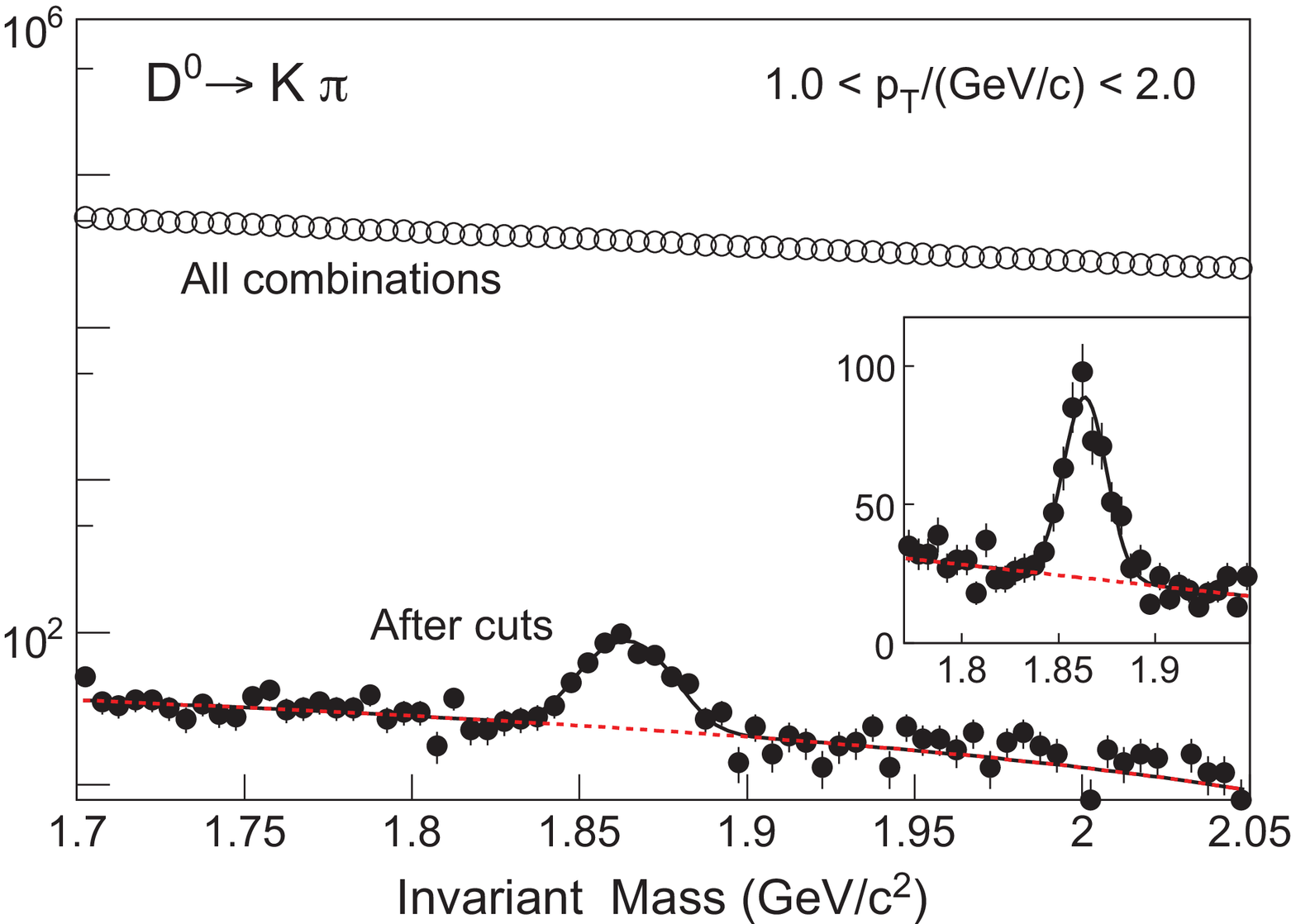}
  \includegraphics[width=0.48\textwidth]{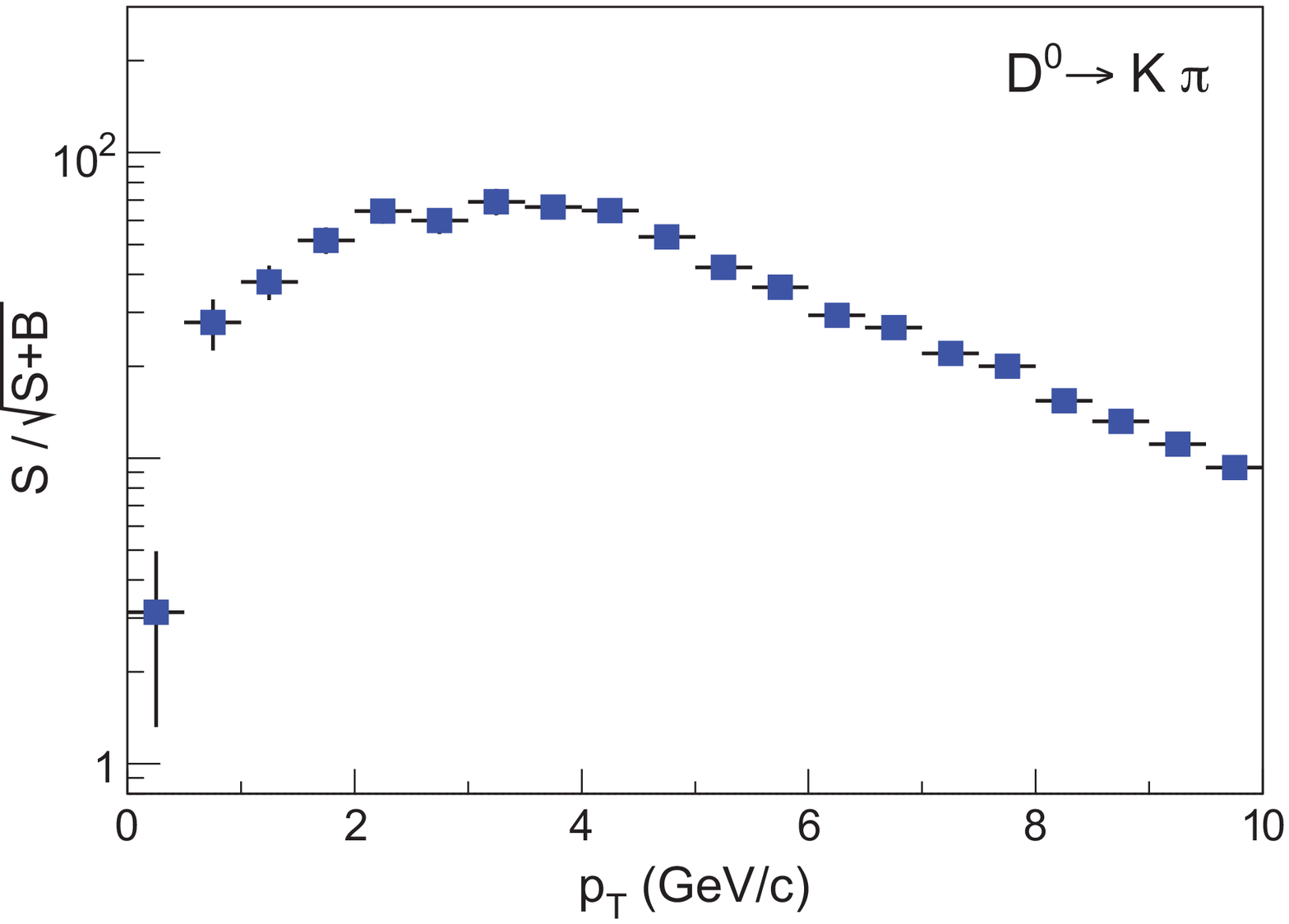}
\caption{Left: Illustration of $D^0 \rightarrow K \pi$ reconstruction
    with the HFT in STAR.  The open symbols depict the invariant $K\pi$ 
    mass of all candidate tracks without any topological cuts
    applied. The closed symbols show the final signal after secondary
    vertex reconstruction. The dashed line shows the background
    fit outside the $D^0$ mass range. The signal peak is
    shown on a linear scale in the inset.  Right: Expected
    significance of the $D^0$ measurement in 200 GeV \AuAu\ collisions
    with 100 M central events as a function of \pT.}
\label{fig:hft-D0-projections}
\end{figure}

The left side of Fig.~\ref{fig:hft-D0-projections} shows the
reconstructed $D^0 \rightarrow K \pi$ channel from STAR
simulations. Note that the signal cannot be distinguished from
the background (random combinations of $K$ and $\pi$ tracks) without
making topological cuts on the pairs. The plot illustrates the need for
high pointing resolution of the two innermost Si-detector layers of the HFT;
tight displaced vertex cuts will eliminate most of the prompt hadron
tracks from the combinatorial background.  
The overall reconstruction efficiency,
taking into account the acceptance, single track efficiency and $D^0$
reconstruction efficiency increases from $\sim 10$\% at low \pT\ to
$\sim 25$\% at high \pT.  The efficiency is small at low \pT\ because the 
minimal boost from the parent particle allows the pool of daughter track 
candidates to be contaminated by primary tracks, diluting the signal and 
increasing the background.

The right-hand side of Fig.~\ref{fig:hft-D0-projections} 
shows the significance of
the STAR $D^0$ signal for $10^8$ central \AuAu\ events
as a function of \pT. With the upgraded STAR data acquisition and TPC readout
electronics, such a sample is expected in two weeks of running.
The signal significance is directly related to the
precision with which one can determine suppression, $R_{AA}$, and flow, $v_2$.

\begin{figure}[tbh]
  \includegraphics[width=0.47\textwidth]{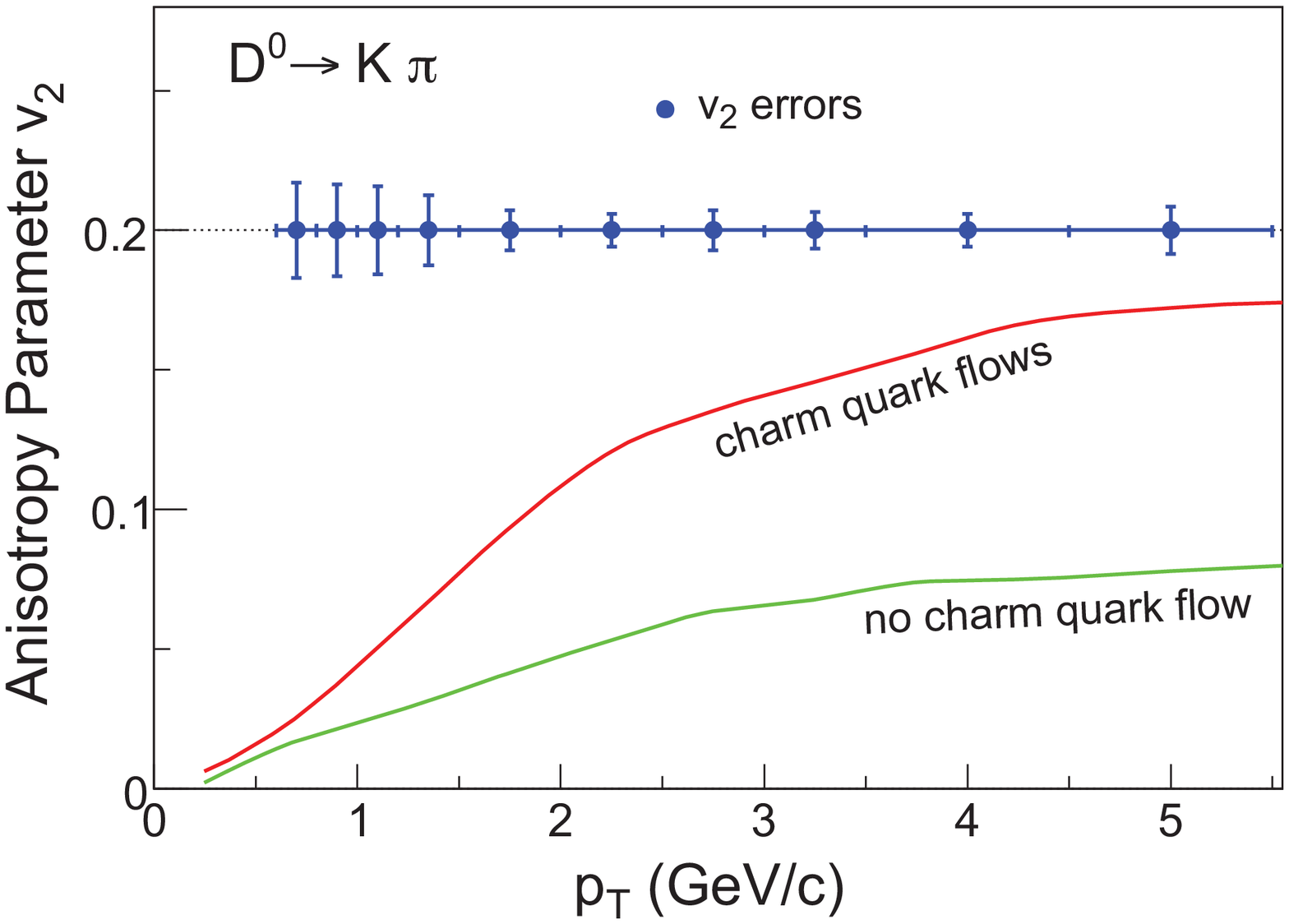}
  \includegraphics[width=0.50\textwidth]{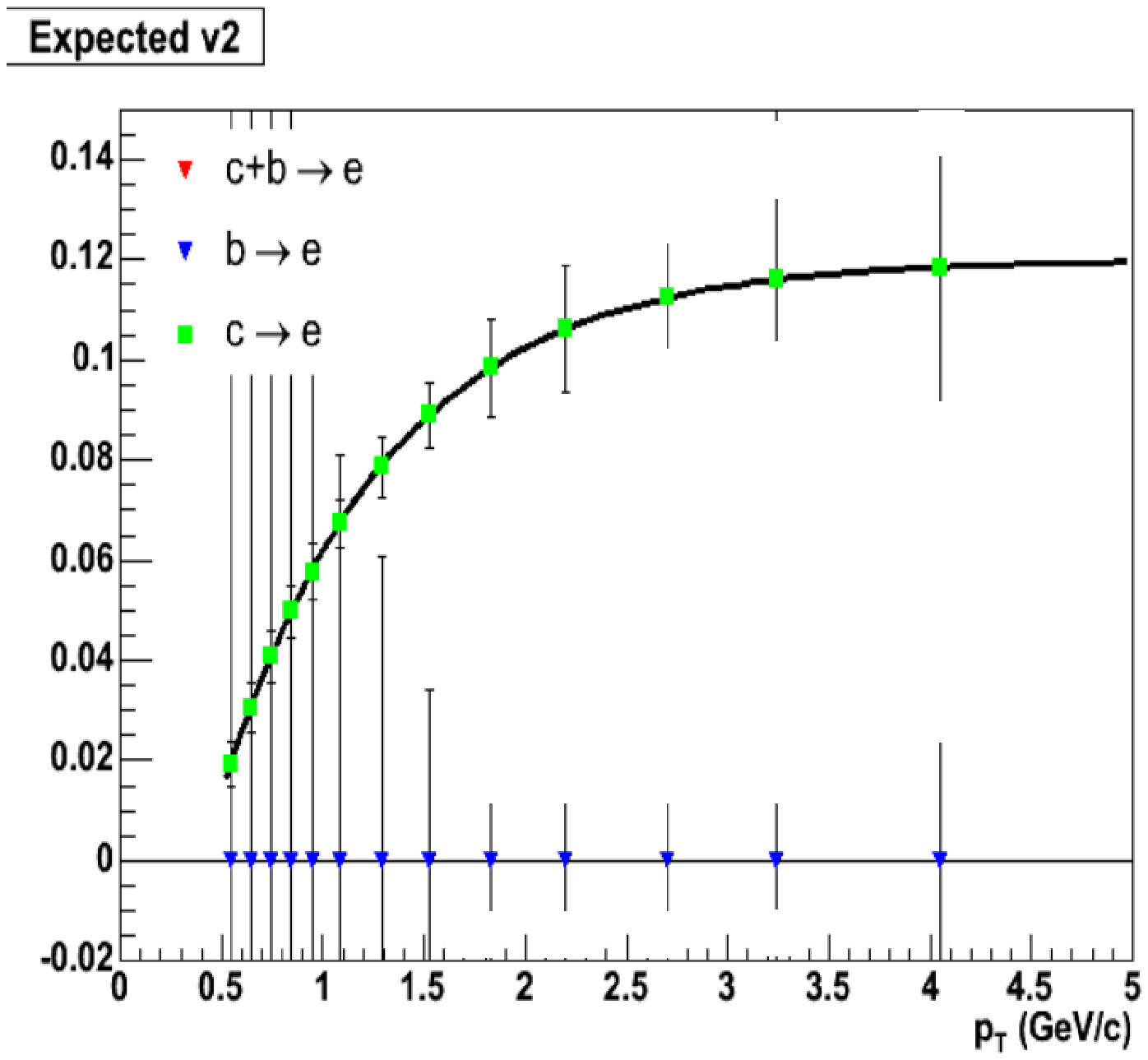}
\caption{Left: Estimate by of the $D^0$ $v_2$ precision measured
    by STAR through $D^0 \rightarrow K \pi$ decays. The error
    bars depict the anticipated errors of the measurement in $10^8$
    \AuAu\ minimum bias events. The lower curve illustrates $v_2$ model
    predictions if only the light quarks in the $D$ mesons flow. The 
    upper curve illustrates the increase in $v_2$ if the charm
    quark also flows \protect\cite{Greco:2003vf}.  Right:
    Estimates \protect\cite{electron_v2_rhic2_estimates} 
    of PHENIX $v_2$ precision for open charm and bottom decay
    electrons using a displaced vertex measurement, calculated for a delivered 
    luminosity of 30 $nb^{-1}$. The uncertainties on
    bottom decays at the lowest $p_T$ are dominated by systematic
    errors.}
\label{fig:electron_v2_rhic2}
\end{figure}

A precision measurement of charm elliptic flow is one of the main
goals for any future RHIC program. In order to simulate
minimum-bias \AuAu\  measurements, it is necessary to make assumptions about
the magnitude and the \pT\ dependence of the anisotropy of the flow
distributions. Models of $D^0$ flow with and without charm quark flow
\cite{Greco:2003vf}, represent the two possible extremes and
thus bracket the expected $v_2$ range, shown on the left-hand
side of Fig.~\ref{fig:electron_v2_rhic2}.  With the known efficiency
and the anticipated signal-to-background ratios, the statistical
errors on a proposed measurement can be estimated. The small error bars
shown (symbols) can be achieved by STAR $10^8$ minimum-bias \AuAu\ events. 
The most important momentum range for determining the thermodynamic
behavior of the $D^0$ is $p_T < 3$ \gevc\ since other dynamical effects 
such as jet correlations will become
important  at higher \pT.

\begin{figure}[tbh]
\begin{center}
  \includegraphics[width=0.45\textwidth]{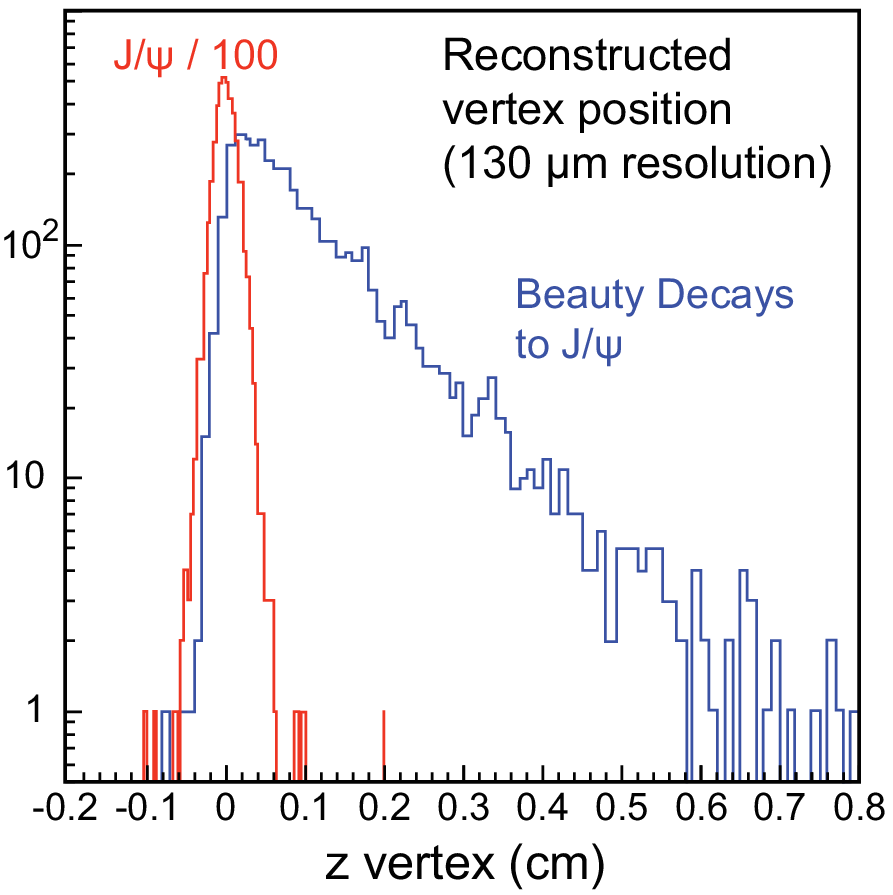}
  \includegraphics[width=0.45\textwidth]{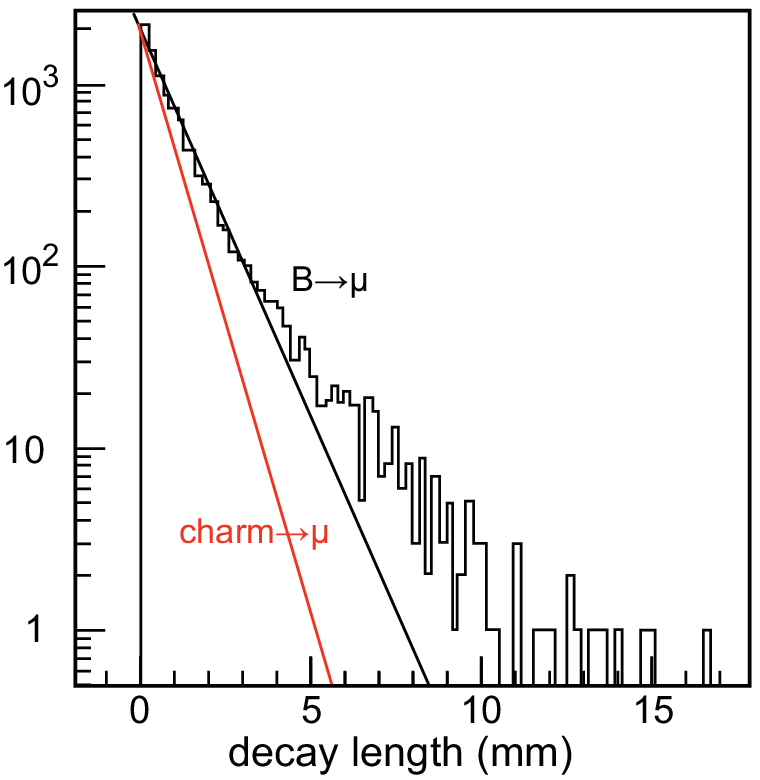}
\end{center}
\caption{Left: Comparison of prompt $J/\psi X$ displaced vertex distribution 
(in cm) with that from $B \rightarrow J/\psi X$ decays 
\protect\cite{phenix_fvtx_ref}.  Note that the prompt 
$J/\psi$ distribution is scaled down by a factor of 100. 
Right: Decay length distributions (in mm) from open charm and 
bottom simulations \protect\cite{phenix_fvtx_ref}.}
\label{fig:decay_lengths}
\end{figure}

Another way to measure open heavy flavor is through 
semileptonic decays. The azimuthal angle of the $D$ and its
decay lepton are closely correlated. With the exception of very low \pT, the 
$v_2$ of the lepton is a good proxy for the $v_2$ of the heavy flavor meson. The
disadvantage of this method, however, is the difficulty of separating
leptons from charm and bottom decays.  With a displaced vertex
measurement and the RHIC II luminosity, separation of the charm and
bottom contributions to the semileptonic decay spectra can be done
statistically using the different decay lengths for charm and bottom
mesons (see the right-hand side of Fig.~\ref{fig:decay_lengths}). By
analyzing data samples with different decay length cuts, the fraction
of the signal due to $b$ quarks can be varied. The addition of a
displaced vertex measurement greatly reduces the background for all
open heavy flavor measurements, improving both statistical and 
systematic errors. The expected precision for separate $v_2$ measurements 
of charm and bottom via electron decays at midrapidity in PHENIX are shown 
on the right-hand side of Fig.~\ref{fig:electron_v2_rhic2}. 

\begin{figure}[tbh]
\begin{center}
  \includegraphics[width=0.45\textwidth]{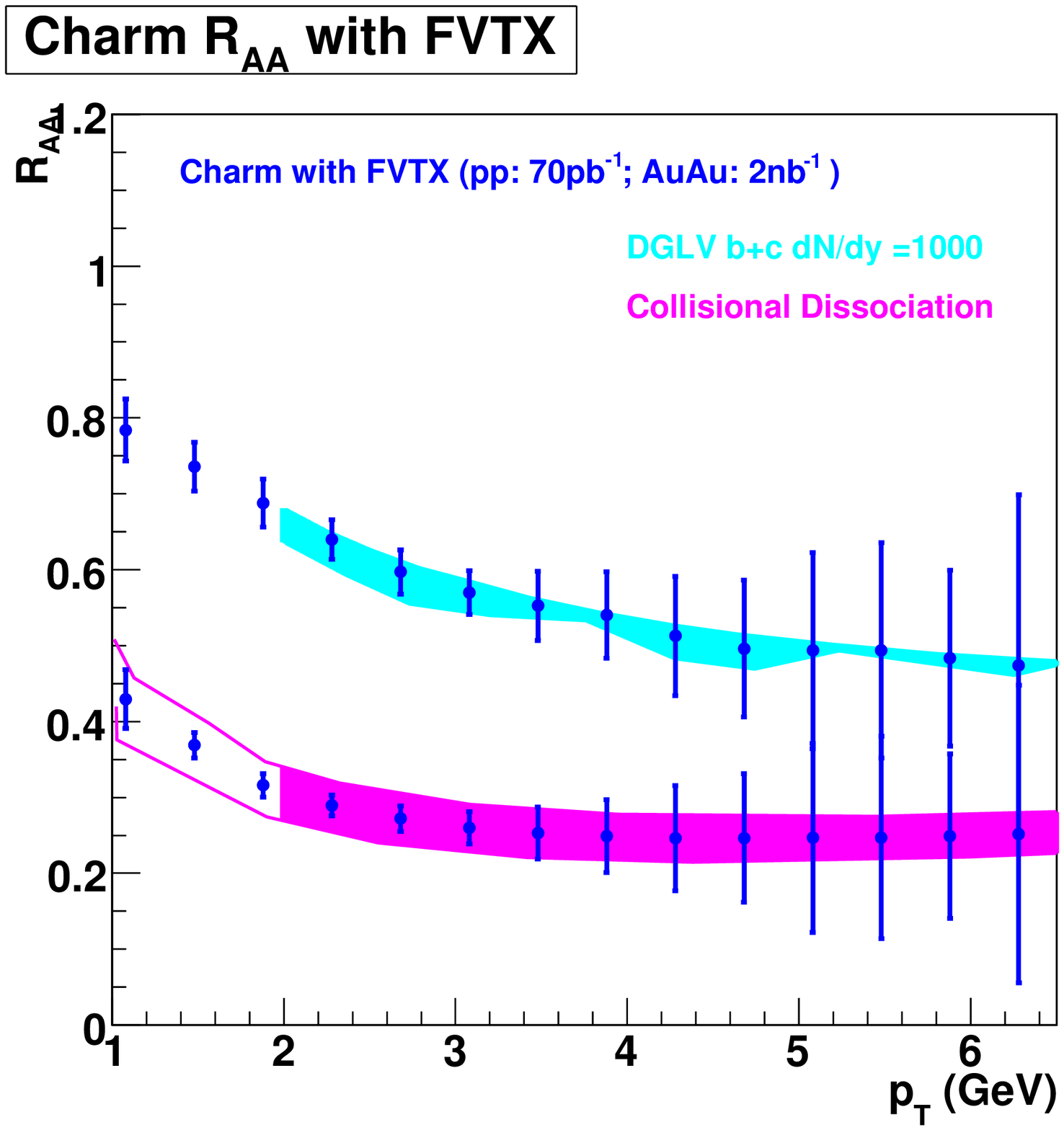}
  \includegraphics[width=0.45\textwidth]{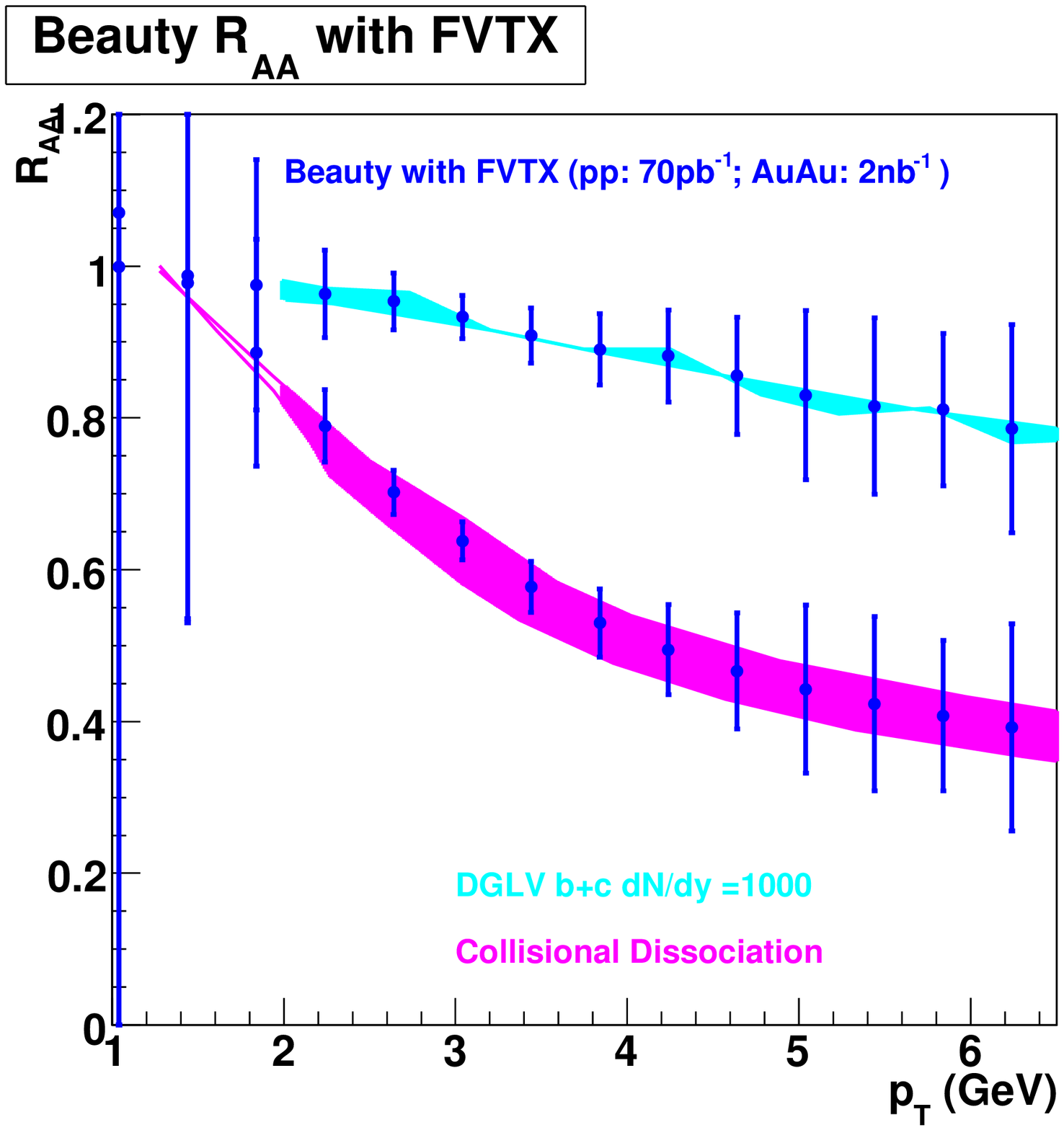}
\end{center}
  \caption{PHENIX estimates of the precision of the heavy flavor semileptonic 
    decay \RAA\ obtained using a displaced vertex measurement, calculated for a
    delivered luminosity of 30 $nb^{-1}$ for (left) open charm 
    and (right) open bottom. }
  \label{fig:fvtx_separated_raa}
\end{figure}

At forward rapidity in PHENIX, displaced vertex measurements can be used 
to measure charm and bottom separately using muon decays. In that case, 
adding a 1 cm displaced vertex cut reduces the muon yields 
from light hadron decays by about one order of magnitude, reducing a major 
background source for single muon measurements, and adding a very tight 
cut around zero vertex displacement also greatly reduces the punch-through 
hadron background.
Figure~\ref{fig:fvtx_separated_raa} shows simulations of the expected precision
for charm and bottom \RAA\ measurements at forward rapidity in PHENIX in
one year of RHIC running.
The right-hand side of Fig.~\ref{fig:electron_v2_rhic2} shows the estimated
$v_2$ precision \cite{electron_v2_rhic2_estimates} from semileptonic charm and 
bottom decays in the PHENIX central arms for a run with a 30 $nb^{-1}$ 
delivered luminosity.

Thus, separate $R_{AA}$ and $v_2$ measurements as a function of $p_T$ and 
$y$ should be possible for both open charm and bottom over a wide 
rapidity range at RHIC II.

With a displaced vertex measurement and RHIC II luminosity, the $B
\rightarrow J/\psi X$ decay channel can provide a very clean
measurement of open bottom production by both PHENIX and STAR (see
Tables~\ref{rhic1_rhic2_yields_phenix} and
~\ref{rhic1_rhic2_yields_star}).  The displaced vertex distributions
for prompt $J/\psi$ and for $B \rightarrow J/\psi X$ decays accepted in the
PHENIX muon arms are compared on the left-hand side of
Fig.~\ref{fig:decay_lengths}. The yields in
Tables~\ref{rhic1_rhic2_yields_phenix} and
~\ref{rhic1_rhic2_yields_star} assume a displaced vertex cut of 1 mm.
A good measurement of the cross section and $R_{AA}$ as a function of 
$p_T$ and $y$ for open bottom production will be possible using $B \rightarrow
J/\psi X$.  Even at RHIC II luminosity, however, the yields are not
expected to be large enough to permit a $v_2$ measurement.

\section{Hidden heavy flavor: quarkonium}

In this section we present a more detailed discussion of 
the theoretical motivation for studying heavy quarkonia in heavy-ion 
collisions.  We also summarize the present experimental and theoretical 
status and describe the proposed RHIC II experimental quarkonia program.

\subsection{Theoretical results}

\subsubsection{Cross sections in \pp\ collisions}

We discuss quarkonium 
production in the color evaporation model (CEM) which can be
used to calculate the total quarkonium cross sections. 
The CEM \cite{Barger:1979js,Barger:1980mg}
has enjoyed considerable phenomenological success. 
In the CEM, the quarkonium production cross section is some fraction $F_C$ of 
all $Q \overline Q$ pairs below the $H \overline H$ threshold where $H$ is
the lowest mass heavy-flavor hadron.  Thus the CEM cross section is
simply the $Q \overline Q$ production cross section with a cut on the pair mass
but without any constraints on the 
color or spin of the final state.  The produced $Q
\overline Q$ pair then neutralizes its color by
interaction with the collision-induced color field---``color evaporation''.
The $Q$ and the $\overline Q$ either combine with light
quarks to produce heavy-flavored hadrons or bind with each other 
to form quarkonium.  The additional energy needed to produce
heavy-flavored hadrons when the partonic center-of-mass energy, 
$\sqrt{\hat s}$, is less than $2m_H$, the heavy hadron
threshold, is obtained nonperturbatively from the
color field in the interaction region.
Thus the yield of all quarkonium states
may be only a small fraction of the total $Q\overline 
Q$ cross section below $2m_H$.
At leading order, the production cross section of quarkonium state $C$ in
an $AB$ collision is
\begin{eqnarray}
  \sigma_C = F_C \sum_{i,j} \int_{4m_Q^2}^{4m_H^2} d\hat{s}
  \int dx_1 dx_2~f_i^A(x_1,\mu^2)~f_j^B(x_2,\mu^2)~ 
  \hat\sigma_{ij}(\hat{s})~\delta(\hat{s} - x_1x_2s)\, 
  \,  \label{sigtil}
\end{eqnarray} 
where $\hat{s}$ is the square of the parton-parton center of mass energy, 
$ij = q \overline q$ or $gg$ and $\hat\sigma_{ij}(\hat s)$ is the
$ij\rightarrow Q\overline Q$ subprocess cross section.  The total $Q \overline
Q$ cross section takes $\hat{s} \rightarrow s$ in the upper limit of the 
integral over $\hat{s}$ in Eq.~(\ref{sigtil}).

The fraction $F_C$ must be universal so that, once it is fixed by
data, the quarkonium production ratios should be constant as a
function of $\sqrt{s}$, $y$ and $p_T$.  The actual value of $F_C$
depends on the heavy quark mass, $m_Q$, the scale, $\mu^2$, the parton
densities, $f_i^A(x,\mu^2)$ and the order of the calculation.  It was
shown in Ref.~\cite{Gavai:1994in} that the quarkonium production
ratios were indeed constant, as expected by the model.

Of course the leading order calculation in Eq.~(\ref{sigtil}) is
insufficient to describe high $p_T$ quarkonium production since the $Q
\overline Q$ pair $p_T$ is zero at LO.  Therefore, the CEM was taken
to NLO \cite{Gavai:1994in,Schuler:1996ku} using the exclusive $Q
\overline Q$ hadroproduction code of Ref.~\cite{Mangano:kq}.  At NLO
in the CEM, the process $gg \rightarrow g Q \overline Q$ is included,
providing a good description of the quarkonium $p_T$ distributions at
the Tevatron \cite{Schuler:1996ku}.  In the exclusive NLO calculation
\cite{Mangano:kq}, both the $Q$ and $\overline Q$ variables are
integrated to obtain the pair distributions.  Thus, although $\mu
\propto m_Q$ in analytic LO calculations, at NLO, $\mu^2 \propto m_T^2 =
m_Q^2 + p_T^2$ where $p_T$ is that of the $Q \overline Q$ pair, $p_T^2
= 0.5(p_{T_Q}^2 + p_{T_{\overline Q}}^2)$.

\begin{figure}[tbh]
\setlength{\epsfxsize=0.95\textwidth}
\setlength{\epsfysize=0.4\textheight}
\centerline{\epsffile{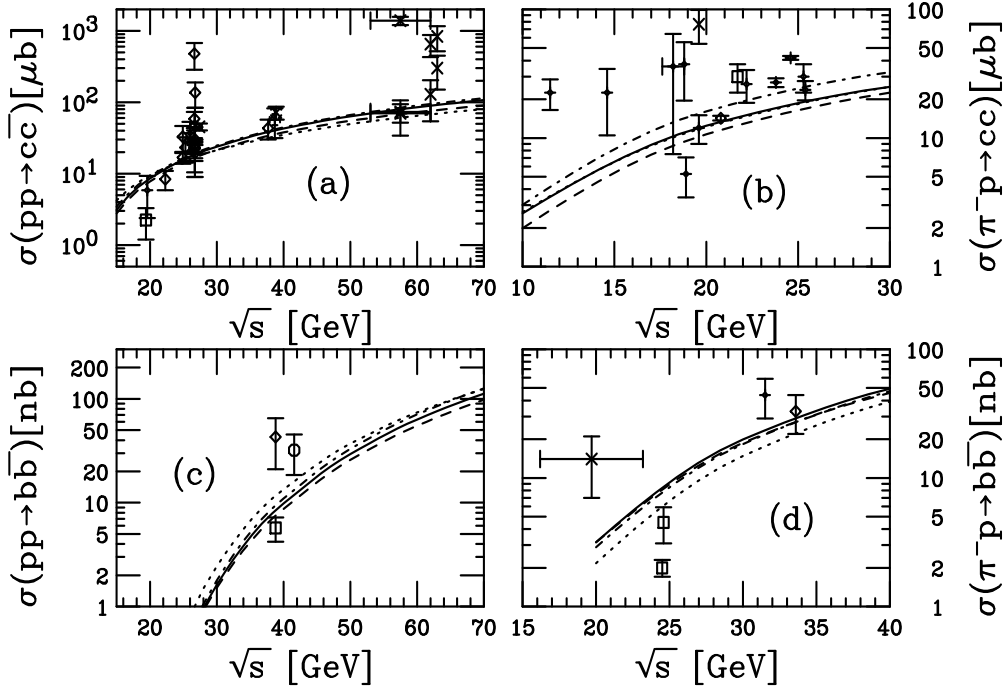}}
\caption{The $c \overline c$, (a) and (b), and $b \overline b$, (c) and (d), 
total cross section data in \pp\ and $\pi^- p$ interactions compared to NLO
calculations.  In (a) and (b), we show $\psi1$ (solid), $\psi2$ (dashed),
$\psi3$ (dot-dashed) and $\psi4$ (dotted).  In (c) and (d), we show
$\Upsilon1$ (solid), $\Upsilon2$ (dashed), $\Upsilon3$ (dot-dashed) and 
$\Upsilon4$ (dotted) \protect\cite{hfYR}.}
\label{qqbfig}
\end{figure}

We use parton densities and parameters that approximately agree with 
the $Q \overline Q$ total cross section data, given in Table~\ref{qqbparams},
to determine $F_C$ for $J/\psi$ and $\Upsilon$ production.  
The fit parameters \cite{Vogt:2002vx,Vogt:2002ve} for the parton densities 
\cite{Martin:1998sq,Lai:1999wy,Gluck:1998xa}, quark
masses and scales are given in
Table~\ref{qqbparams} while the $Q \overline Q$ cross sections calculated with
these parameters are 
compared to $pp \rightarrow Q \overline Q$ and $\pi^- p \rightarrow Q \overline
Q$ data in Fig.~\ref{qqbfig}.

\begin{table}[tbh]
\begin{center}
\begin{tabular}{|c|c|c|c|c||c|c|c|c|c|} \hline
\multicolumn{5}{|c||}{$c \overline c$} & \multicolumn{5}{c|}{$b 
\overline b$} \\ \hline
 Set & PDF & $m_c$ & $\xi_T$ & $F_{J/\psi}$ & Set & PDF & $m_b$ & 
$\xi_T$ & $F_{\Upsilon}$ \\ \hline
$\psi1$ & MRST HO & 1.2 & 2   & 0.0144 
 & $\Upsilon1$ & MRST HO & 4.75 & 1   & 0.0276 \\
$\psi2$ & MRST HO & 1.4 & 1   & 0.0248 
 & $\Upsilon2$ & MRST HO & 4.5  & 2   & 0.0201 \\
$\psi3$ & CTEQ 5M & 1.2 & 2   & 0.0155 
 & $\Upsilon3$ & MRST HO & 5.0  & 0.5 & 0.0508 \\
$\psi4$ & GRV 98 HO & 1.3 & 1 & 0.0229 
 & $\Upsilon4$ & GRV 98 HO & 4.75 & 1 & 0.0225 \\ \hline
\end{tabular}
\end{center}
\caption{The parton densities (PDFs), quark masses, $m_Q$, and scale to
mass ratios, $\xi_T = \mu/m_T$, used to obtain 
the `best' agreement to the $Q \overline Q$
cross sections. The quark mass is given in~\gevcc.  The inclusive $J/\psi$
production fraction, $F_{J/\psi}$, and the inclusive $\Upsilon$ production
fraction, $F_\Upsilon$, obtained from the data are also given 
\protect\cite{hfYR}.}
\label{qqbparams}
\end{table}

\begin{figure}[tbh]
\setlength{\epsfxsize=0.95\textwidth}
\setlength{\epsfysize=0.4\textheight}
\centerline{\epsffile{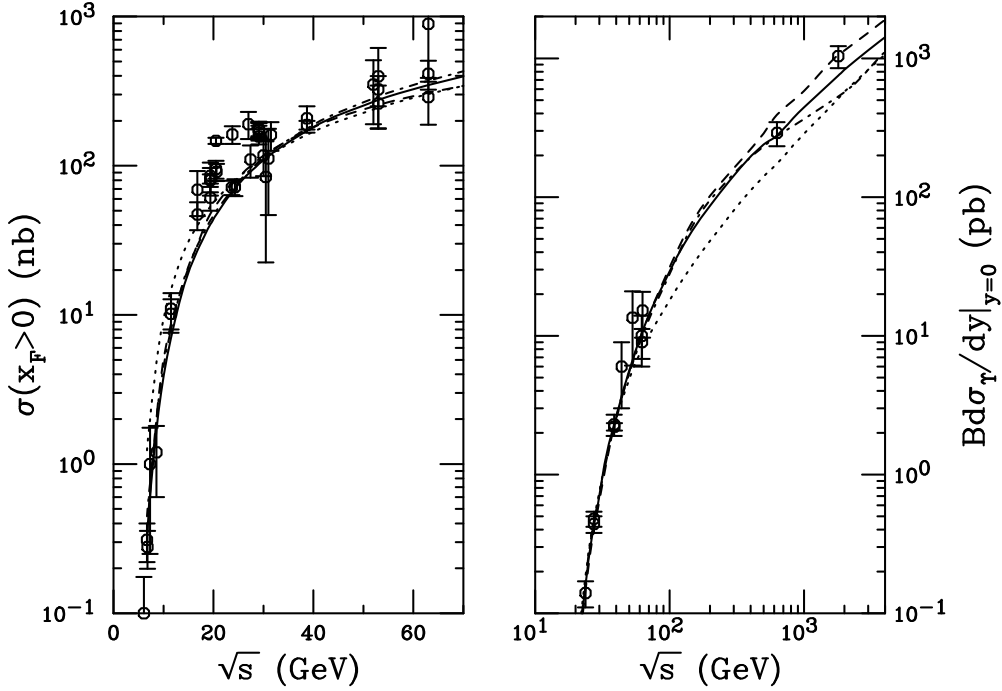}}
\caption{Forward $J/\psi$ (left) and combined $\Upsilon + \Upsilon' +
\Upsilon''$ inclusive (right) cross sections calculated to NLO in the CEM.  On
the left-hand side, we show $\psi1$ (solid), $\psi2$ (dashed),
$\psi3$ (dot-dashed) and $\psi4$ (dotted).  On the right-hand side, we show
$\Upsilon1$ (solid), $\Upsilon2$ (dashed), $\Upsilon3$ (dot-dashed) and 
$\Upsilon4$ (dotted) \protect\cite{hfYR}.}
\label{psiupsfixt}
\end{figure}

We now describe the extraction of $F_C$ in Eq.~(\ref{sigtil}) for the 
individual quarkonium states.
This is done using $J/\psi$ cross sections measured in \pp\ and \pA\ 
interactions up to $\sqrt{s} =
63$ \gev.  The data are of two types: the forward cross section, $\sigma(x_F >
0)$, and the cross section at zero rapidity, $d\sigma/dy|_{y=0}$.  All the
cross sections are inclusive with feed down from $\chi_c$ and $\psi'$
decays.  To obtain $F_{J/\psi}$ for inclusive $J/\psi$ production, the 
normalization of Eq.~(\ref{sigtil}) is obtained from a fit using the 
$c \overline c$ parameters
in Table~\ref{qqbparams}.  The comparison of $\sigma_{J/\psi}$ to the
$x_F > 0$ data for all four fits is shown on the left-hand side of 
Fig.~\ref{psiupsfixt}.  The ratios of the direct production cross sections to
the inclusive $J/\psi$ cross section can be determined from data on inclusive
cross section ratios and branching fractions.  These direct
ratios, $R_C$, given 
in Table~\ref{ratios}, are multiplied by the inclusive fitted $F_{J/\psi}$,
also shown in Table~\ref{qqbparams}, to
obtain the direct production fractions, $F^{\rm dir}_C = F_{J/\psi} R_C$, 
for the various quarkonium states.

\begin{table}[tbh]
\begin{center}
\begin{tabular}{|c|c|c|c|c|c|c|c|c|c|} \hline
& $J/\psi$ & $\psi'$ & $\chi_{c1}$ & $\chi_{c2}$ & $\Upsilon$ & $\Upsilon'$
& $\Upsilon''$ & $\chi_b(1P)$ & $\chi_b(2P)$ \\ \hline
$R_C$ & 0.62 & 0.14 & 0.60 & 0.99 & 0.52 & 0.33 & 0.20 & 1.08 & 0.84 \\ \hline
\end{tabular}
\end{center}
\caption{Direct quarkonium production ratios, $R_C = \sigma^{\rm 
dir}_C/\sigma_{C'}^{\rm inc}$ where $C' = J/\psi$ and $\Upsilon$.
From Ref.~\protect \cite{Digal:2001ue}.}
\label{ratios}
\end{table}

The same procedure, albeit somewhat more complicated due to the larger number 
of bottomonium states below the $B \overline B$ threshold, is followed for 
bottomonium.  For most data below $\sqrt{s} = 100$ \gev, 
the three bottomonium $S$ states were either not separated or their sum was 
reported.  No  $x_F$-integrated cross sections were available so that  
the CEM $\Upsilon$ cross section were fitted to the effective lepton pair 
cross section at $y=0$ for the three $\Upsilon(nS)$ states.  The extracted 
fit fraction, $F_{{\small \sum}
\Upsilon}$, combined with $\sigma_\Upsilon$ and 
compared to the data for all parameter sets in Table~\ref{qqbparams}, 
is shown on the
right-hand side of Fig.~\ref{psiupsfixt}.  Using the individual branching
ratios of the $\Upsilon$, $\Upsilon'$ and $\Upsilon''$ to lepton pairs and the 
total cross sections reported by CDF \cite{Affolder:1999wm}, 
it is possible to extract 
the inclusive $\Upsilon$ fit fraction, $F_\Upsilon$, given in
Table~\ref{qqbparams}.  The direct production 
ratios obtained in Ref.~\cite{gunvogt} have been updated in 
Ref.~\cite{Digal:2001ue} using recent CDF $\chi_b$ data.  
The resulting direct to inclusive $\Upsilon$ ratios, $R_C$, are also
given in Table~\ref{ratios}.  The sub-threshold $b \overline b$ cross section
is then multiplied by $F_C^{\rm dir} = F_\Upsilon R_C$ to obtain the direct
bottomonium cross sections.

The total cross sections for the charmonium and bottomonium states 
in \pp\ collisions at $\sqrt{s}
= 200$ \gev are shown in Tables~\ref{sigpsi} and \ref{sigups} respectively.
\begin{table}[tbh]
\begin{center}
\begin{tabular}{|c|c|c|c|c|c|} \hline
 & $\sigma_{J/\psi}^{\rm inc}$ ($\mu$b) &
 $\sigma_{J/\psi}^{\rm dir}$ ($\mu$b) & $\sigma_{\chi_{c1}}$ ($\mu$b)
& $\sigma_{\chi_{c2}}$ ($\mu$b) & $\sigma_{\psi'}$ ($\mu$b)  \\ \hline
$\psi$1  & 2.35 & 1.46 & 1.41 & 2.33 & 0.33 \\
$\psi$2  & 1.76 & 1.09 & 1.06 & 1.74 & 0.25 \\
$\psi$3  & 2.84 & 1.76 & 1.70 & 2.81 & 0.40 \\
$\psi$4  & 2.10 & 1.31 & 1.26 & 2.08 & 0.29 \\ \hline
\end{tabular}
\end{center}
\caption{The charmonium cross sections
for 200 \gev \pp\ collisions.
The inclusive and direct $J/\psi$ cross sections are both given.}
\label{sigpsi}
\end{table}
\begin{table}[tbh]
\begin{center}
\begin{tabular}{|c|c|c|c|c|c|c|} \hline
 & $\sigma_{\Upsilon}^{\rm inc}$ (nb) & $\sigma_{\Upsilon}^{\rm dir}$ (nb) &
 $\sigma_{\Upsilon'}$ (nb) & $\sigma_{\Upsilon''}$ (nb)
& $\sigma_{\chi_b(1P)}$ (nb) & $\sigma_{\chi_b(2P)}$ (nb) \\ \hline
$\Upsilon$1 & 6.60 & 3.43 & 2.18 & 1.32 & 7.13 & 5.54  \\
$\Upsilon$2 & 7.54 & 3.92 & 2.49 & 1.51 & 8.15 & 6.34  \\
$\Upsilon$3 & 5.75 & 2.99 & 1.90 & 1.15 & 6.21 & 4.83  \\
$\Upsilon$4 & 4.31 & 2.24 & 1.42 & 0.86 & 4.66 & 3.62  \\ \hline
\end{tabular}
\end{center}
\caption{The direct bottomonium cross sections
for \pp\ collisions at 200 \gev.  The production
fractions for the total $\Upsilon$
are multiplied by the appropriate ratios determined from data.}
\label{sigups}
\end{table}

The energy dependence, shown for both states in Fig.~\ref{psiupsfixt}, is well
reproduced by the NLO CEM.  All the fits are equivalent for $\sqrt{s} = 100$
\gev\ but differ by up to a factor of two at 2 TeV.  
The high energy $\Upsilon$ data seem to agree 
best with the energy dependence of $\Upsilon1$ and $\Upsilon2$ although
$\Upsilon1$ underestimates the Tevatron result by a factor of $\approx 1.4$.

\subsubsection{Cold nuclear matter effects on quarkonium production at RHIC}

It is essential that the $A$ dependence be
understood in cold nuclear matter to set a proper baseline for quarkonium
suppression in \AA\ collisions.  The NA50 collaboration
has studied the $J/\psi$ $A$ dependence and attributed its behavior to
dissociation by nucleons in the final state, referred to as nuclear absorption.
However, the parton distributions are modified in
the nucleus relative to free protons.  This modification, referred to here as
shadowing, is increasingly important at higher energies, as emphasized in
Ref.~\cite{psidaprl}.  We now discuss
the interplay of shadowing and absorption in \dAu\ and \AA\ collisions at
RHIC.
 
Shadowing, the modification of the parton densities in
the nucleus with respect to the free nucleon, is taken into account by
replacing $f_j^p(x,\mu^2)$ in Eq.~(\ref{sigtil}) by
$F_j^A(x,\mu^2,\vec b,z) = \rho_A(\vec b,z) S^j(A,x,\mu^2,\vec b,z)
f_j^p(x,\mu^2)$ and adding integrals over the spatial coordinates.  Here $S^j$
is the shadowing parameterization.
The density distribution of the deuteron is also included
in these calculations but the small effects of shadowing in deuterium are
ignored.  The PHENIX $J/\psi$ \dAu\ data as a function of rapidity show a
dependence consistent with nuclear shadowing plus a small absorption
cross section, common to all quarkonium states \cite{ksoct}, of $1-3$ mb. 
The $J/\psi$ production cross section is calculated 
in the CEM using Eq.~(\ref{sigtil}) with the same mass and scale as in $c
\overline c$ production.  The calculations of the \dAu/\pp\ and \AA/\pp\ 
ratios are done at LO to simplify the calculations since the LO and NLO 
ratios are equivalent \cite{prcda}.
 
To implement nuclear absorption in
\dAu\ collisions, the per nucleon production cross section is weighted by the
survival probability, $S^{\rm abs}$ \cite{rvherab},
\begin{eqnarray}
S^{\rm abs}(\vec b,z) = \exp \left\{
-\int_{z}^{\infty} dz^{\prime}
\rho_A (\vec b,z^{\prime})
\sigma_{\rm abs}(z^{\prime} - z)\right\} \, \, ,
\label{nsurv}
\end{eqnarray}
where $z$ is the longitudinal production point and $z^{\prime}$
is the point at which the state is absorbed.
The nucleon absorption cross section, $\sigma_{\rm abs}$, typically
depends on where the
state is produced in the medium and how far it travels through nuclear matter.
If absorption alone is active, {\it i.e.}\ no shadowing so that
$S^j \equiv 1$, then an effective minimum bias $A$ dependence
is obtained after integrating $S^{\rm abs}$ over the spatial coordinates.
If $S^{\rm abs} = 1$ also, $\sigma_{{\rm d}A} = 2A \sigma_{pN}$.
When $S^{\rm abs} \neq 1$, $\sigma_{{\rm d}A} \sim 2A^\alpha \sigma_{pN}$ 
where, if $\sigma_{\rm abs}$ is a constant, independent of the production 
mechanism for a hard sphere nucleus of constant density inside $R_A$, 
{\it ie.} $\rho_A = \rho_0 \theta(R_A - b)$,
$\alpha = 1 - 9\sigma_{\rm abs}/(16 \pi r_0^2)$
with $r_0 = 1.2$ fm.  The contribution to the full $A$
dependence of $\alpha$ from absorption alone is only constant if
$\sigma_{\rm abs}$ is constant and independent of the production mechanism
\cite{rvherab}.
The observed $J/\psi$ yield includes feed down from $\chi_{c}$ and $\psi'$
decays, 
\begin{eqnarray}
S_{J/\psi}^{\rm abs}(b,z) = 0.6 S_{J/\psi, \, {\rm dir}}^{\rm abs}(b,z)
+ 0.3 S_{\chi_{c}}^{\rm abs}(b,z) + 0.1 S_{\psi'}^{\rm abs}(b,z)
\, \, . \label{psisurv}
\end{eqnarray}

\begin{figure}[tbh]
  \setlength{\epsfxsize=0.95\textwidth}
  \setlength{\epsfysize=0.3\textheight}
  \centerline{\epsffile{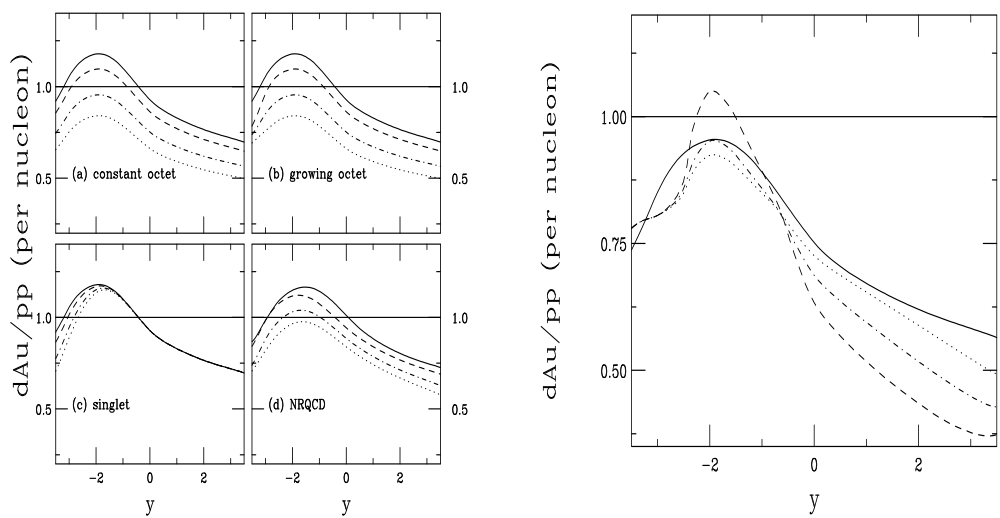}}
  \caption{Left-hand side:  The $J/\psi$ \dAu/\pp\ ratio with EKS98 at 200 \gev\
      as a function of rapidity for (a) constant octet, (b) growing
      octet, (c) singlet, all calculated in the CEM and (d) NRQCD.
      For (a)-(c), the curves are no absorption (solid), $\sigma_{\rm
          abs} = 1$ (dashed), 3 (dot-dashed) and 5 mb (dotted).  For
      (d), we show no absorption (solid), 1 mb octet/1 mb singlet
      (dashed), 3 mb octet/3 mb singlet (dot-dashed), and 5 mb octet/3
      mb singlet (dotted).  Right-hand side: The $J/\psi$ \dAu/\pp\ 
      ratio at 200 \gev\ for a growing octet with $\sigma_{\rm abs} =
      3$ mb is compared for four shadowing parameterizations.  We show
      the EKS98 (solid), FGSo (dashed), FGSh (dot-dashed) and FGSl
      (dotted) results as a function of rapidity.  From
      Ref.~\protect\cite{prcda}, reprinted with permission from APS.
  }
  \label{abs_da}
\end{figure}
 
\begin{figure}[tbh]
  \setlength{\epsfxsize=0.95\textwidth}
  \setlength{\epsfysize=0.4\textheight}
  \centerline{\epsffile{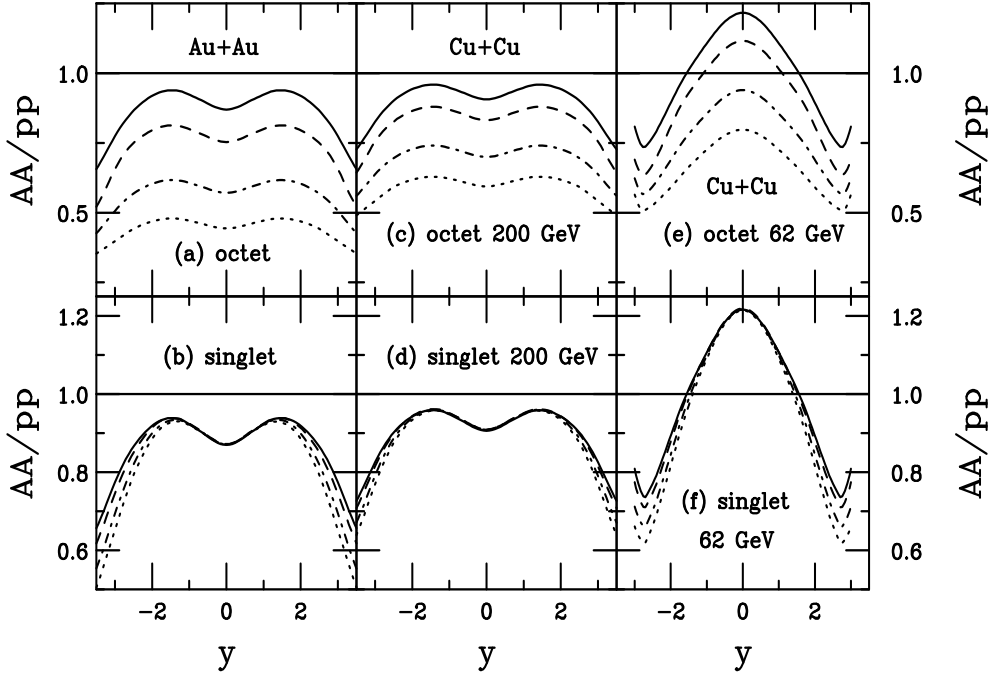}}
  \caption{The $AA/pp$ ratio with the
      EKS98 parameterization as a function of $y$ for octet (upper)
      and singlet (lower) absorption.  In (a) and (b) we show the
      \AuAu\ results at 200 \gev\ while the \CuCu\ results are shown
      at 200 \gev\ (c) and (d) as well as at 62 \gev\ (e) and (f).
      The curves are $\sigma_{\rm abs} = 0$ (solid), 1 (dashed), 3
      (dot-dashed) and 5 mb (dotted). From Ref.~\protect\cite{rvhip},
      reprinted with permission from Acta Physica Hungarica.  }
  \label{abs_aa}
\end{figure}

The $J/\psi$ may be produced as a color singlet, a color octet or in a
combination of the two.  In color singlet production, the 
absorption cross section depends on the size of the $c \overline c$
pair as it traverses the nucleus, allowing absorption to be effective
only while the cross section is growing toward its asymptotic size
inside the target.  On the other hand, if the $c \overline c$ is 
produced as a color octet, hadronization will occur only after the
pair has traversed the target except at very backward rapidity.  We
have considered a constant octet cross section (hadronization 
outside the medium at all rapidity), 
as well as a growing octet cross section which hadronizes to a color singlet 
inside the medium at backward rapidities.  
For singlets,
$S_{J/\psi, \, {\rm dir}}^{\rm abs} \neq S_{\chi_{c}}^{\rm abs} \neq
S_{\psi'}^{\rm abs}$ but, with octets, one assumes that $S_{J/\psi, \,
    {\rm dir}}^{\rm abs} = S_{\chi_{c}}^{\rm abs} = S_{\psi'}^{\rm
    abs}$.  If this assumption is relaxed and the octet absorption
cross sections depend on the final-state size, then
$\sigma^{\psi^\prime}_{\rm abs} > \sigma^{\chi_c}_{\rm abs} >
\sigma^{J/\psi \, {\rm dir}}_{\rm abs}$ and $S_{J/\psi, \, {\rm
        dir}}^{\rm abs} > S_{\chi_{c}}^{\rm abs} > S_{\psi'}^{\rm
    abs}$.  The feed down contributions then effectively lower the
value of $\sigma_{\rm abs}$ needed to describe the data.  As can be
seen in Fig.~\ref{abs_da}, the difference between the constant and
growing octet assumptions is quite small at large \sqrtsNN\ with only
a small singlet effect at $y< -2$.  Singlet absorption is also
important only at similarly negative rapidities. At other rapidities singlet
production is dominated by shadowing alone.  Finally, we have also considered 
a combination
of octet and singlet absorption in the context of the NRQCD approach,
see Ref.~\cite{rvherab} for more details.  The combination of
nonperturbative singlet and octet parameters changes the shape of the
shadowing ratio slightly.  Including the singlet contribution weakens
the effective absorption.  The results are shown integrated over
impact parameter.  The calculations use the EKS98 shadowing
parameterization \cite{EKS} since it gives good agreement with the
trend of the PHENIX data.  For results with other shadowing
parameterizations, see Refs.~\cite{prcda,rvhip}.

Several values of the asymptotic absorption cross section,
$\sigma_{\rm abs} = 1$, 3 and 5 mb, corresponding to $\alpha = 0.98$,
0.95 and 0.92 respectively using Eqs.~(\ref{nsurv}) and
(\ref{psisurv}), are shown in Figs.~\ref{abs_da} and \ref{abs_aa} for
\dAu\ and \AA\ collisions respectively.  These values of $\sigma_{\rm
    abs}$ are somewhat smaller than those obtained for the sharp
sphere approximation.  The diffuse surface of a real nucleus and the
longer range of the density distribution result in a smaller value of
$\sigma_{\rm abs}$ than a spherical nucleus.  As will be seen later,
there is good agreement with the trend of the PHENIX data
\cite{PHENIX_jpsi_pp&dau_run3} for $\sigma_{\rm abs} = 0-3$ mb.  Work
is in progress to quantify the shadowing parameterization and
absorption cross section more precisely over a range of energies
\cite{mikeandme}.

The current RHIC data are not sufficiently precise to distinguish
between $J/\psi$ production and absorption in the CEM relative to that
in the NRQCD approach.  However, a measurement of the $\chi_c$ $A$
dependence may be able to clarify the situation \cite{rvherab}.  In
the CEM, the $J/\psi$ and $\chi_c$ distributions differ only in the
value of $F_C$.  In the NRQCD approach, the $J/\psi$ is produced
primarily in a color octet state while the $\chi_c$ is produced as a
color singlet state.  Thus while the production of both states would
exhibit the same shadowing effect, a difference in the $J/\psi$ and
$\chi_c$ \dAu/\pp\ ratios due to octet relative to singlet absorption
may be measurable.

We now turn to the centrality dependence of $J/\psi$ production in
\dAu\ and \AA\ collisions.  In central collisions, inhomogeneous
(spatially dependent) shadowing is stronger than the homogeneous
(averaged over centrality) result.  The stronger the homogeneous shadowing, the
larger the inhomogeneity.  In peripheral collisions, inhomogeneous
effects are weaker than the homogeneous results but some shadowing is
still present.  Shadowing persists in peripheral collisions in part
because the density in a heavy nucleus is large and approximately
constant except close to the surface, and because the deuteron wave
function has a long tail.  Absorption is also expected to be stronger
in central collisions.
  
\begin{figure}[tbh]
    \setlength{\epsfxsize=0.95\textwidth}
    \setlength{\epsfysize=0.5\textheight}
    \centerline{\epsffile{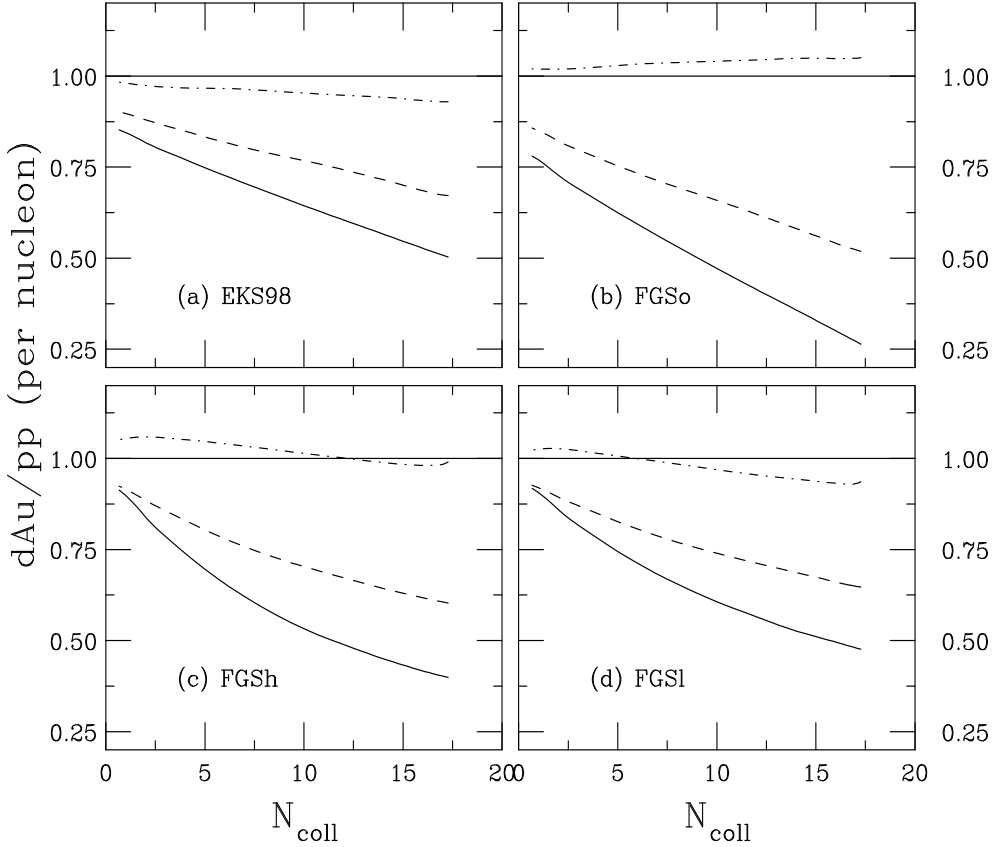}}
  \caption{The ratio \dAu/\pp\ as a function
      of \nbin\ for the EKS98 (a), FGSo (b), FGSh (c) and FGSl (d)
      shadowing parameterizations.  The calculations with EKS98 and
      FGSo use the inhomogeneous path length parameterization while
      that obtained by FGS is used with FGSh and FGSl.  Results are
      given for $y=-2$ (dot-dashed), $y=0$ (dashed) and $y=2$ (solid)
      at 200 \gev\ for a growing octet with $\sigma_{\rm abs} = 3$ mb.
      From Ref.~\protect\cite{prcda}, reprinted with permission from
      APS.  }
  \label{ncoll_da}
\end{figure}

\begin{figure}[tbh]
    \setlength{\epsfxsize=0.95\textwidth}
    \setlength{\epsfysize=0.25\textheight}
    \centerline{\epsffile{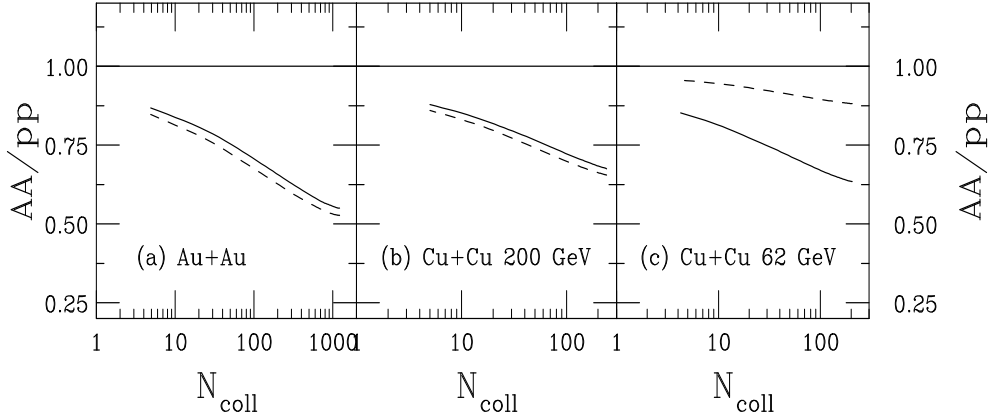}}
  \caption{The ratio $AA/pp$ as a function of \nbin\
      for a 3 mb octet absorption cross section and the EKS98
      parameterization at $y = 0$ (dashed) and $y=2$ (solid) for
      \AuAu\ at 200 \gev\ (a) and \CuCu\ at 200 \gev\ (b) and 62 \gev\ 
      (c). From Ref.~\protect\cite{rvhip}, reprinted with permission
      from Acta Physica Hungarica.  }
  \label{ncoll_aa}
\end{figure}

To study the centrality dependence of shadowing and absorption, the
\dAu/\pp\ and \AA/\pp\ ratios are presented as a function of \nbin,
\begin{eqnarray}
\nbin(b) = \sigma_{NN}^{\rm in} \int d^2s T_A(s)
T_B (|\vec b - \vec s|) \, \, , \nonumber
\end{eqnarray}
where $T_A$ and $T_B$ are the nuclear thickness functions and the
inelastic nucleon-nucleon cross section, $\sigma_{NN}^{\rm in}$, is 42
mb at 200 \gev.  In Figs.~\ref{ncoll_da} and \ref{ncoll_aa}, the
\nbin\ dependence is shown for several representative rapidities, $y =
-2$, 0 and 2 for RHIC.  The inhomogeneous shadowing parameterization
is chosen to be proportional to the path length of the parton through
the nucleus \cite{psidaprl}.  For more results, see
Refs.~\cite{prcda,rvhip}.

The dependence of the RHIC ratios on \nbin\ is almost linear, as seen
in Figs.~\ref{ncoll_da} and \ref{ncoll_aa}. The weakest \nbin\ dependence 
occurs in the
antishadowing region, illustrated by the $y = -2$ result (dot-dashed
curve).  The overall dependence on \nbin\ is stronger than that
obtained from shadowing alone, described in Ref.~\cite{psidaprl},
where inhomogeneous shadowing effects depend strongly on the amount of
homogeneous shadowing.  Relatively large effects at low $x$ are
accompanied by the strongest impact parameter, $b$, dependence.  In
the transition region around midrapidity at RHIC, the $b$ dependence
of the ratio \dAu/\pp\ due to shadowing is nearly negligible and
almost all of the \nbin\ dependence at $y \sim 0$ can be attributed to
absorption.  The $y=-2$ results for color singlet production and
absorption, in the antishadowing region, are fairly independent of
\nbin.  The stronger $AA/pp$ effect at $y=0$ in Fig.~\ref{ncoll_aa}(a)
and (b) can be described by the convolution of antishadowing at $y\sim -2$
in $R_{\rm dAu}$ and shadowing at $y\sim 2$, resulting in the dip in
$AA/pp$ at $y=0$ shown in Fig.~\ref{abs_aa}.
  
\subsubsection{Models of quarkonium production in heavy-ion collisions} 
\hspace*{50mm}

{\noindent\textit{In-medium properties of quarkonium from lattice QCD:}}

\begin{figure}[tbh]
  \begin{center}
    \epsfig{file=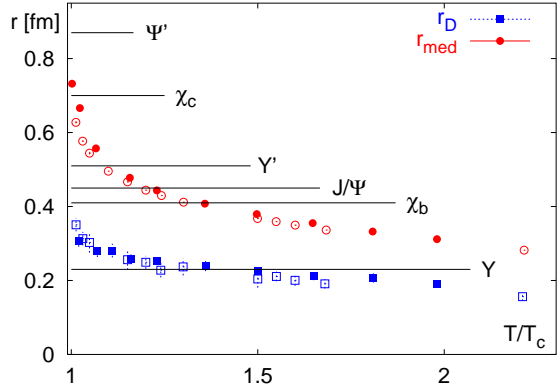,width=8.0cm}
  \end{center}
  \caption{Mean squared charge radii of some charmonium and bottomonium states 
    compared to the Debye screening radius, $r_D\equiv 1/m_D$, and a related 
    scale, $r_{\rm med}$, an estimate of the distance beyond
    which the force between a static $Q \overline Q$ pair is strongly modified
    by temperature effects \protect\cite{karsch}.
    Open (closed) symbols correspond to SU(3) (2-flavor QCD) calculations. 
    From Ref.~\protect\cite{freeenergy}, reprinted 
    with permission from Springer-Verlag.
  }
  \label{fig:screen}
\end{figure}

Properties of heavy quarks have been used to characterize ``thermal
properties of the QCD vacuum'' ever since the first lattice
calculations at non-zero temperature \cite{first}. Modifications of
the interactions between heavy, static quarks in a thermal heat bath
are clearly reflected by changes of the free energy which, in the zero
temperature limit, reduces to the heavy quark potential
\cite{freeenergy}. To use this information to analyze thermal
modifications of quarkonia requires an intermediate, phenomenological
step: the construction of a temperature-dependent effective potential
which then can be used in a nonrelativistic Schr\"odinger equation
\cite{Digal:2001iu,Wong:2004zr,alberico} or a more refined
coupled-channel analysis \cite{shuryak,blaschke}. Quite generically,
the potential model analyses suggest a sequential suppression pattern
where heavy quark bound states dissociate at temperatures at which
their bound state radii become comparable to the Debye screening
radius, illustrated in Fig.~\ref{fig:screen}. Table~\ref{satz_tdtc}
shows quarkonium dissociation temperatures from
Ref.~\cite{satz_0512217}.

\begin{table}[tbh]
  \begin{center}
    \begin{tabular}{|c||c|c|c||c|c|c|c|c|} \hline
      State & $J/\psi(1S)$ & $\chi_c(1P)$ & $\psi^\prime(2S)$ & $\Upsilon(1S)$ 
      & $\chi_b(1P)$ & $\Upsilon(2S)$ & $\chi_b(2P)$ & $\Upsilon(3S)$ \\ \hline
      $T_d/T_c$  & 2.10 & 1.16 & 1.12 & $>$ 4.10 & $<$ 1.76 & 1.60 & 1.19 & 
      1.17 \\ \hline
    \end{tabular}
  \end{center}
  \caption{Quarkonium dissociation temperatures \protect\cite{satz_0512217}, 
    illustrating the effects of binding energy on the dissociation temperature.}
  \label{satz_tdtc}
\end{table}

More recently, the calculation of thermal
hadron correlation functions and their spectral analysis \cite{hatsuda}
eliminated some of the ambiguities inherent in the potential model approach.
The spectral analysis, at least in principle, provides an ab-initio approach
to the calculation of in-medium properties of heavy quark bound states.
Its predictive power is reduced only by the application of statistical
tools like the Maximum Entropy Method (MEM) which, however, can be steadily
improved with further development of the available computing resources
and numerical techniques. Predictions based on potential model calculations
as well as the spectral analysis have been reviewed in recent studies that
have been performed to analyze prospects for quarkonium studies at the 
LHC \cite{hfYR,Brambilla:2004wf}. 
In the following, the analysis of thermal hadron correlation functions and the 
extracted spectral functions are discussed.

The finite temperature, Euclidean time correlation functions,
\begin{equation}
G_H(\tau,\vec{r}, T) =
\langle J_H(\tau, \vec{r}) J_H^\dagger (0, \vec{0}) \rangle \quad , 
\label{def2pt}
\end{equation}
of hadronic
currents, $J_H=\bar{q}(\tau,\vec{r})\Gamma_H q(\tau,\vec{r})$, where
$\Gamma_H$ denotes a suitable product of gamma matrices that
projects onto the appropriate quantum numbers of hadron $H$, 
are directly related to the spectral functions, $\sigma_H(\omega, T)$.  These 
spectral functions encompass all information about 
thermal modifications of the hadron spectrum in channel $H$ so that 
\begin{equation}
G_H (\tau,\vec{r},T) = 
\int_{0}^{\infty} d\omega\; \frac{d^3\vec{p}}{(2\pi)^3}
\sigma_H (\omega, \vec{p}, T) \;{\rm e}^{i\vec{p} \cdot \vec{r}}
{\cosh (\omega (\tau - 1/2T)) \over \sinh ( \omega /2T )}\quad 
\label{correlator}
\end{equation}
are directly related to experimental observables.
In particular, the spectral function in the vector channel, 
$\sigma_V(\omega,\vec{p},T)$, is directly related to the differential 
cross section for thermal dilepton production,
\begin{equation}
\frac{dW}{d\omega d^3p} =
{5 \alpha^2 \over 27 \pi^2} {\sigma_V(\omega,\vec{p},T)
\over \omega^2 ({\rm e}^{\omega/T} - 1)} \quad .
\label{rates}
\end{equation}
Note that the rates obtained using this method do not include any 
contributions arising from the feed down of other channels 
into the vector channel \cite{Digal:2001ue,Petronzio}.

Some general aspects of the influence of a thermal medium on
states with different quantum numbers can be deduced
from the temperature dependence of the thermal correlation functions
themselves and does not require the additional step of applying the
MEM analysis which is based on probabilistic assumptions. 
Such comparisons show that zero-momentum, thermal hadron correlation functions
in the ground state channels, {\it i.e.} the vector ($J/\psi$, $\Upsilon$)
and pseudoscalar ($\eta_c$, $\eta_b$) channels show little modification
in a thermal medium up to temperatures $T \, \gsim \, 1.5~T_c$. 
Correlation functions corresponding to radially
excited charmonium states ($\chi_c$), however, are already strongly modified
close to or at $T_c$. 

\begin{figure*}
\begin{center}
\epsfig{file=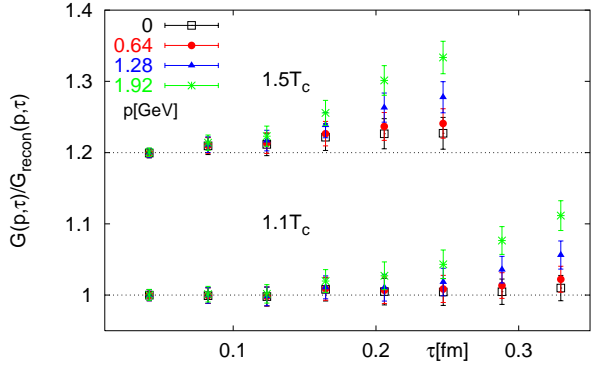,width=8.0cm}
\end{center}
\caption{The ratio of pseudoscalar correlation functions at non-zero momentum
above and below $T_c$ \cite{datta05}. Deviations from the horizontal
dotted lines
for increasing $p$ indicate stronger temperature-dependent modifications
of the spectral functions.
}
\label{fig:momentum}
\end{figure*}

These generic features are reflected by the spectral
functions. Although results from different groups currently still differ 
in detail, there are some general trends.
\begin{figure*}
\begin{center}
\includegraphics[width=0.7\textwidth]{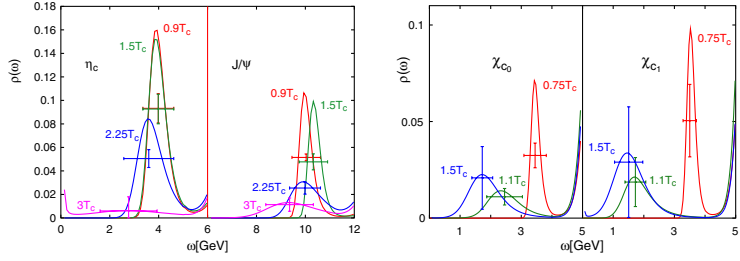}
\end{center}
\caption{Spectral functions of $\eta_c$, $J/\psi$ (left) and 
and $\chi_{c0},~\chi_{c1}$ (right) at various temperatures below and
above $T_c$.
The vertical bars give the error on the average value of the spectral function
in the bin indicated by the horizontal bars.  
}
\label{fig:charmonium}
\end{figure*}
The $J/\psi$ and $\eta_c$ remain unaffected by the thermal medium up to 
$T=1.5~T_c$, shown on the left-hand side of Fig.~\ref{fig:charmonium}.  
At higher temperatures it is unclear whether
the $J/\psi$ already disappears at $\simeq 1.9~T_c$ 
\cite{hatsuda2} or persists as a strongly modified resonance up to 
$2.25~T_c$ \cite{datta}.  The $\chi_{c0}$ and $\chi_{c1}$ both disappear at 
$T \, \lsim \, 1.1~T_c$, see the right-hand side of Fig.~\ref{fig:charmonium}.
Finite momentum $J/\psi$ states show statistically significant but
still small modifications for $T \, \lsim \, 1.5~T_c$ \cite{datta05} due to 
collisional broadening by higher momentum gluons seen by bound states moving 
relative to the heat bath, see Fig.~\ref{fig:momentum}
\cite{gale}.
Strong $J/\psi$ binding above $T_c$ is also supported
by the analysis of spatial correlations \cite{Umeda} and the observed
insensitivity of the thermal vector and pseudoscalar correlation functions 
to spatial boundary conditions \cite{doi}. 

Bottomonium studies are considerably more difficult since a larger
lattice cut off is required to properly resolve these states, particularly
for temperatures well above $2~T_c$ where the $\Upsilon$ states 
are expected to be dissolved.
First exploratory finite temperature results on bottomonium have been 
reported for temperatures up to $\sim 1.5~T_c$. At this temperature, no 
thermal modifications of the $\Upsilon$ and $\eta_b$ have been observed, as
expected. The $\chi_b$ correlation functions are, however,
modified at $T \sim 1.5~T_c$, similar to the scalar charmonium case
at $T\simeq 1.1~T_c$. 
To firmly establish the onset of medium modifications in bottomonium
states, however, requires further refined studies of the bottomonium system
at lower and higher temperatures.

More recent calculations of quarkonia screening by Mocsy and Petreckzy
\cite{Mocsy:2007jz} suggest that the quarkonium states break up at lower
temperatures.  They use $2+1$ flavor lattice results on color screening to fix
the parameters of the quarkonium potential used in the calculation of the
quarkonium correlators.  The most extreme potential compatible with the
lattice results is used to derive upper bounds on the quarkonium dissociation
temperatures in the quark-gluon plasma.  In this case, although resonance
structures appear in the spectral functions, the binding energy drops while
the thermal width increases.  When the width is larger than the binding energy,
the state can no longer effectively be observed.  These calculations suggest that
the $\chi_c$ and $\psi'$ break up at or below $T_c$ while the $J/\psi$ and
higher $\Upsilon$ family resonances dissolve around $1.2 T_c$.  Only the
$\Upsilon(1S)$ state would persist to $\sim 2T_c$.  Further calculations are
needed to resolve the issue. \\

{\noindent \textit{Dynamical Coalescence:}}

The production of multiple $c \overline c$ pairs in a single collision 
introduces a new charmonium production mechanism \cite{Thews:2000rj}.  
Charmonium states can be formed in-medium by coalescence of a $c$ and a
$\overline c$ quark from independently-produced $c \overline c$ pairs.  

In the plasma phase, there are two basic approaches: statistical and dynamical
coalescence.  Both these approaches depend on being able to measure the
quarkonium rate relative to total $Q \overline Q$ production.
The first calculations in the statistical approach assumed an
equilibrated fireball in a grand canonical ensemble \cite{pbmjs1,pbmjs2}.
This approach could be reasonable at the high energies of the LHC, where the
number of produced $c \overline c$ pairs is large. But, at lower energies, 
charm conservation is required since a $c \overline c$ pair is not produced 
in every event.  More recent calculations assumed a canonical ensemble only 
for charm production \cite{goren1,goren2,goren3}. 
Dynamical coalescence models assume that some of the produced $Q \overline Q$
pairs which would not otherwise do so can also form quarkonium. This 
coalescence can take place in the QGP \cite{Thews:2000rj,grandch1} or at 
hadronization \cite{grrapp}.
The model includes the rapidity differences, $|\Delta y|$, 
between the $Q$ and $\overline Q$ and shows
that the enhancement is smaller for large $|\Delta y|$.  
The impact parameter dependence of statistical and dynamical coalescence
is quite different.  Statistical coalescence gives the largest 
enhancement in peripheral collisions where the volume of the plasma is small, 
with only a minor enhancement in central collisions.  Dynamical
coalescence produces a larger enhancement in central collisions where the
number of $Q \overline Q$ pairs per event is greatest but still produces a 
significant effect in peripheral collisions \cite{thewslhc}.  

Much smaller enhancements are predicted for secondary quarkonium production
in the hadron gas, particularly for the $J/\psi$ where the additional 
production is either small ($20 - 60$\%) \cite{krpbm} or 
about a factor of two \cite{kzwz} at LHC energies and smaller still for RHIC.  
Larger enhancements may be expected for the
$\psi'$ \cite{krpbm}.  The predictions depend strongly on the
$J/\psi + \pi(\rho)$ cross sections, typically not more than $1-2$ mb 
\cite{psicomo}.  

These secondary production models are already testable at RHIC where
factors of $2-3$ enhancement are expected from coalescence 
\cite{Thews:2000rj,goren3}.  Hard scattering of produced particles 
from different interactions is related
to the idea of crosstalk between unrelated interactions \cite{cross}.  
Important crosstalk effects were predicted in
$e^+ e^-$ collisions at LEP \cite{cross} but were not observed.  If secondary 
quarkonium production is found, it would indicate the relevance of such 
effects.  Secondary quarkonium may be separated from primary
quarkonium, which is subject to suppression, by kinematic cuts that 
take advantage of the fact that different mechanisms dominate production in 
different kinematic regimes.  

Predictions of $J/\psi$ production by dynamical
coalescence suffer from substantial uncertainties due to the dependence on
the charm quark distributions in the medium.  In fact, it is possible to turn
this uncertainty into an advantage and probe the medium properties using
the observed $J/\psi$ momentum distributions.
Two extremes can be considered \cite{BobMic}: either the initial
$c$ quark distributions are unchanged by the presence of the medium or
the $c$ quarks are thermalized. If the charm quark distributions
in the medium are identical to those of the initial production process,
the interactions of charm quarks with the medium would be very weak.  In
this case, both the $J/\psi$ rapidity and $p_T$ distributions will be narrower
than if no plasma is formed simply because the center of mass energy of 
secondary $J/\psi$ production is lower than that of the initial nucleon-nucleon
interactions. The lower energy results in a reduced $\langle p_T^2 \rangle$
and a narrower rapidity distribution.  Thus, instead of the transverse momentum
broadening expected from initial-state multiple scattering going from \pp\
to \pA\ to \AA, the average $p_T^2$ in \AA\ would no longer exhibit the 
monotonic increase seen in \pp\ and \pA\ interactions for increasing $A$.
On the other hand, if the charm quarks are assumed to be in thermal equilibrium
with the surrounding medium, the charm interaction with the medium would
be very strong.  Any $J/\psi$'s produced from thermalized charm quarks
flowing with the medium would have a $p_T$ distribution with a slope 
characteristic of the temperature of the system at the time they were formed,
resulting in even narrower rapidity and $p_T$ distributions.
In either case, the effect would be largest in central collisions, 
reverting to ``normal''
broadening in peripheral collisions where on the order of one or fewer
$c \overline c$ pairs will be produced since the number of $c \overline c$
pairs scales approximately with the number of collisions.

The $\langle p_T^2 \rangle$ of coalescence and suppression has been calculated
by several groups \cite{yan_zhuang_xu_2006,zhao_rapp_2007,thews_vienna}.  
No realistic calculations of the rapidity distribution from models including 
\jpsi\ coalescence are available for heavy-ion collisions so far
but, based on the assumption of an underlying open charm distribution peaked at
$y$ = 0, it is predicted \cite{BobMic} that 
a strong charm coalescence contribution to $J/\psi$ production will lead to a 
narrowing of the rapidity distribution, similar to the $p_T$ distribution.

In order to extract the medium properties from secondary $J/\psi$ production,
systematic studies of $J/\psi$ production in \pp, \pA\ and \AA\ interactions
are necessary.  The \pp\ data determine the intrinsic transverse momentum scale
for a particular energy while the \pA\ results fix the level of 
broadening due to cold nuclear matter effects which would then apply to
\AA\ interactions at the same energy.

Models of coalescence, of course, also include $J/\psi$ suppression.
In addition to the screening effects discussed previously, the $J/\psi$
can scatter with quarks and gluons in the plasma which may break it up
more efficiently than screening effects alone, especially if temperatures
significantly above $T_c$ are needed for screening to dissociate the directly 
produced $J/\psi$, as discussed in Ref.~\cite{grrapp}.
At low temperatures, relevant for SPS energies, $g J/\psi \rightarrow c
\overline c$ is effective for $J/\psi$ breakup by a thermal gluon.  However,
at higher temperatures where the $J/\psi$ should be more loosely bound, 
inelastic parton scattering, $g (q, \overline q) J/\psi \rightarrow g (q, 
\overline q) c \overline c$, calculated using the leading order matrix
elements for $gc$ and $gq$ scattering, is more effective.

\begin{figure}[tbh]
  \centering\includegraphics[width=0.5\textwidth]{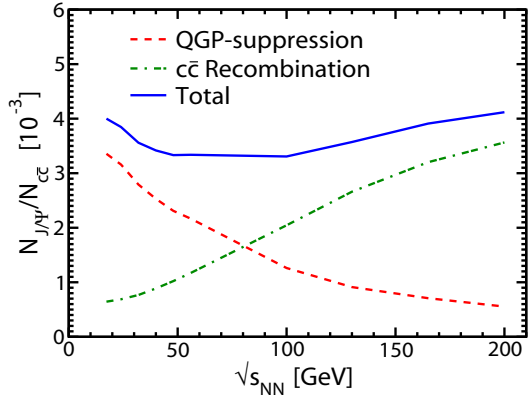}
  \caption{Excitation function of the ratio of produced $J/\psi$ to
    the number of  $c \overline c$ pairs in central heavy-ion collisions
    for $N_{\rm part} = 360$. From Ref.~\protect\cite{rapp2}, reprinted with 
    permission from Elsevier.}
  \label{fig:rapp}
\end{figure}

The relative importance of $J/\psi$ suppression and coalescence will change
as a function of energy, as shown in Fig.~\ref{fig:rapp} for central 
collisions of heavy nuclei with $N_{\rm part} = 360$, from Ref.~\cite{rapp2}.  
The $J/\psi$ yield is dominated by primordial production at SPS energies and 
by coalescence at the full RHIC energy.\\

{\noindent \textit{Other Models:}}\\

{\noindent \textit{Comovers:}}
The comover model, proposed in Refs.~\cite{ggj,vkph,roman} and further 
developed by Capella {\it et al.}, see most recently Ref.~\cite{capella_2007}, 
describes $J/\psi$
suppression in nucleus-nucleus collisions by final-state interactions with
the dense medium created in the collision, the ``comovers''.  The SPS
$J/\psi$ data are reproduced with a small effective dissociation cross section,
$\sigma_{\rm co} = 0.65$ mb \cite{acf}.  At RHIC energies, comover interactions
alone would suggest stronger suppression than at the CERN SPS since the density
of produced particles is larger at RHIC.  Since the produced particle density
is largest at midrapidity, comover interactions alone would predict stronger 
suppression at $y=0$ than at forward rapidity \cite{ferriero_2006}.  

However, although the comover
model is based on rate equations, the gain term was only previously considered
in Ref.~\cite{vkph} and reintroduced by Capella {\it et al.} in an attempt to
account for the stronger observed suppression at forward rapidity at RHIC
\cite{capella_2007}.  The open charm cross section at forward rapidity, 
where RHIC data are not yet very precise, was obtained from the shape of 
PYTHIA simulations of the rapidity distribution.  As expected, coalescence 
effects are stronger at midrapidity so that the behavior calculated in 
Ref.~\cite{ferriero_2006} is reversed in
Ref.~\cite{capella_2007} and forward suppression is found to be larger.\\

{\noindent \textit{Sequential Suppression:}}
As discussed previously, interpretation of the quarkonium dissociation 
temperatures in Table~\ref{satz_tdtc} suggests a sequential suppression pattern
in which the quarkonium states dissolve when the system is at or above $T_d$.
Applied to RHIC, the $\psi'$ and $\chi_c$ would be completely suppressed above
a critical energy density, $\epsilon(T_d)$ while the temperature at RHIC is
too high for direct $J/\psi$ suppression by color screening.  Since the turn on
of suppression at $\epsilon(T_d)$ would result in a step-wise suppression
pattern, the suppression is typically smeared over a range of impact parameters.

Several models of sequential suppression have been proposed 
\cite{Karsch_Kharzeev_Satz,bando96,chaudhuri_2006,gunji_hirano_2007}.  
Most include some level
of normal cold nuclear matter effects but neglect gluon dissociation of 
final-state $J/\psi$'s.  The $\langle p_T^2 \rangle$ of $J/\psi$ suppression,
is typically flat or decreasing with centrality in these scenarios.  

In Ref.~\cite{Karsch_Kharzeev_Satz}, the normal absorption cross section is 
fixed by the RHIC d+Au data in each rapidity region and then applied to the 
\AuAu\ data.  The threshold energy density model, originally proposed in 
Ref.~\cite{bando96}, suppresses the $J/\psi$ yield above a threshold energy 
density, neglecting $\chi_c$ and $\psi'$ feed down.  It was recently applied 
to RHIC but included no rapidity dependence \cite{chaudhuri_2006}.  The most 
dynamically complete model of sequential suppression describes suppression in 
a dynamically-expanding quark-gluon fluid using relativistic hydrodynamics 
\cite{gunji_hirano_2007}.\\

{\noindent \textit{Conformal Field Theory:}} Recently, ${\cal N} = 4$ super
Yang-Mills theory, a conformally-invariant field theory, has been applied to
$J/\psi$ suppression \cite{rajagopal_2006}.  
The screening length is described by the 
dynamics of a color single $Q \overline Q$ dipole of velocity $v$ moving through
the hot plasma.  In its rest frame, the dipole feels the medium as a hot wind
moving past it.  The screening length in this medium decreases as 
$\gamma^{-1/2}$ or $(1 - v^2)^{-1/4}$ so that the quarkonium dissociation 
temperature would decrease with increasing $v$ ($p_T$).  Thus high $p_T$
$J/\psi$'s could be suppressed by the plasma while low $p_T$ $J/\psi$'s are
unaffected.  Such novel suppression patterns are only observable at RHIC for
sufficiently high luminosities \cite{rajagopal_2006}.

\subsection{Status of Quarkonium Physics at the CERN SPS}

\begin{figure}[tbh]
  \centering\includegraphics[width=0.5\textwidth]{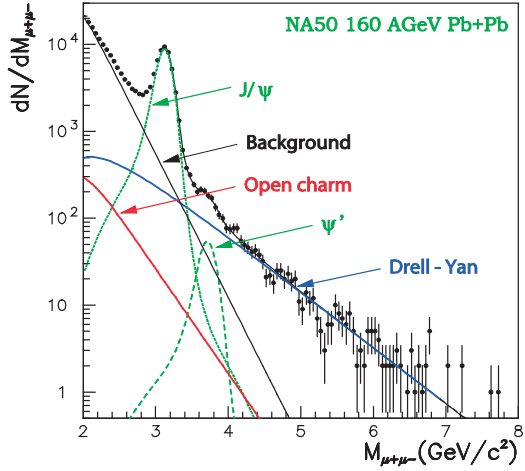}
  \caption{Dimuon invariant mass spectrum from 158 $A$\gev\ \PbPb\
    collisions at NA50. Modified from Ref.~\cite{na50}, reprinted with 
    permission from Elsevier.}
  \label{fig:na50spectra}
\end{figure}

The prospects of a ``clean" QGP signature, destruction of the $J/\psi$ by
color screening, was discussed in the landmark paper
by Matsui and Satz in 1986 \cite{MS}. This triggered an extensive
experimental program at the CERN SPS. HELIOS-III \cite{helios} and NA38
\cite{na38a} (subsequently NA50 \cite{na50a} and currently NA60
\cite{na60a}) made detailed measurements of the dimuon invariant
mass spectrum around midrapidity. Despite early enthusiasm and enormous
statistics (see Fig.~\ref{fig:na50spectra}), the picture that evolved
is still rather ambiguous. The SPS measurements must also be
understood in light of the many results on quarkonium production in \pA\
collisions from fixed target experiments. 
The status of the SPS program is summarized in this section.

\begin{figure}[tbh]

    \centering\includegraphics[width=0.75\textwidth]{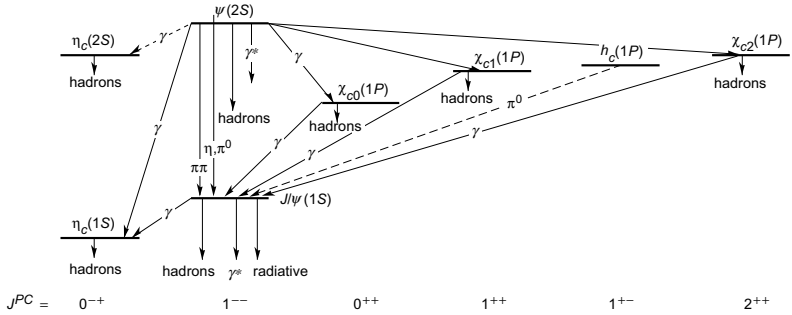}
    \caption{Charmonium mass levels and spin states.  Common feed down
        channels are indicated.}
    \label{fig:feeddown}
\end{figure}

\begin{figure}[tbh]

    \centering\includegraphics[width=0.5\textwidth]{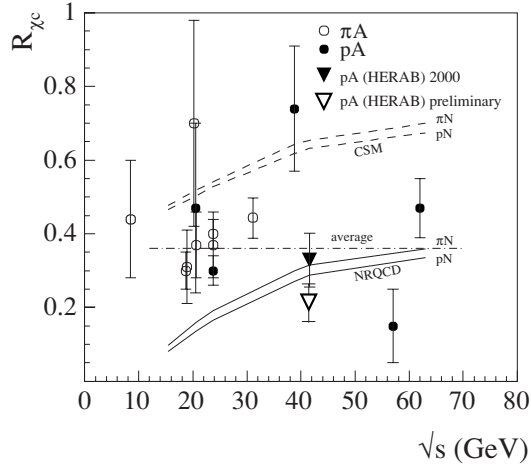}
  \caption{The fraction, $R_{\chi_c}$, of observed $J/\psi$'s originating 
      from radiative $\chi_{c1,2} \rightarrow J/\psi \gamma$ decays as
      a function of energy for proton and pion beams. From
      Ref.~\cite{heraBa}, reprinted with permission from IOP.}
  \label{fig:chic-herab}
\end{figure}

Feed down contributions (see Fig.~\ref{fig:feeddown}) from
higher charmonium states, $\chi_c \rightarrow J/\psi \gamma$ ($\sim 30$\%)
and $\psi' \rightarrow J/\psi \pi \pi$ ($\sim 10$\%), are
important \cite{heraBa,heraBb}.  The $\chi_c$ has not yet
been measured by the heavy-ion detectors at the SPS although it has
been seen in $pp$ and $p+A$ experiments there.  These measurements are 
extremely difficult and the large scatter of available data depicted in
Fig.~\ref{fig:chic-herab} indicates that better measurements are
desperately needed.  Although the NA60 experiment is planning to conduct this
analysis, the feasibility with the present
data sets still has to be verified.

The $J/\psi$ and $\psi^\prime$ have substantially modified cross sections 
when produced in normal (cold) nuclear matter. Considerable effort has gone into
studying \pA\ data from six targets covering a wide range of nuclei, from Be to
Pb, measured with proton beam energies of 
400 and 450 GeV \cite{Alessandro:2006jt}. The \pA\ \jpsi\ and \psip\ 
data were fitted assuming a single effective absorption cross section for each
charmonium state ({\it ie.} no explicit treatment of feed down effects), 
and without shadowing (analyses including shadowing effects are now 
underway \cite{lvw}) or comover absorption \cite{capella}. 
It was concluded that absorption cross sections of 
$4.2 \pm 0.5$ mb and $7.7 \pm 0.9$ mb for the \jpsi\ and \psip\ 
respectively should be used to predict 
the cold nuclear matter baseline suppression for \PbPb\ collisions at 158$A$ 
GeV beam energy. 

\begin{figure}[tbh]
  \centering\includegraphics[width=0.5\textwidth]{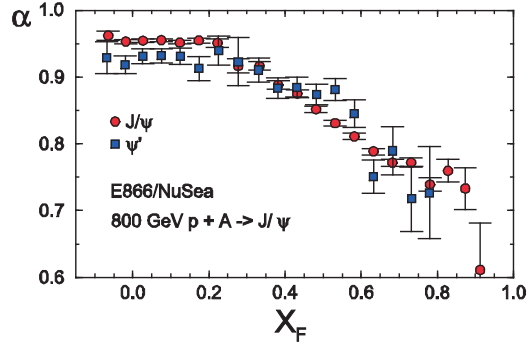}
  \caption{Measurement of $J/\psi$ and $\psi^\prime$ absorption in 800 \gev\
    \pA\ collisions as a function of $x_F$ at the Tevatron \cite{e866}, 
    reprinted with permission from APS. }
  \label{fig:e866}
\end{figure}

Studies of the $A$ dependence in cold matter were also made at Fermilab over a 
larger $x_F$ range, albeit at higher \sqrtsNN\ \cite{e866}. A very strong 
$x_F$ dependence was observed for $x_F > 0.2$, 
as depicted in Fig.~\ref{fig:e866}. Effects such as
shadowing, absorption and energy loss play roles at different $x_F$, 
resulting in the observed dependence \cite{prcda}.

\begin{figure}[tbh]
  \centering\includegraphics[width=0.5\textwidth]{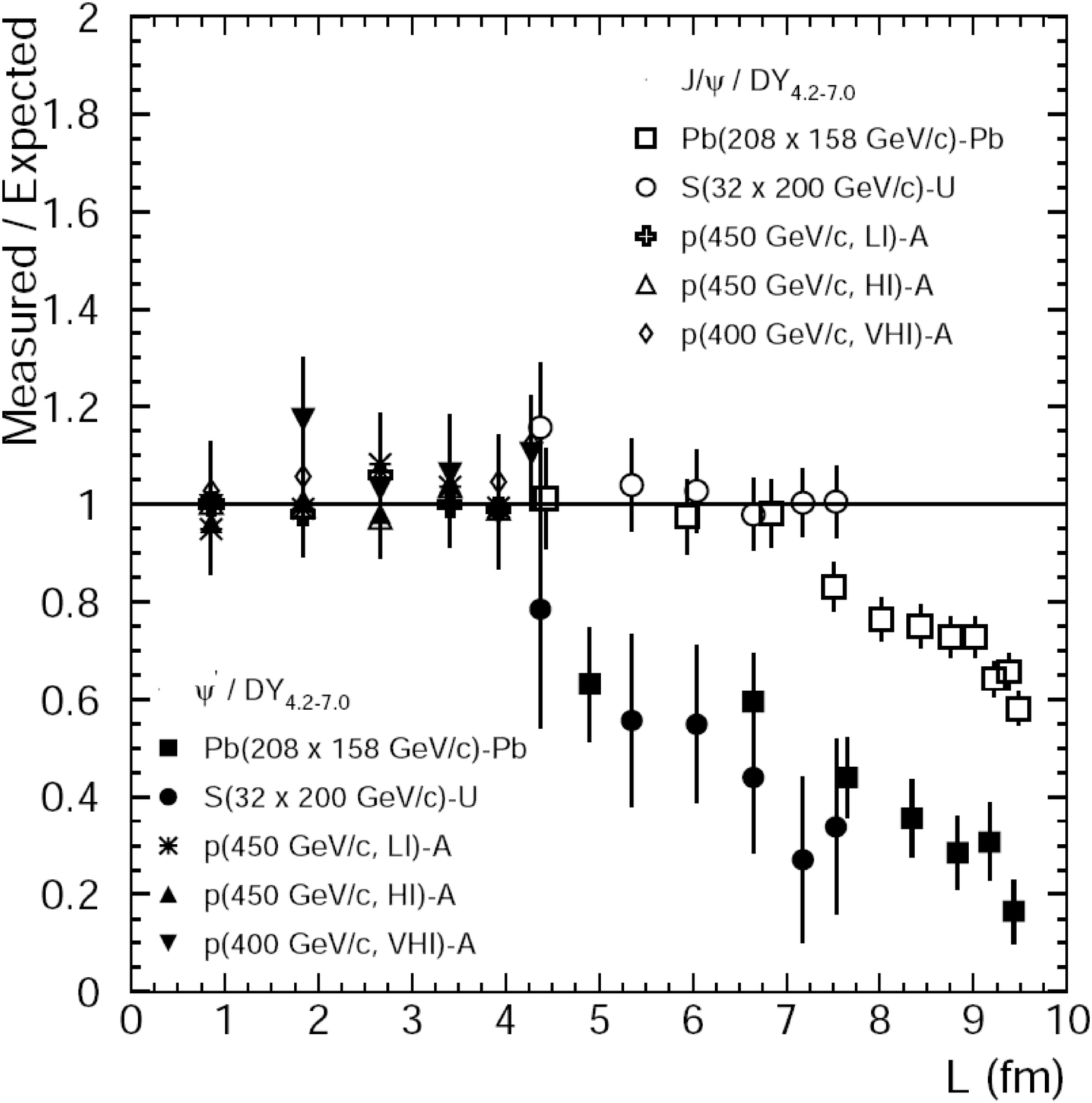}
  \caption{The ratio of measured \jpsi\ and \psip\ yields to those expected 
    from cold nuclear matter effects in heavy-ion and \pA\ collisions as a
    function of the \ccbar\ path length through the colliding 
    nuclei~\protect\cite{Alessandro:2006ju}. The measured and expected yields
    are both normalized to the Drell-Yan yields in the mass range $4.2 < M <
    7$ GeV. The lighter systems are scaled to $\sqrt{s_{NN}} = 17.3$ GeV for 
    comparison with the \PbPb\ data and isospin
    corrections are applied to the Drell-Yan yields. }
  \label{fig:na50supp}
\end{figure}

Figure~\ref{fig:na50supp} summarizes the SPS results on \jpsi\ and \psip\ 
suppression in \PbPb\, \SU\ and \pA\ 
collisions~\cite{Alessandro:2006ju,Alessandro:2004ap}. 
The figure shows measured  \jpsi\ and \psip\ yields (normalized to the 
Drell-Yan dimuon yield since it is proportional to the number of collisions) 
as a function of the average path length of the \ccbar\ pair through the 
nucleus, $L$, a measure of centrality. The measured yields are divided by 
those expected from cold nuclear matter effects, calculated using the effective
nuclear absorption cross sections, fit to the \pA\ data, mentioned earlier. 
The \pA\ and \SU\ data in Fig.~\ref{fig:na50supp} have been rescaled to 
$\sqrt{s_{NN}} = 17.3$ for comparison with the \PbPb\ data.  Isospin 
corrections were applied to the Drell-Yan cross sections.
Strikingly, the heavy-ion \jpsi\ data show no suppression beyond that
expected from cold nuclear matter up to $L \sim  7$ fm, after which 
suppression beyond cold nuclear matter expectations becomes significant. 

Alternatives to QGP models of suppression, such as the comover model, 
are able to describe the observed
$J/\psi$ suppression by assuming breakup of the bound state by
comoving matter \cite{capella}.  Although these approaches make some 
unrealistic assumptions about the hadron density, it is possible 
that some fraction of the observed suppression is due to
comover absorption.

The \psip\ suppression in Fig.~\ref{fig:na50supp} is similar to that of
the $J/\psi$ although suppression beyond cold nuclear matter 
effects becomes significant at a much shorter path length, $L \sim 4$ fm, 
{\it e.g.} in peripheral S+U collisions. Thus the \psip\ appears to be more
readily suppressed in the final state.  This may be because the
$\psi'$, lying 50 MeV below the $D \overline D$ threshold, can be 
more easily broken up by interactions in the medium.  
The strong $\psi'$ suppression measured by NA50 has been interpreted as both
total suppression of the $\psi'$ by color screening \cite{perc} and a larger 
interaction cross section for comovers \cite{klns}.

The NA60 In+In $J/\psi$ measurements exhibit a level of suppression similar to
the NA50 Pb+Pb data \cite{na60}. The \jpsi\ suppression observed
in \AA\ interactions at the SPS can, for the most part, be accounted for
by the assumption that the more loosely bound $\psi'$ and $\chi_c$ states
are both suppressed by plasma production, eliminating their contribution to
the inclusive $J/\psi$ measurement.  The direct $J/\psi$ contribution is
assumed to not be suppressed at the SPS \cite{perc,dinh}. However this 
indirect $J/\psi$ suppression picture 
does not seem to be easily reconciled with the strong observed \psip\ 
suppression for path lengths corresponding 
to rather peripheral collisions, see Ref.~\cite{Alessandro:2004ap}, 
well before any reduction in the \jpsi\ yield is observed.

Charm production was not measured in heavy-ion experiments
at the SPS until very recently. Figure~\ref{fig:na50spectra} shows that 
the open charm contribution to the dilepton continuum in the $J/\psi$ mass 
region is negligible at the SPS. Open charm measurements are, however, key to 
understanding the intermediate mass dilepton region.  The NA60 experiment 
has used displaced vertices to separate charm decays from prompt dileptons.
They have presented preliminary In+In results which show that while
the enhancement in the intermediate mass region is confirmed, it is 
inconsistent with enhanced open charm. Instead, the enhancement is consistent 
with a prompt dilepton source \cite{shahoyan}.

The picture emerging from SPS studies is still somewhat ambiguous.
In spite of the good precision of the heavy ion data for both
\jpsi\ and \psip, and the effort that went into determining the 
cold nuclear matter baseline from \pA\ data, the centrality dependence 
of the suppression pattern is insufficient to draw unique conclusions.
On the other hand, the vast experience gained at
the SPS can and should be taken into account at RHIC.  The main lesson
learned is that a simple $J/\psi$ measurement in \AA\ collisions as a
function of centrality is insufficient to draw unique
conclusions, even when a cold nuclear matter baseline is available. 
Rather, a systematic and detailed study of all related
aspects, \ie, a systematic study of open charm, $J/\psi$, $\psi^\prime$, and
$\chi_c$ production in \pp, \pA, and \AA\ collisions is required.
Centrality, rapidity, and $A$ dependence studies are mandatory so that competing
models based on different mechanisms can be differentiated.

\subsection{Quarkonium measurements to date at RHIC}

All of the published \jpsi\ results from RHIC to date are from PHENIX. 
Some preliminary $J/\psi$ results from STAR were presented at 
Quark Matter 2008. 
PHENIX measures quarkonium yields by reconstructing their invariant mass
from dilepton decays. Dielectrons are used in the central arms 
($|\eta|<0.35$) and dimuons are used in the muon arms ($1.2<|\eta|<2.2$).
STAR uses dielectrons within the TPC acceptance ($|\eta|<1$).

PHENIX has reported $J/\psi$ results at 200 \gev\ from \pp\ 
\cite{PHENIX_jpsi_pprun5,PHENIX_jpsi_pp&dau_run3,PHENIX_jpsi_pp_run2}, \dAu\  
\cite{PHENIX_jpsi_dau_new,PHENIX_jpsi_pp&dau_run3}, \AuAu\ 
\cite{PHENIX_jpsi_auau} and \CuCu\ collisions \cite{PHENIX_jpsi_cucu}. 
STAR reported preliminary high \pT\ \jpsi\ measurements in \pp\ and \CuCu\ 
collisions at Quark Matter 2008 \cite{star_tang_qm08}.

PHENIX reported observation of $\Upsilon \rightarrow \mu^+ \mu^-$
for $1.2 < |y| < 2.2$ in 200 \gev\ \pp\ collisions from RHIC Run 5 
\cite{PHENIX_upsilon_QM05}. This very low statistics
measurement (27 counts in both muon arms combined) was used to make a
preliminary cross section estimate of BR$(d\sigma/dy)_{y=1.7} = 45.2
\pm 9.5 \, ({\rm stat}) \, \pm 6.3 \, ({\rm sys})$ nb.  STAR observed
about 50 $\Upsilon \rightarrow e^+e^-$ at $|y| < 1$, leading to a
preliminary cross section of BR$(d\sigma/dy)_{y=0} = 91
\pm 28 \, ({\rm stat}) \, \pm 22 \, ({\rm sys})$ nb
\cite{STAR_upsilon_QM06}.  In both cases, the cross section is for the
lowest three \ups\ states combined. It would be difficult to make
definitive $\Upsilon$ measurements at present RHIC luminosities, but crude
measurements of the \ups\ nuclear modification factor by both PHENIX
and STAR are expected to be possible at RHIC with about 10 times the Run 4 
integrated luminosity, which will most likely be available after the next 
long \AuAu\ run.

\subsubsection{Baseline quarkonium measurements at RHIC}

\begin{figure}[tbh]
  \vspace{0.5cm}
  \centering
  \includegraphics[width=0.6\textwidth]{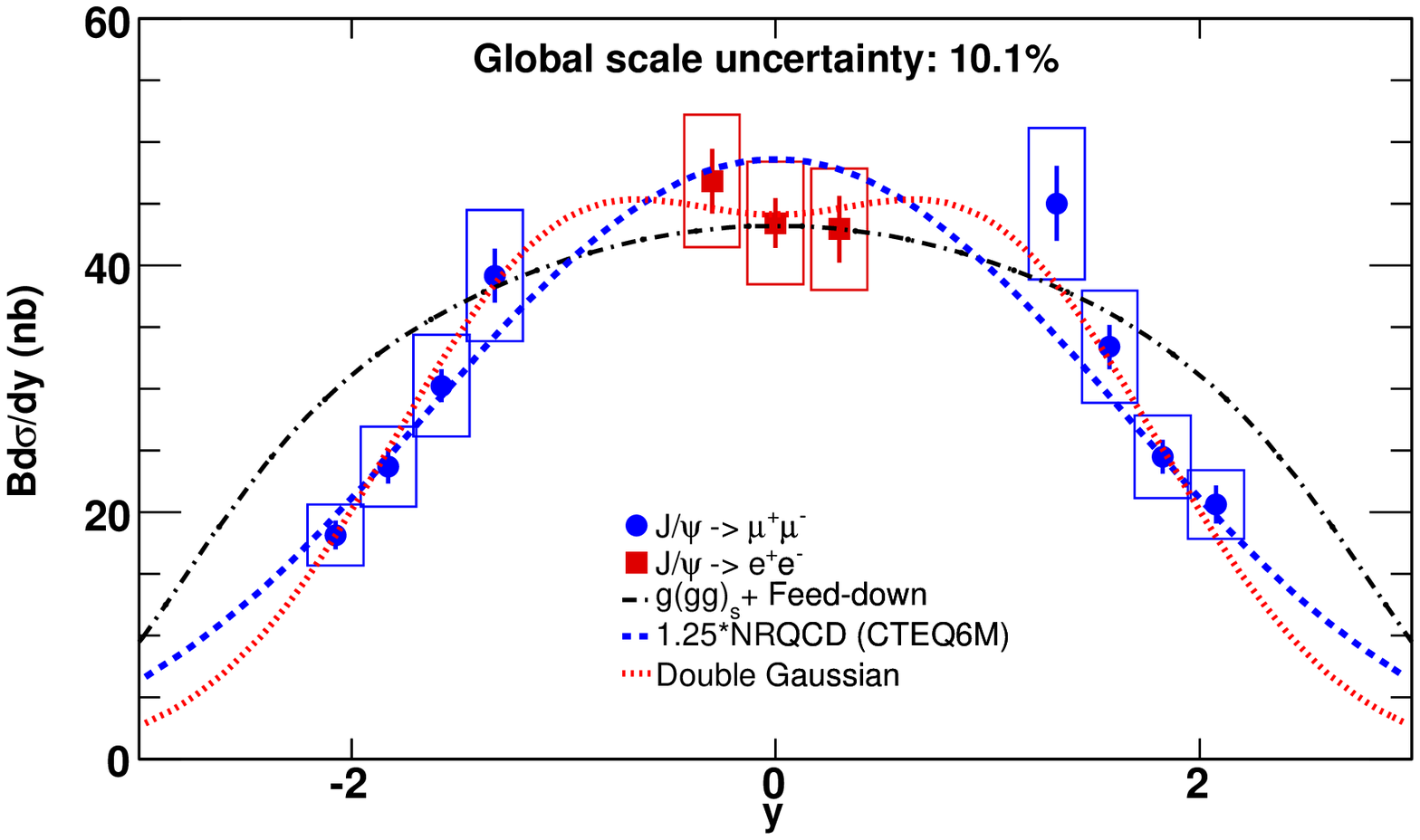}
  \includegraphics[width=0.34\textwidth]{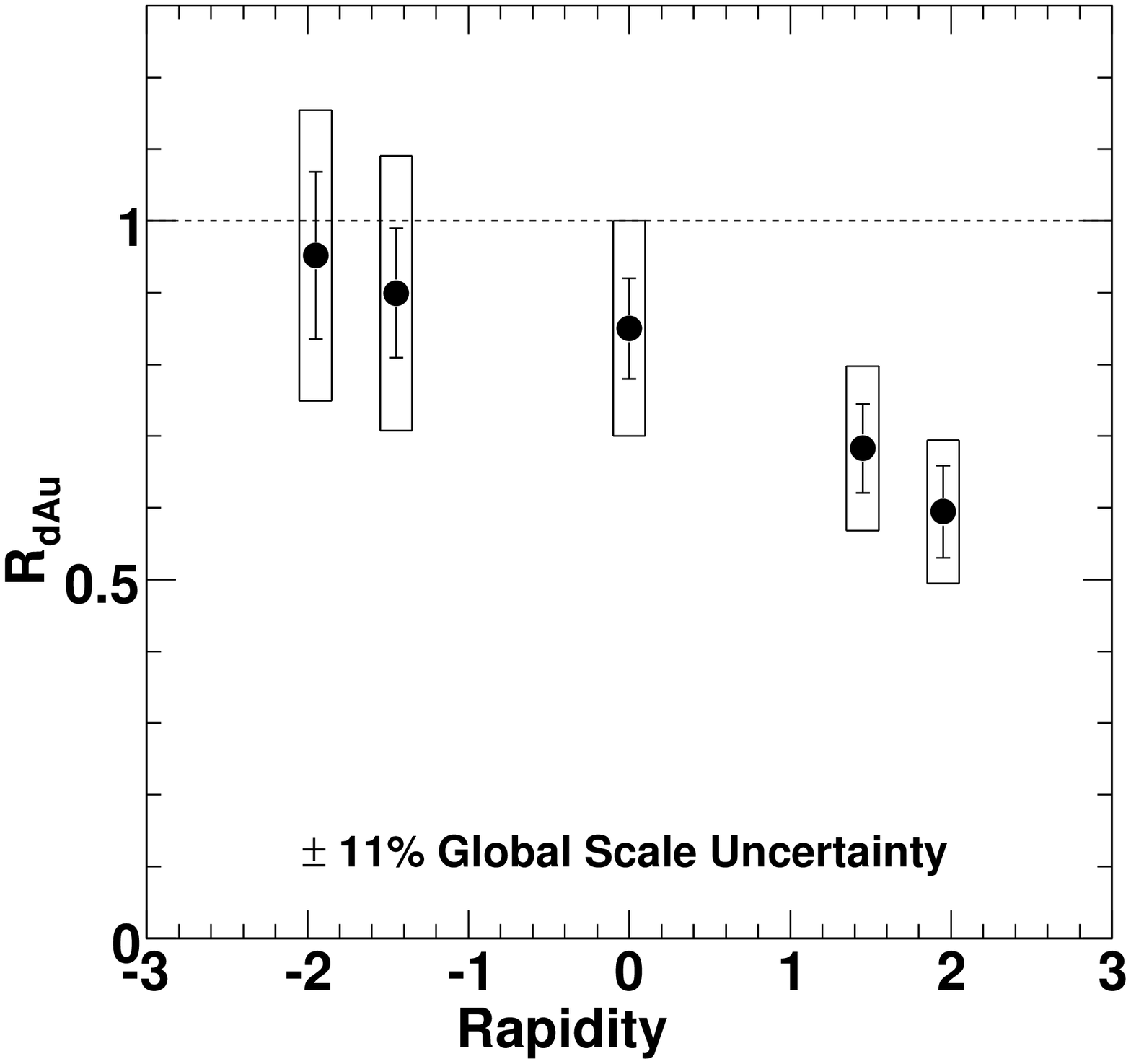}
  \caption{The rapidity dependence of the \pp\ $J/\psi$ cross section at 
    200 \gev\ \protect\cite{PHENIX_jpsi_pprun5} (left-hand side, 
    reprinted with permission from APS) and the nuclear 
    modification factor for minimum bias \dAu\ collisions 
    \protect\cite{PHENIX_jpsi_dau_new} 
    (right-hand side). The curves on the left-hand side are fits used to 
    extract the total cross section and estimate the systematic error. 
    The deuteron is defined to be moving toward positive rapidity.}
  \label{fig:PHENIX_pp&dau_rap}
  \vspace{0.5cm}
\end{figure}

PHENIX has measured $J/\psi$ cross sections in \pp\ \cite{PHENIX_jpsi_pprun5} 
and \dAu\ collisions 
\cite{PHENIX_jpsi_dau_new,PHENIX_jpsi_pp&dau_run3} at 200 \gev. The 
rapidity dependence is summarized in Fig.~\ref{fig:PHENIX_pp&dau_rap}. 
The left side shows the invariant $J/\psi$ cross section 
in \pp\ collisions while the right side shows the 
nuclear modification factor, $R_{\rm dAu}$, for minimum bias \dAu\ collisions. 

\begin{figure}[tbh]
  \vspace{0.5cm}
  \centering
  \includegraphics[width=0.48\textwidth]{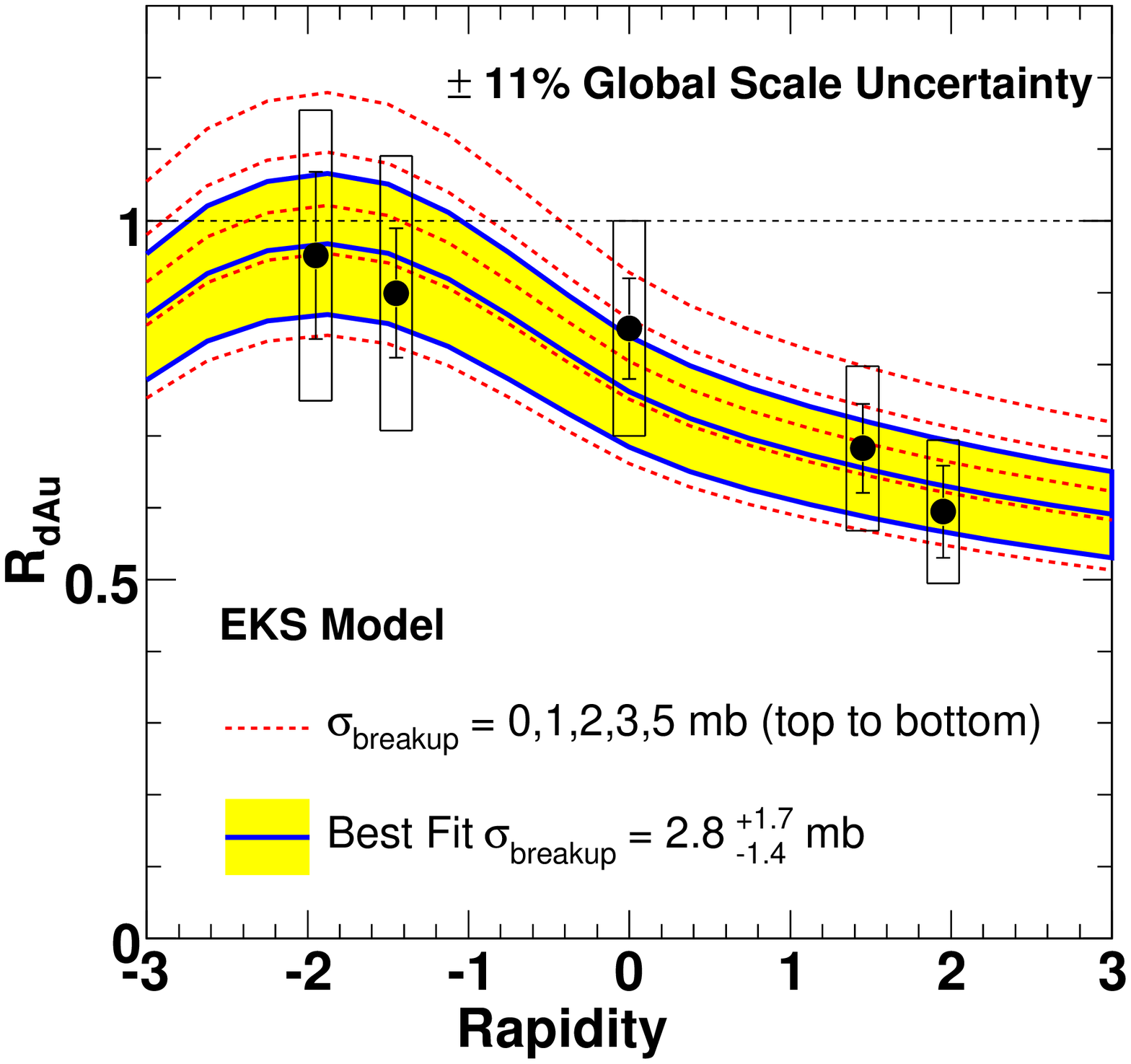}
  \includegraphics[width=0.48\textwidth]{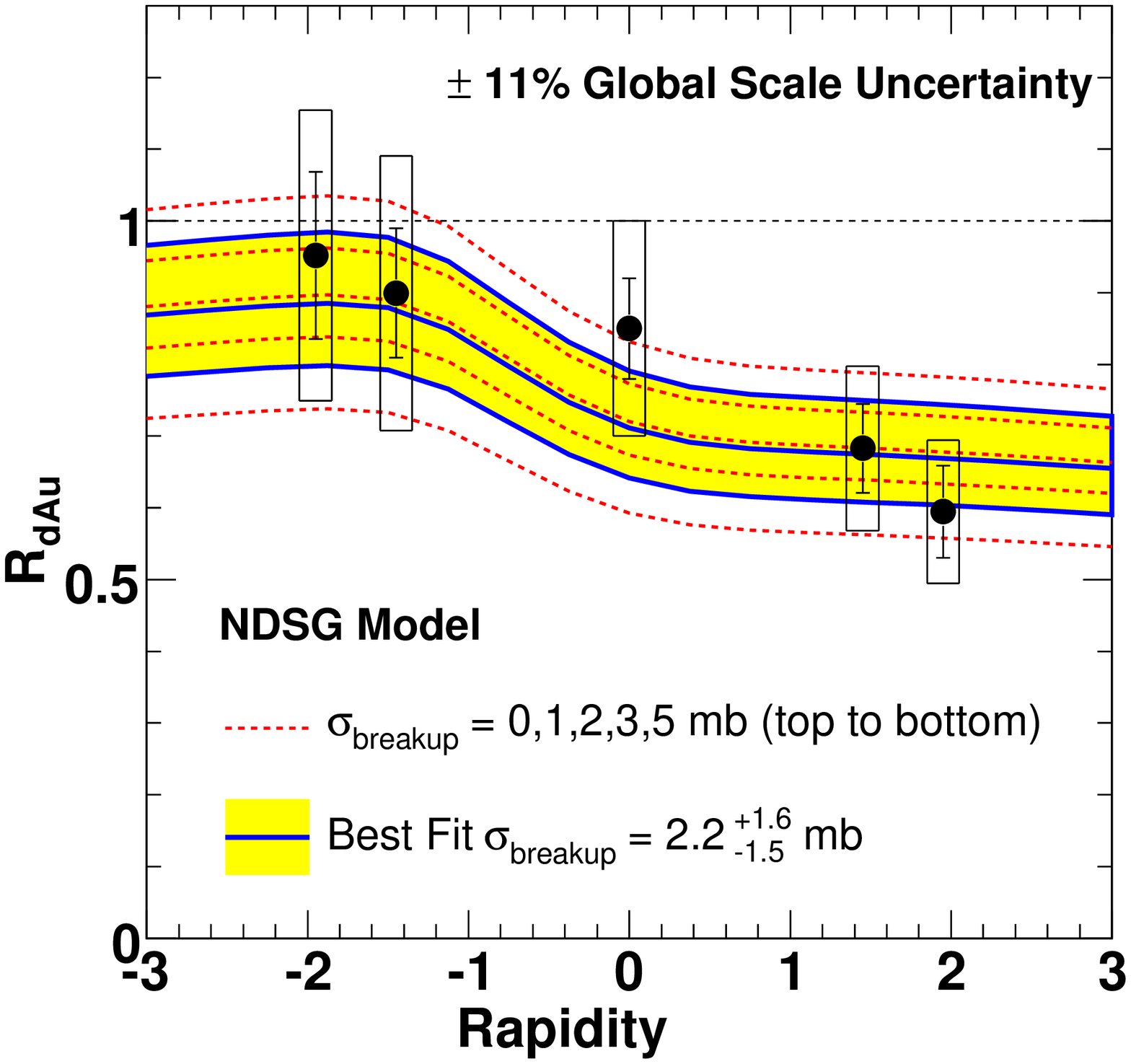}
  \caption{The rapidity dependence of the \dAu\ $J/\psi$ invariant yield at 
    200 \gev\ \protect\cite{PHENIX_jpsi_dau_new} compared with calculations 
    \protect\cite{prcda} including shadowing and a $J/\psi N$ 
    breakup cross section. The left-hand side compares the data with 
    calculations using the EKS98 parameterization \cite{EKS} while the
    right-hand side employs the nDSg parameterization \cite{NDSG}.}
  \label{fig:PHENIX_dau_rap_theory}
  \vspace{0.5cm}
\end{figure}

\begin{figure}[tbh]
  \vspace{0.5cm}
  \centering
  \includegraphics[width=0.5\textwidth]{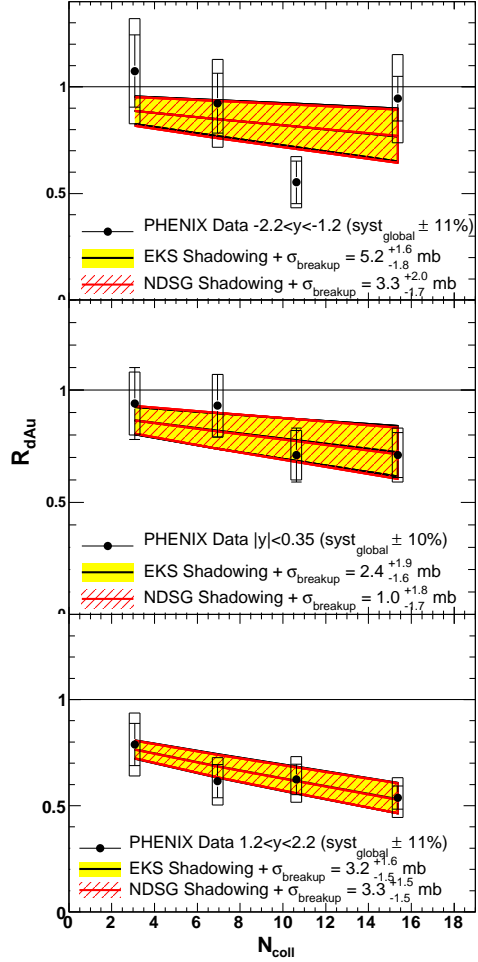}
  \caption{The nuclear modification factor in \dAu\ collisions measured at 
    forward (bottom), central (middle) and 
    backward (top) rapidity  as a function of centrality. The deuteron is 
    moving toward forward rapidity. The curves, including both shadowing and 
    nuclear breakup, are discussed in the text. 
    From Ref.~\protect\cite{PHENIX_jpsi_dau_new}, 
    reprinted with permission from APS.}
  \label{fig:PHENIX_dau_rda_cent}
  \vspace{0.5cm}
\end{figure}

The data shown on the right-hand side of Fig.~\ref{fig:PHENIX_pp&dau_rap} are 
compared~\cite{PHENIX_jpsi_dau_new} with \dAu\ calculations \cite{prcda} 
that include shadowing 
(EKS98 \cite{EKS} and nDSg \cite{NDSG}) and $J/\psi N$ breakup in 
Fig.~\ref{fig:PHENIX_dau_rap_theory}. The shaded bands show the one standard 
deviation limits of fitting the $J/\psi N$ breakup cross section. The extracted 
values are consistent, within large 
uncertainties, with the $4.2 \pm 0.5$ mb cross section determined 
at the CERN SPS \cite{Alessandro:2006jt,NA50_breakup_sigma}. 

The \jpsi\ nuclear modification factor in \dAu\ collisions
is shown as a function of centrality in 
Fig.~\ref{fig:PHENIX_dau_rda_cent} \cite{PHENIX_jpsi_dau_new} 
for the forward, central and backward rapidity regions covered by the 
three PHENIX arms. The shaded bands show the same calculation as in 
Fig.~\ref{fig:PHENIX_dau_rap_theory} but the breakup cross section is now fit
to the data in each rapidity region.  The $N_{\rm coll}$ dependence is the
same because the impact parameter dependence of the shadowing is assumed to be
proportional to the $J/\psi$ path length through the medium in both cases
\cite{prcda}. It is evident that 
the parameters are not well constrained by the current \dAu\ data.

\subsubsection{Quarkonium measurements in heavy-ion collisions at RHIC}

Measurements with the statistical precision needed 
to provide a strong test of models of $J/\psi$ production in heavy-ion 
collisions are now available from RHIC. These PHENIX measurements are from the 
Run 4 \AuAu\ and the Run 5 \CuCu\ data sets. The main features are summarized
here. 

\begin{figure}[tbh]
  \vspace{0.5cm} \centering
  \includegraphics[width=0.75\textwidth]{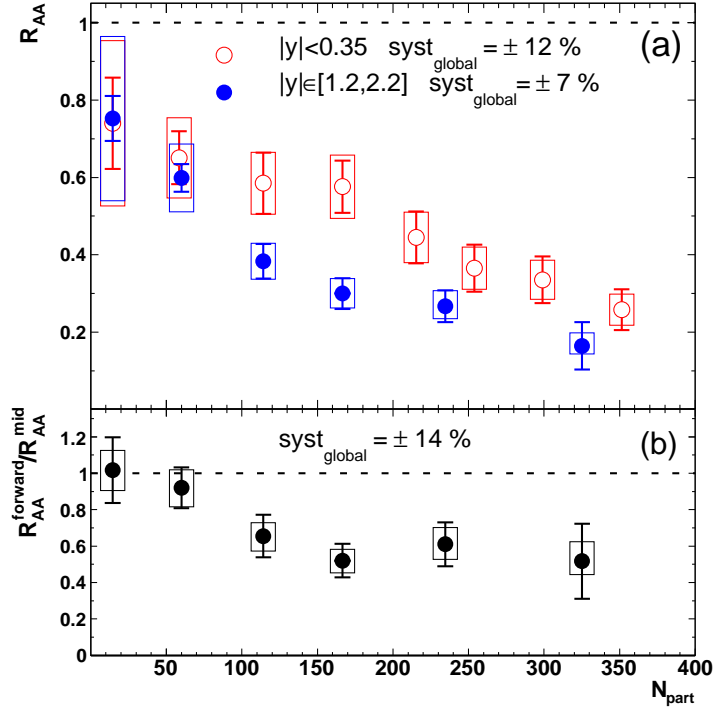}
  \caption{
    The nuclear modification factor as a function of
    centrality for 200 \gev\ \AuAu\ collisions measured at central and
    forward/backward rapidity \cite{PHENIX_jpsi_auau}.  }
  \label{fig:PHENIX_auau_raa}
  \vspace{0.5cm}
\end{figure}

Figure~\ref{fig:PHENIX_auau_raa} shows the $J/\psi$ nuclear 
modification factor, $R_{AA}$, measured in 200 \gev\ \AuAu\ 
collisions  at both central and forward/backward rapidities 
\cite{PHENIX_jpsi_auau}, along with the ratio of forward to
midrapidity \RAA\ values, $R_{AA}^{\rm forward}/R_{AA}^{\rm mid}$. 
Some systematic uncertainties cancel in the ratio.
The ratio $R_{AA}^{\rm forward}/R_{AA}^{\rm mid}$, in the lower panel of 
Fig.~\ref{fig:PHENIX_auau_raa}, shows that
when $N_{\rm part} > 150$ the \AuAu\ nuclear modification factor 
at forward rapidity is only about 60\% of that at midrapidity. 
Thus the \AuAu\ data show that the 
suppression is weaker at the maximum transverse
energy density (midrapidity).  The weaker suppression could be due to either 
entrance channel effects (shadowing) or final-state effects such as \jpsi\ 
formation by coalescence of uncorrelated charm quarks.

\begin{figure}[tbh]
  \vspace{0.5cm} \centering
  \includegraphics[width=0.75\textwidth]{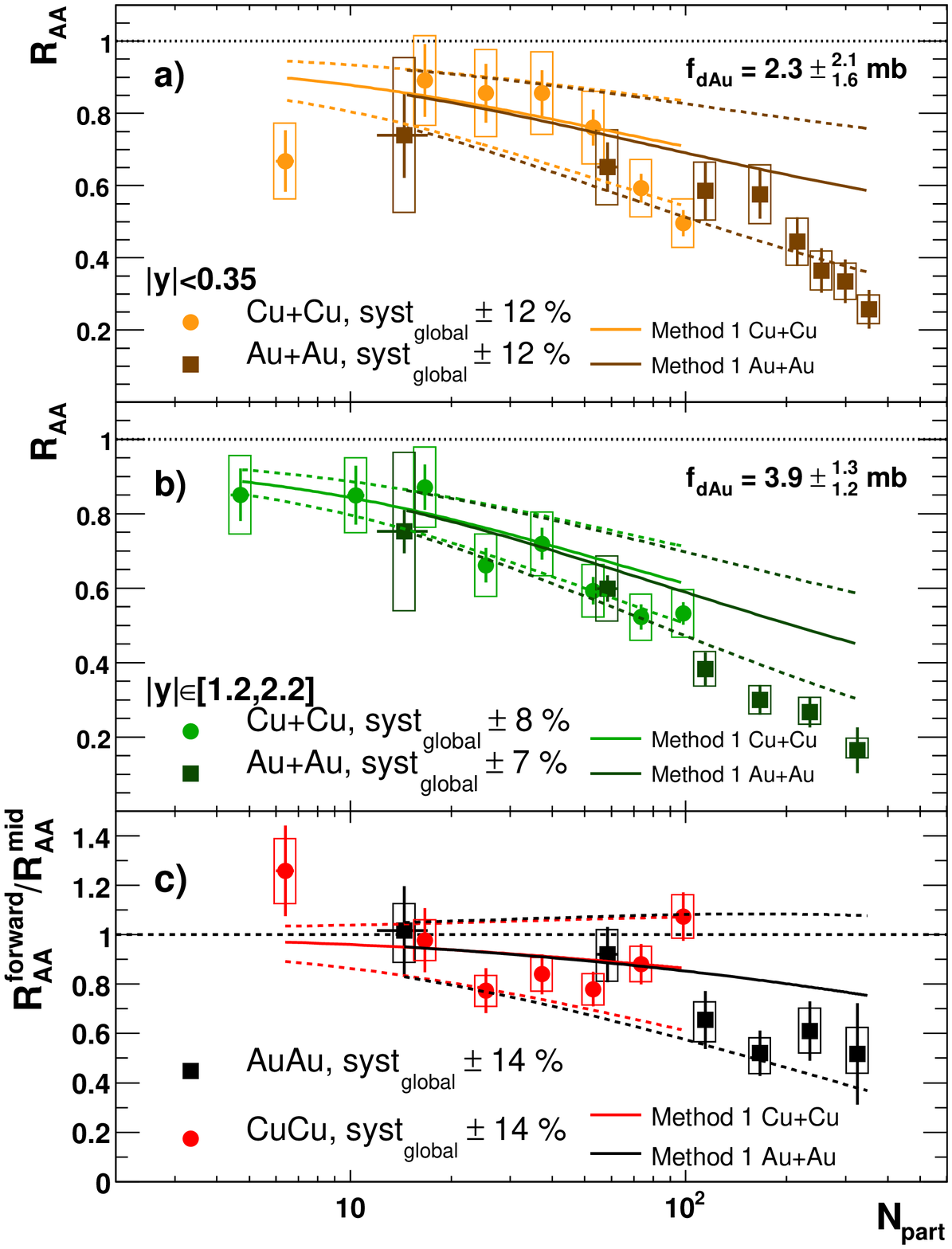}
  \caption{
    The nuclear modification factor as a function of
    centrality for 200 \gev\ \AuAu\ and \CuCu\ collisions measured at central 
    and 
    forward/backward rapidity \cite{PHENIX_jpsi_auau,PHENIX_jpsi_cucu}.  }
  \label{fig:PHENIX_auau&cucu_raa}
  \vspace{0.5cm}
\end{figure}

\begin{figure}[tbh]
  \vspace{0.5cm} \centering
  \includegraphics[width=0.75\textwidth]{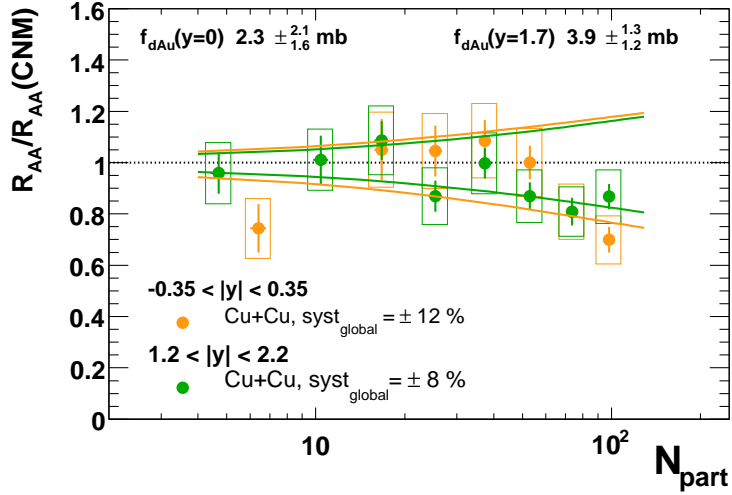}
  \caption{The ratio of measured \RAA\ to the cold nuclear matter \RAA\ 
    calculated from the ad hoc fits to d+Au data, shown in 
    Fig.~\ref{fig:PHENIX_auau&cucu_raa}, for \CuCu\ collisions. 
    The one standard deviation uncertainties in the cold nuclear matter
    reference are indicated by bands around unity.
  }
  \label{fig:survival_eks_ad_hoc}
  \vspace{0.5cm}
\end{figure}

Figure~\ref{fig:PHENIX_auau&cucu_raa} shows the nuclear modification factor for 
\AuAu\ and \CuCu\ at central and forward/backward rapidity, along with 
$R_{AA}^{\rm forward}/R_{AA}^{\rm mid}$. Within uncertainties, 
the data for \CuCu\ and \AuAu\ agree where they overlap 
in \npart. It is of considerable interest to see what constraints the \dAu\ 
data place on the cold nuclear matter contribution to the heavy-ion \RAA\ by 
using the fits discussed in the previous section. However,
$R_{AA}^{\rm forward}/R_{AA}^{\rm mid}$ is independent of the breakup cross 
section in those models because the rapidity dependence of the absorption
model is small in the range of the data \cite{prcda}.
Thus the model calculation of $R_{AA}^{\rm forward}/R_{AA}^{\rm mid}$
reflects only the effects of the assumed shadowing model. To explore how
the \dAu\ data constrain the forward to midrapidity ratio for heavy ions, 
an ad hoc model was used 
to parameterize the \dAu\ data. In the ad hoc model, the breakup cross 
section is allowed to vary independently at $y = 0$ and $|y| = 1.7$. 
The resulting parameterizations of the \dAu\ data
can be used to predict the cold nuclear matter contribution to the heavy-ion 
\RAA\ independently at each rapidity, with independent uncertainties, and 
allow the constraints on $R_{AA}^{\rm forward}/R_{AA}^{\rm mid}$ to be 
quantified. The calculations for EKS98 are compared with the heavy-ion data 
in Fig.~\ref{fig:PHENIX_auau&cucu_raa}. The nDSg
results are similar. The cold nuclear matter \RAA\ predicted by the ad hoc 
fits to the \dAu\ data are in good agreement with the \CuCu\ \RAA\ measured
in more peripheral collisions.  
This is illustrated in Fig.~\ref{fig:survival_eks_ad_hoc}, which shows the 
survival probability relative to cold nuclear matter effects on the Cu+Cu data, 
calculated using the ad hoc fits to the d+Au data shown
in Fig.~\ref{fig:PHENIX_auau&cucu_raa}. Below \npart~$\sim~50$ the survival 
probability is consistent with unity within the $\sim$ 15\% uncertainties. 
Beyond that, the uncertainty in the cold nuclear matter
reference rapidly worsens.

The bottom panel of Fig.~\ref{fig:PHENIX_auau&cucu_raa} shows that the 
existing \dAu\ data allow the additional suppression from hot nuclear matter 
at forward rapidity  to be anywhere between zero and a factor of two for 
central \AuAu\ collisions. Clearly, better \dAu\ data are required before the 
effects from cold nuclear matter are quantified for \AuAu\ 
collisions. The Run 8 d+Au data set will provide a greatly improved cold nuclear
matter reference.

\begin{figure}[tbh]
  \vspace{0.5cm}
  \centering
  \includegraphics[width=0.95\textwidth]{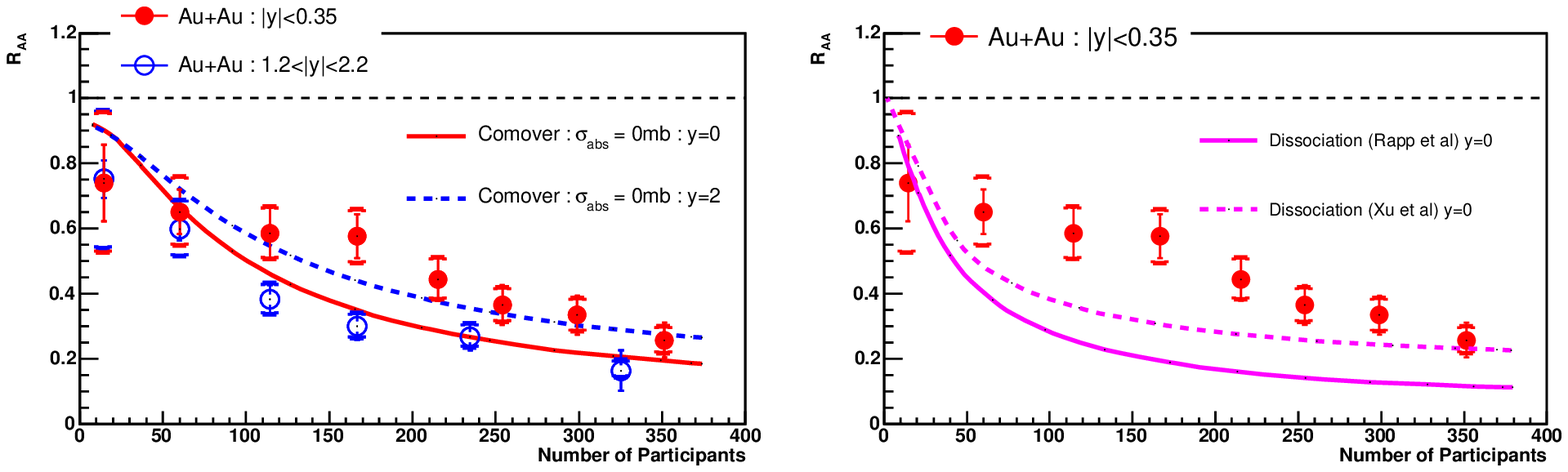}
  \caption{Left: Nuclear modification factor compared with a comover model
    calculation at central and forward/backward rapidity. Note that the 
    model predicts maximum suppression at $y = 0$. Right: Comparison with two 
    calculations of gluon dissociation at midrapidity 
    \protect\cite{gunji_2006}.}
  \label{fig:PHENIX_auau_raa_models_1}
  \vspace{0.5cm}
\end{figure}

\begin{figure}[tbh]
    \vspace{0.5cm} \centering
    \includegraphics[width=0.65\textwidth]{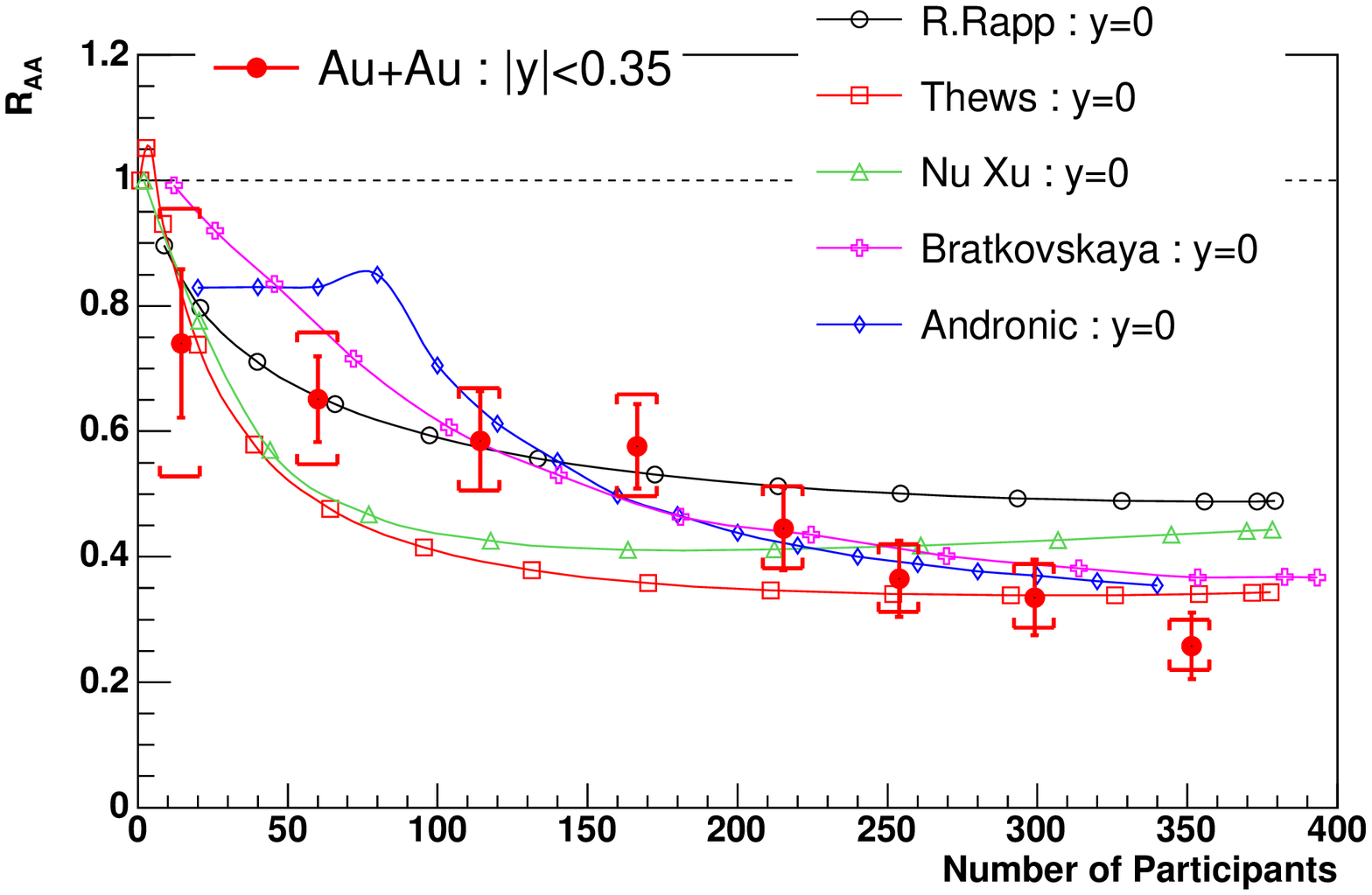}
  \caption{Nuclear modification factor compared with models that include 
      \jpsi\ formation by coalescence.  The calculations are for $y = 0$
      only \protect\cite{gunji_2006}. See the text for details.}
  \label{fig:PHENIX_auau_raa_models_2}
  \vspace{0.5cm}
\end{figure}

The left-hand side of Fig.~\ref{fig:PHENIX_auau_raa_models_1} 
\cite{gunji_2006} compares the PHENIX \AuAu\ data to the comover model
\cite{ferriero_2006}, with shadowing and comover interactions,
where the suppression is predicted to be larger at $y =0$ than at $y = 1.7$, 
in disagreement with the data.  The right-hand side of the figures compares
the $y = 0$ \AuAu\ data 
with calculations containing charmonium dissociation by thermal gluons with no
coalescence \cite{rapp_epj_2005,yan_zhuang_xu_2006}.  All of these
calculations overestimate midrapidity \jpsi\ suppression.
Figure~\ref{fig:PHENIX_auau_raa_models_2} \cite{gunji_2006} compares the
midrapidity data to model calculations
\cite{rapp_epj_2005,yan_zhuang_xu_2006,BobMic,andronic_0611023,cassing_2000}
including \jpsi\ formation by coalescence. These models generally
do a better job of describing the magnitude, if not the detailed centrality
dependence, of the suppression. However they rely
heavily on the rapidity density of charm production as input to the
coalescence calculations and the charm distributions are currently not 
very precisely
defined by the RHIC data, especially at forward rapidity.

\begin{figure}[tbh]
    \vspace{0.5cm} \centering
    \includegraphics[width=0.45\textwidth]{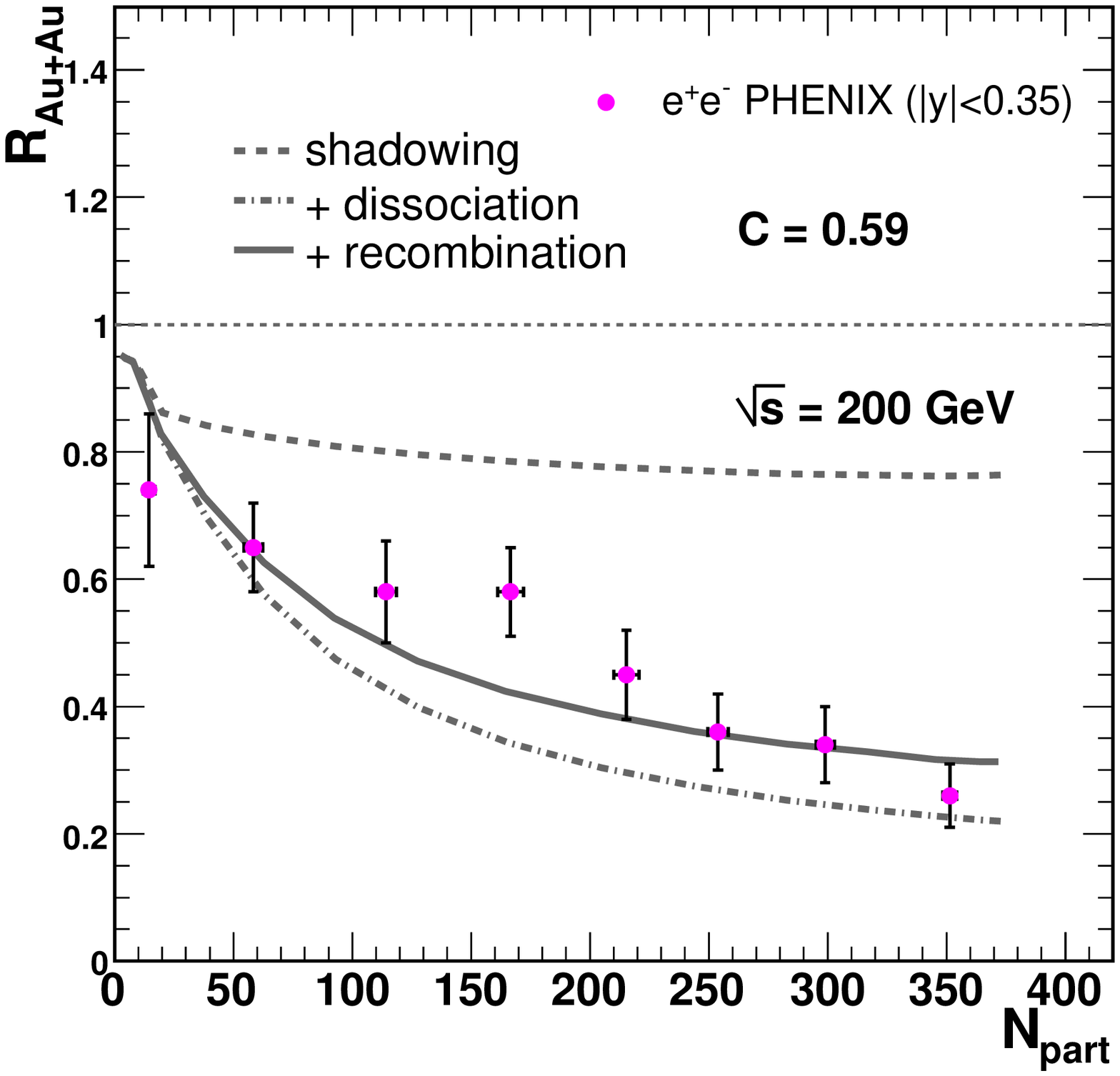}
    \includegraphics[width=0.45\textwidth]{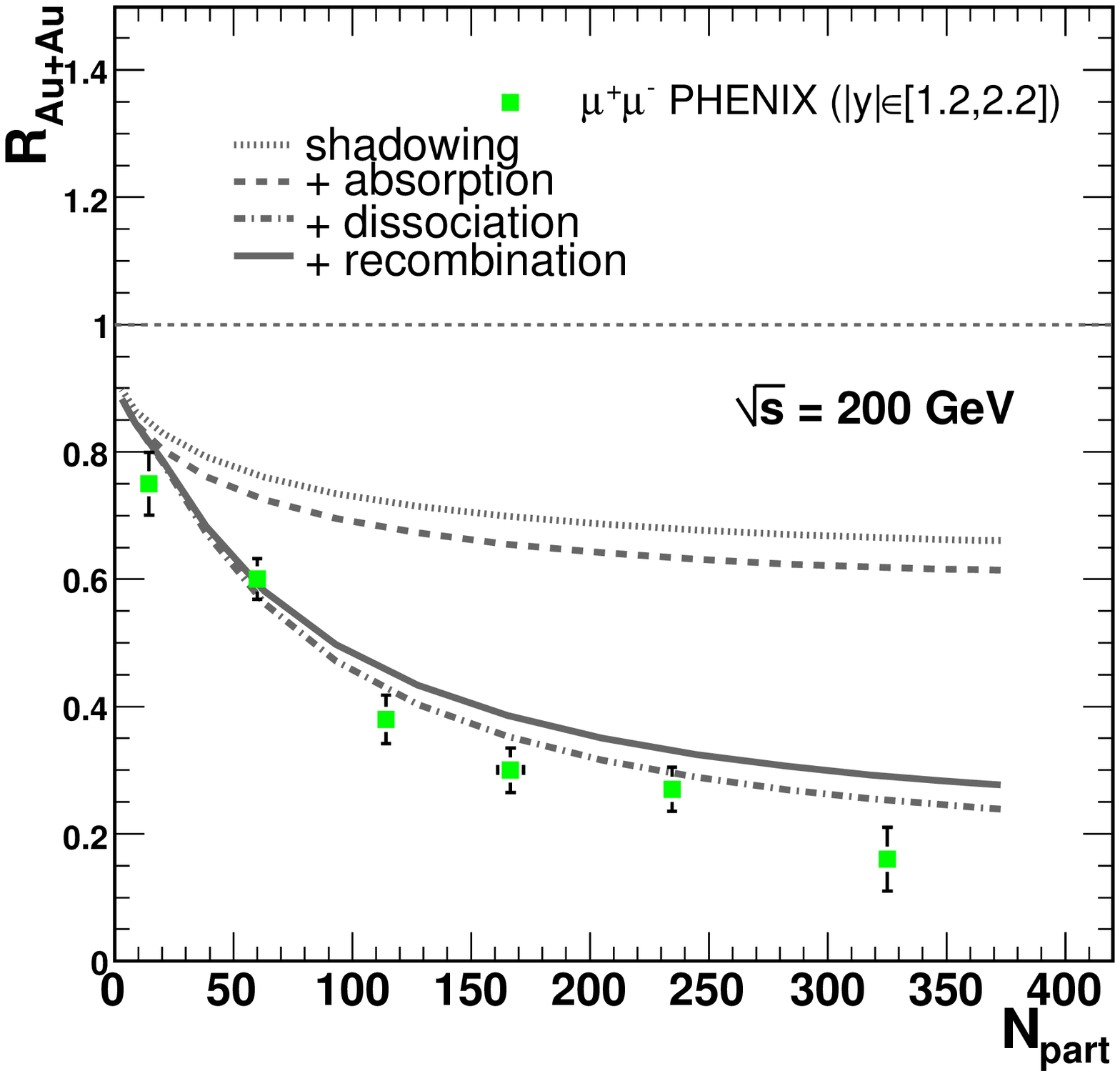}
  \caption{Calculation of \jpsi\ \RAA\ at central and forward rapidities 
    in the comover model with coalescence \protect\cite{capella_2007}.}
  \label{fig:PHENIX_auau_raa_models_3}
  \vspace{0.5cm}
\end{figure}

The extended comover interaction model \cite{capella_2007}, including \jpsi\ 
coalescence as well as suppression, shown in 
Fig.~\ref{fig:PHENIX_auau_raa_models_3}, has slightly stronger suppression 
at forward rapidity than seen in Fig.~\ref{fig:PHENIX_auau_raa_models_1}, 
in better agreement with the data. 
A midrapidity calculation by Zhao and Rapp \cite{zhao_rapp_2007} extends the
suppression and coalescence model \cite{rapp2} to include momentum-dependent 
dissociation rates.  The $J/\psi N$ cross section was reduced 
to 1.5 mb, closer to the PHENIX \dAu\ data, and a longer charm 
quark thermalization time, 7 fm/$c$  was assumed to better match the magnitude 
of the observed suppression in the most central \AuAu\ collisions. 
The calculated centrality dependence of \RAA\ at midrapidity is in 
good agreement with the data.  Reference~\cite{zhao_rapp_2007} also
presents calculations of  ${\langle}p_T^2{\rangle}$ and \RAA\  as a function
of \pT, discussed later.

PHENIX also showed a measurement of the nuclear modification factor
for 62 \gev\ \CuCu\ collisions at Quark Matter 2005. While this
measurement has relatively low statistics, the 62 \gev\ data exhibit
similar, and perhaps slightly stronger, suppression in the most
central collisions than that seen in 200 \gev\ \CuCu\ collisions,
consistent with predictions of color screening and coalescence
\cite{rapp2}.

\begin{figure}[tbh]
  \vspace{0.5cm}
  \centering
  \includegraphics[width=0.75\textwidth]{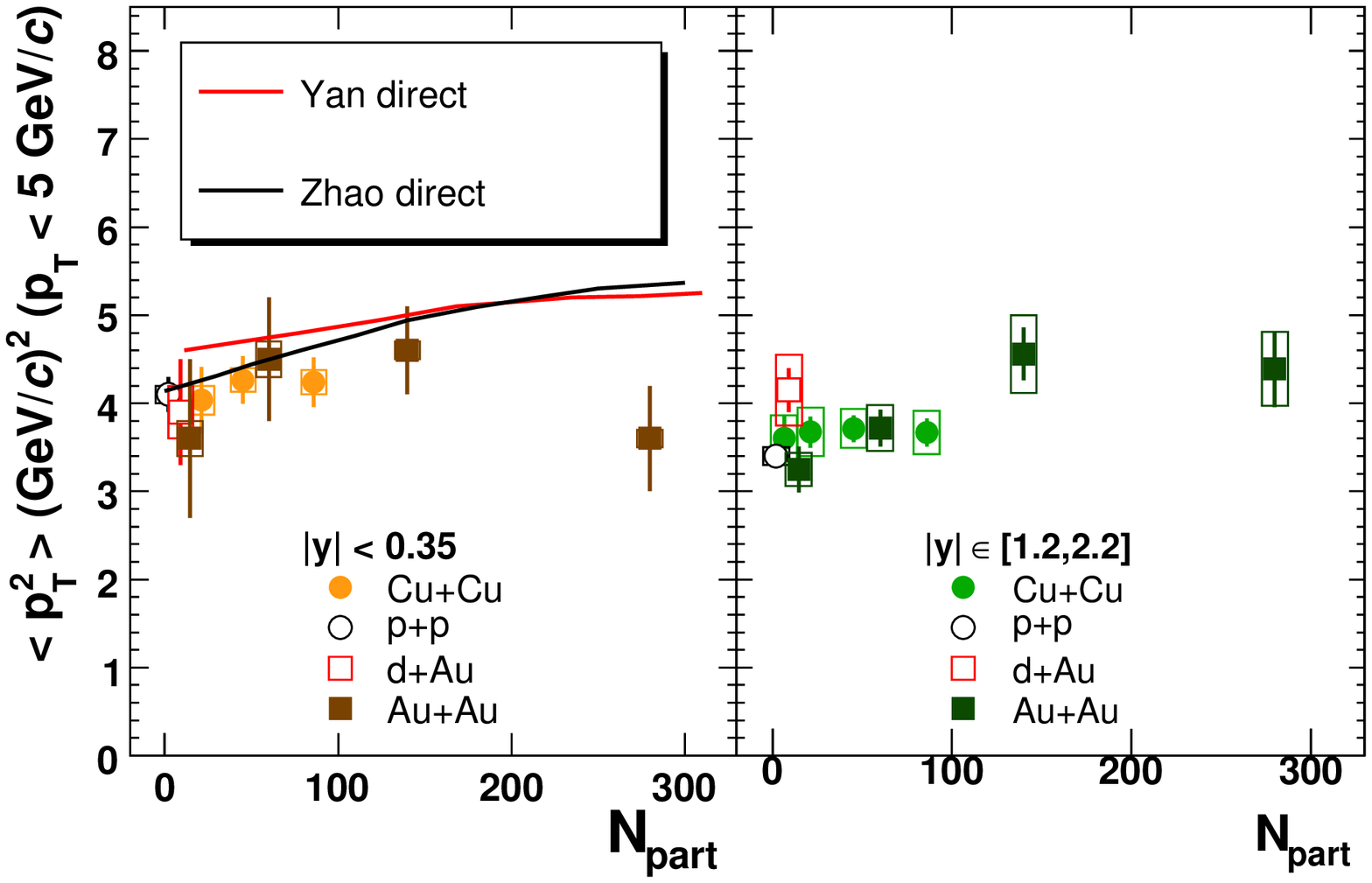}
  \includegraphics[width=0.75\textwidth]{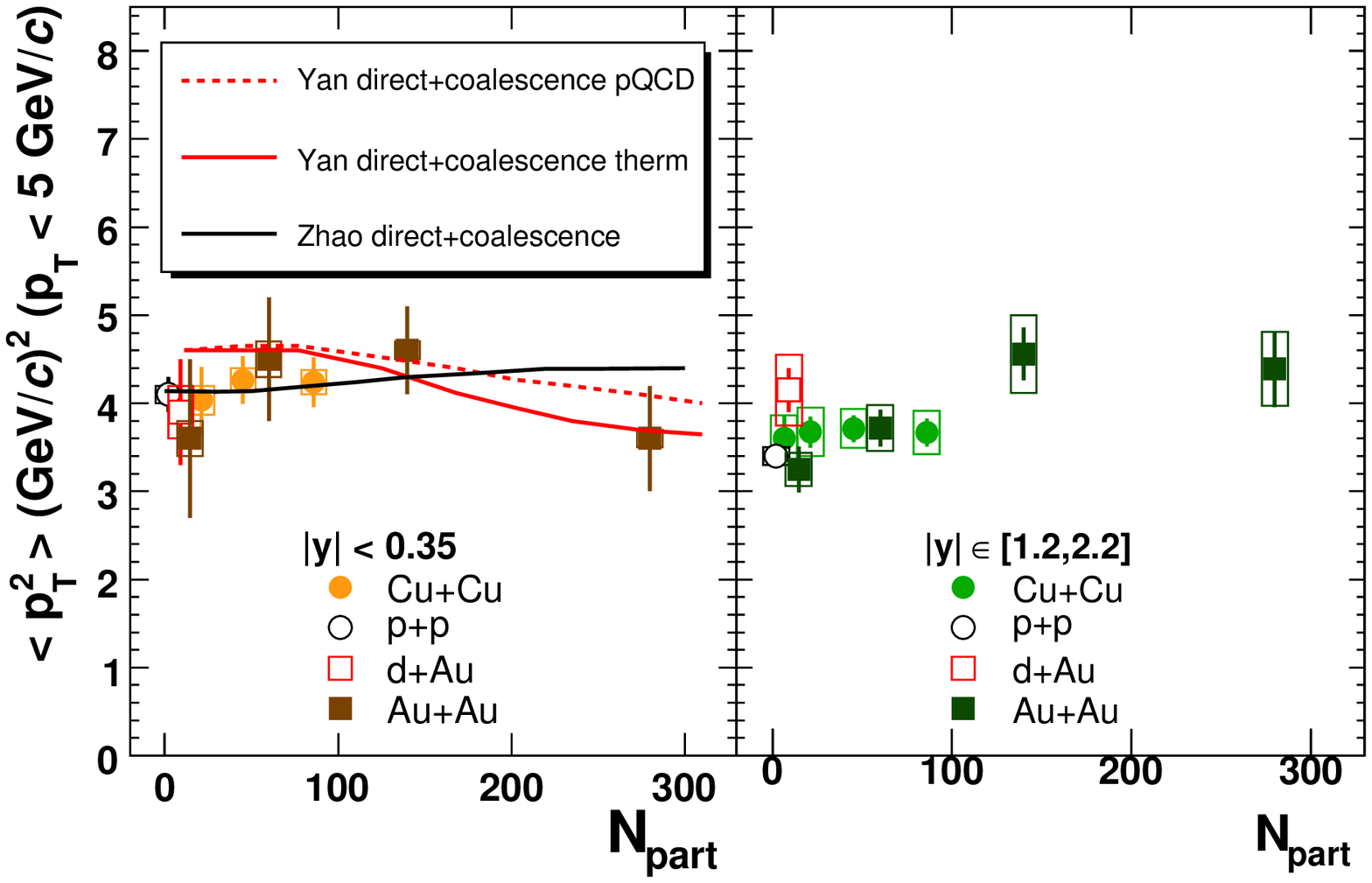}
  \caption{The $\langle p_T^2 \rangle$ for 200 \gev\ \AuAu\ and
    \CuCu\ collisions measured by PHENIX at midrapidity (left) and 
    forward/backward rapidity (right) \cite{PHENIX_jpsi_cucu} compared with 
    calculations without coalescence (top) and including coalescence (bottom)
    \protect\cite{yan_zhuang_xu_2006,zhao_rapp_2007}. The $p_T$ 
    integrals are restricted to less than 5 \gevc\ to reduce the 
    systematic errors.}
  \label{fig:PHENIX_auau&cucu_meanpt}
  \vspace{0.5cm}
\end{figure}

There have been recent predictions of
$c\overline{c}$ coalescence effects on the rapidity and $p_T$ dependence of the
$J/\psi$ yield \cite{thews_vienna,yan_zhuang_xu_2006,zhao_rapp_2007}.
Figure~\ref{fig:PHENIX_auau&cucu_meanpt} shows the average $J/\psi$
$p_T^2$, $\langle p_T^2 \rangle$, obtained from data below 5 \gevc, as
a function of \npart\ for \AuAu\ and \CuCu\ collisions in the two
rapidity regions covered by PHENIX \cite{PHENIX_jpsi_cucu,PHENIX_jpsi_auau}.  
The \AuAu\ and \CuCu\ data are in very good agreement at the same \npart\ and
rapidity.  Also shown are calculations by
Yan, Zhuang and Xu \cite{yan_zhuang_xu_2006} and 
Zhao and Rapp \cite{zhao_rapp_2007} 
that incorporate the effect of coalescence on the centrality dependence of
${\langle}p_T^2{\rangle}$.  An earlier calculation by Thews 
\cite{thews_vienna} predicts a larger and more steeply rising $J/\psi$ $\langle
p_T^2 \rangle$ than shown in Fig.~\ref{fig:PHENIX_auau&cucu_meanpt}. However
all calculations with \jpsi\ coalescence agree fairly well with the
$\langle p_T^2 \rangle$ data, even though coalescence is a smaller
fraction of the \jpsi\ yield in Refs.~\cite{yan_zhuang_xu_2006,zhao_rapp_2007}.

\begin{figure}[tbh]
    \vspace{0.5cm} \centering
    \includegraphics[width=0.45\textwidth]{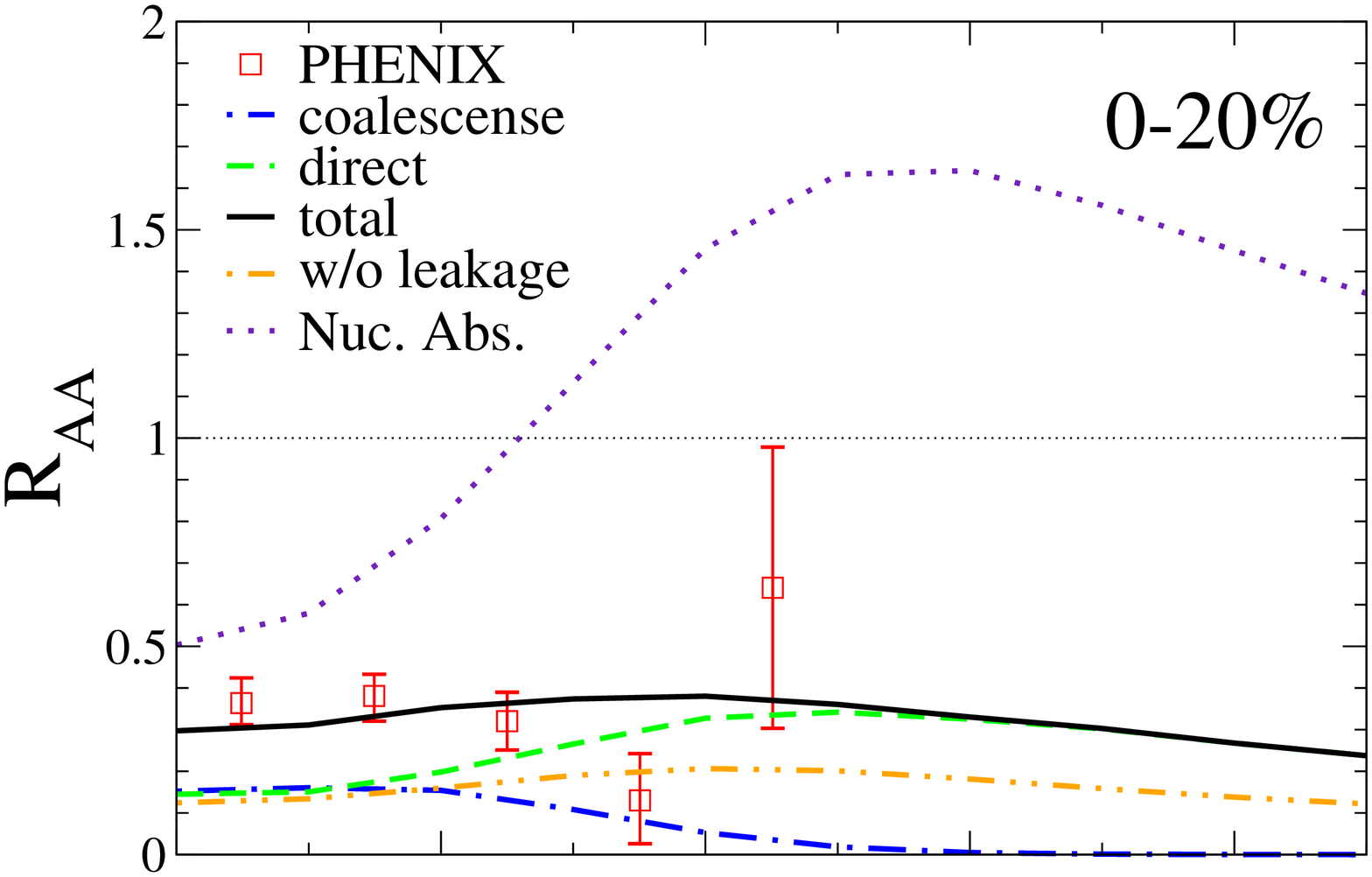}\\[1mm]
    \includegraphics[width=0.45\textwidth]{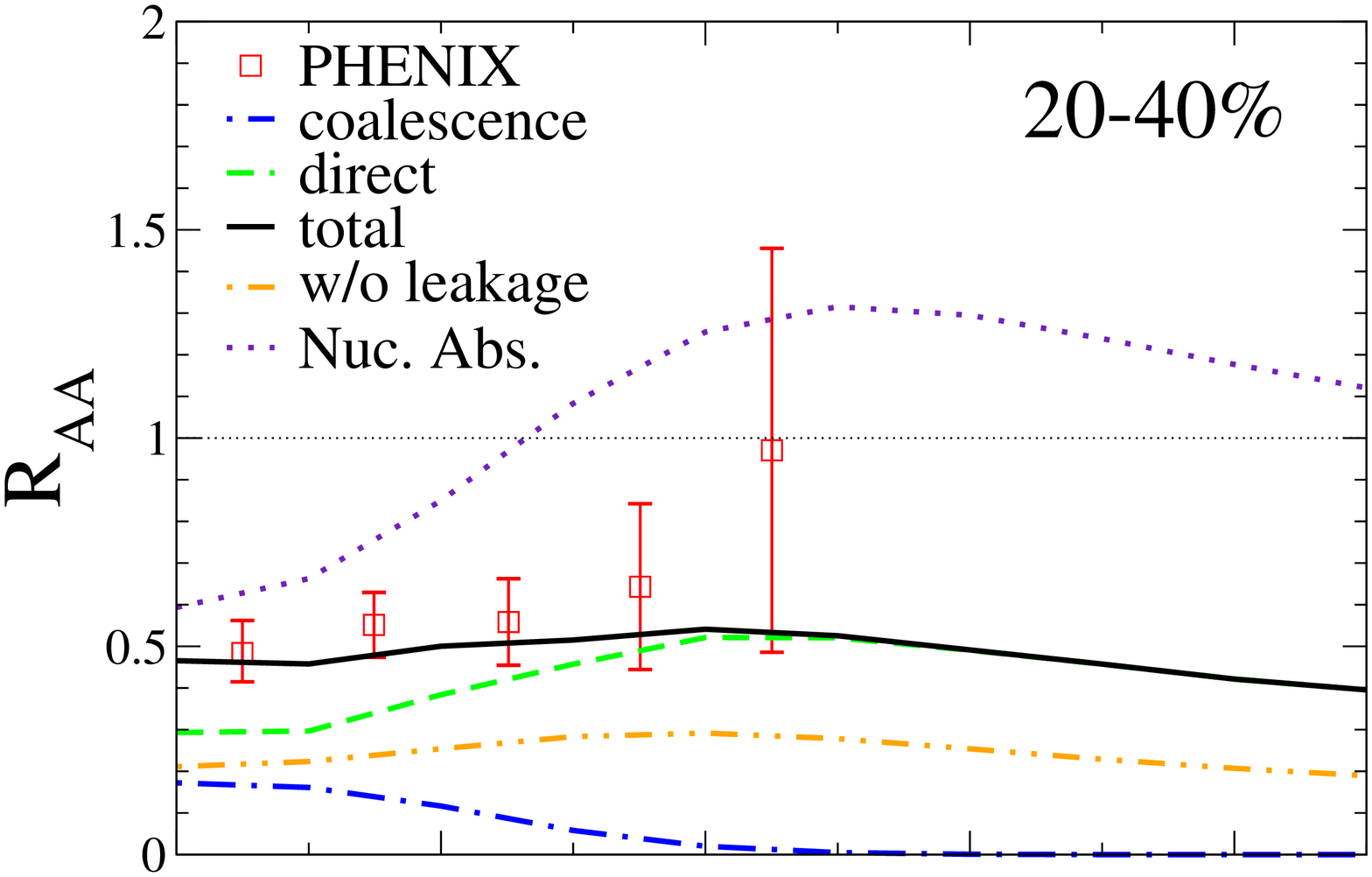}\\[1mm]
    \includegraphics[width=0.45\textwidth]{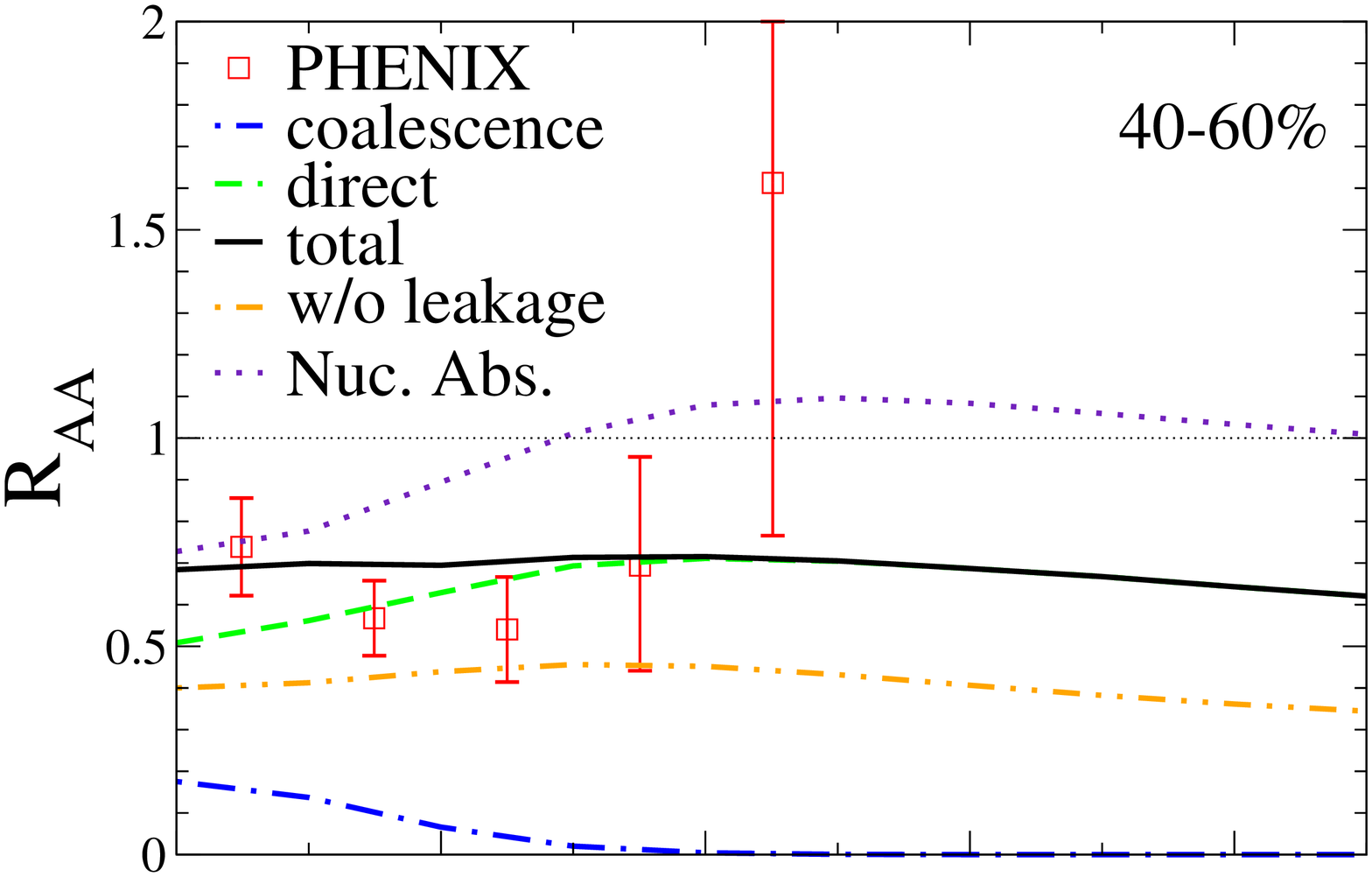}\\[1mm]
    \includegraphics[width=0.45\textwidth]{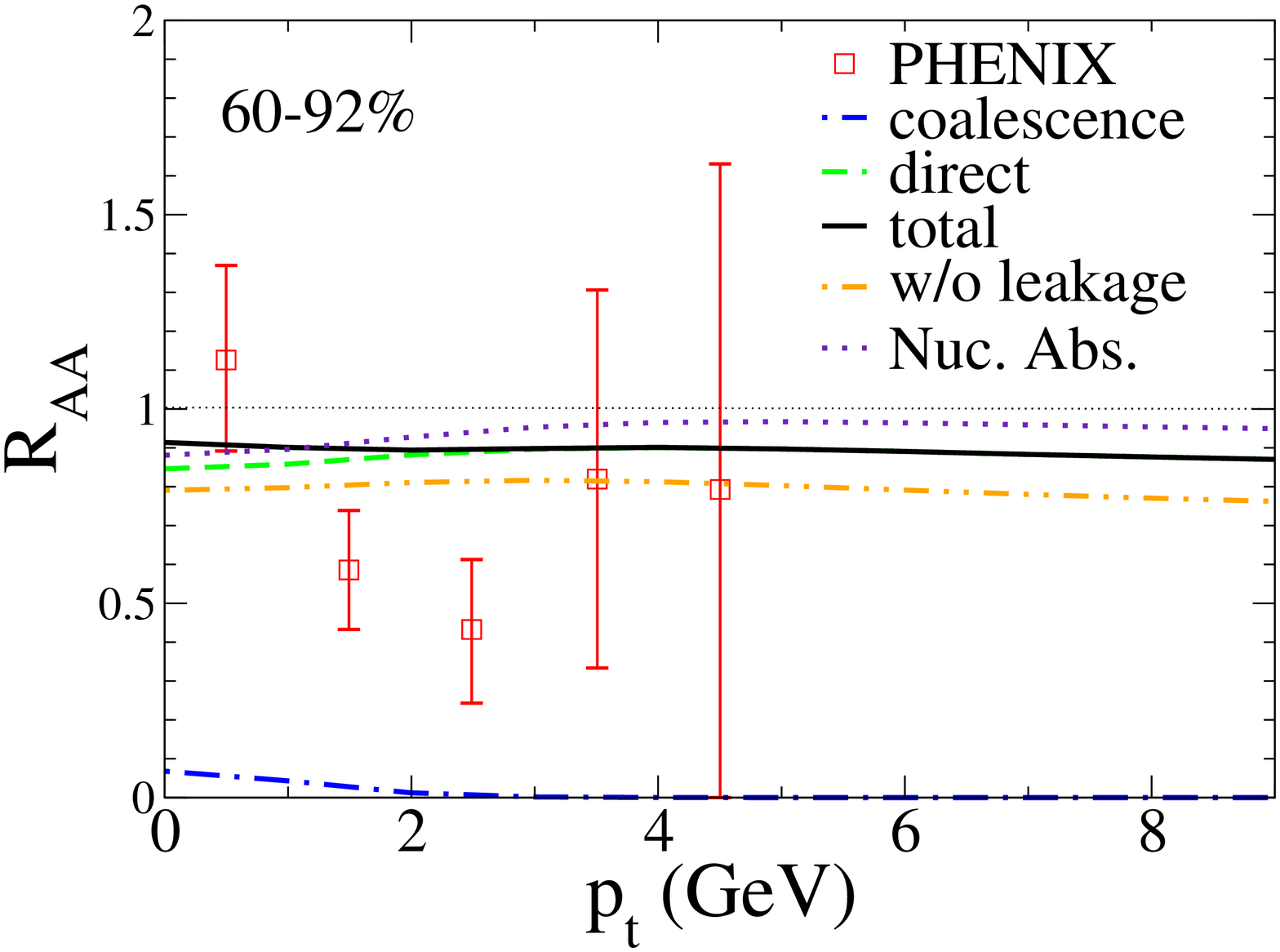}
    \caption{The $J/\psi$ $R_{AA}$ at midrapidity as a function of $p_T$ 
      in \AuAu\ collisions in four centrality
      bins, compared with calculations including formation by coalescence 
      \protect\cite{rapp_pt_2007}.}
  \label{fig:rapp_phenix_raa_pt}
  \vspace{0.5cm}
\end{figure}

There has been considerable interest in what might be learned about the 
\jpsi\ production mechanism from $R_{AA}(p_T)$, since 
coalescence formation of the $J/\psi$ is generally expected to decrease at 
high $p_T$.  PHENIX has published data on the $p_T$ dependence of 
$R_{AA}$ in \AuAu\ and \CuCu\ collisions 
\cite{PHENIX_jpsi_auau,PHENIX_jpsi_cucu}.
The measured $R_{AA}$ for central collisions is flat within errors over 
the range of the data, extending only to 5 \gevc.
Zhao and Rapp \cite{zhao_rapp_2007} have calculated the \pT\ dependence of the 
\jpsi\ \RAA\ at midrapidity and compared it with the \pT\ dependence 
in \AuAu\ collisions measured at four different centralities. The calculation 
agrees well with the data, as seen in Fig. \ref{fig:rapp_phenix_raa_pt}. 
Coalescence is most important for more central 
collisions  at low $p_T$.  It is a negligible effect above 4 \gevc. 
An important goal of future measurements is to improve the poor statistical 
precision of the current data beyond about 3 \gevc.
Any AdS/CFT effects \cite{rajagopal_2006} would become important for 
$p_T > 5$ \gevc, beyond the current range of precision. 
Extending the $R_{AA}$ measurement with good precision to much higher $p_T$ 
will require RHIC II luminosity, as discussed later.

A strong charm coalescence contribution to $J/\psi$ production will result in
a narrower rapidity distribution, similar to the $p_T$ distribution. 
PHENIX extracted the RMS of the rapidity distribution versus centrality
from their \AuAu, \CuCu\ and \pp\ data 
\cite{PHENIX_jpsi_auau,PHENIX_jpsi_cucu,PHENIX_jpsi_pprun5}.  The extracted \CuCu\ data are
independent of centrality within $2-3$\% uncertainties.  The extracted \AuAu\ 
values decrease by 10\% with increasing centrality, roughly a two 
standard deviation reduction.  

In summary, while there is evidence favoring a coalescence contribution to 
the $J/\psi$ yield in central collisions, it does not seem to be definitive. 
The suppression as a function 
of centrality shown in Figs.~\ref{fig:PHENIX_auau_raa_models_2} and 
\ref{fig:PHENIX_auau_raa_models_3} is reasonably consistent with models that 
include coalescence, as is the $\langle p_T^2 \rangle$ in 
Fig.~\ref{fig:PHENIX_auau&cucu_meanpt} and $R_{AA}(p_T)$ shown in 
Fig.~\ref{fig:rapp_phenix_raa_pt}. The observed narrowing of the rapidity 
distribution is weak, and its implication will not be clear until the open 
charm rapidity distribution is better determined experimentally
and the effects of shadowing and other initial state phenomena are understood.
The initial-state effects should be better constrained by the more precise 
\dAu\ data obtained in Run 8. Precise open charm rapidity distributions 
will require upgrades to the PHENIX and STAR detectors to identify 
open charm and bottom decays by displaced decay vertices.
On the other hand, the predicted narrowing of the $p_T$ distribution relative 
to broadening of the primordial \jpsi\ \pT\ distribution expected 
in heavy-ion collisions is based on a steeply falling charm $p_T$
distribution, well established by the existing open charm data, and the 
calculations seem to reproduce the data reasonably well. 

PHENIX is attempting to extract the $J/\psi$ $v_2$ from Run 7 \AuAu\ data. 
The current restricted heavy-ion data sets place major limitations on the 
precision of a $v_2$ measurement.
A statistically meaningful $J/\psi$ polarization measurement is also 
not feasible with the present data sets. Precise $v_2$ and polarization 
measurements will need RHIC II luminosity.

\begin{figure}[tbh]
    \vspace{0.5cm} \centering
    \includegraphics[width=0.95\textwidth]{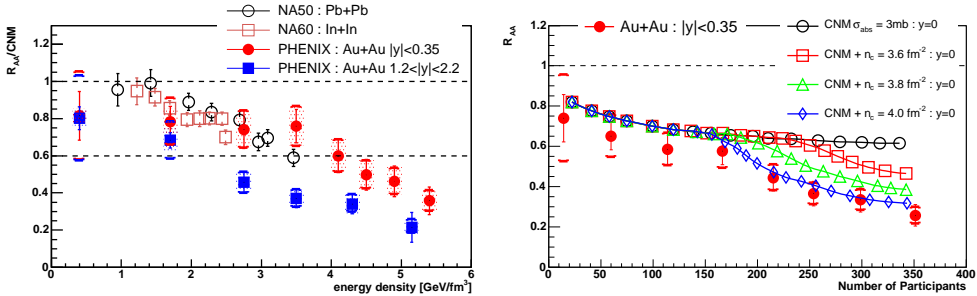}
  \caption{Left: The RHIC and SPS $R_{AA}$ data compared 
      \protect\cite{gunji_2006} with sequential 
      charmonium suppression.  The two dashed lines indicate no
      suppression ($R_{AA}$/CNM = 1) and complete suppression of
      $\psi'$ and $\chi_c$ ($R_{AA}$/CNM = 0.6).  Right: The PHENIX
      midrapidity data compared with a model including a threshold
      density for \jpsi\ suppression \protect\cite{chaudhuri_2006}.  }
  \label{fig:sequential_melting_models}
  \vspace{0.5cm}
\end{figure}

The sequential charmonium suppression model of Ref.~\cite{Karsch_Kharzeev_Satz}
has been applied to the preliminary PHENIX data and to the SPS data.
Normal nuclear absorption is
parameterized by an effective absorption cross section that accounts for all
cold nuclear matter effects.  The SPS $p + A$ data give a larger effective
$J/\psi$ absorption cross section than the RHIC \dAu\ data, as also implied by
Ref.~\cite{mikeandme}.  Three values
of $\sigma_{\rm abs}$ are extracted from the \dAu\ data, one for each rapidity
bin, similar to the ad hoc model applied to the PHENIX data in 
Figs.~\ref{fig:PHENIX_auau&cucu_raa} and \ref{fig:survival_eks_ad_hoc}.  
These values are used to obtain  the survival probability in $A+A$
collisions.  When $N_{\rm part}$ is converted to energy density, $\epsilon$,
and the survival probabilities for color screening and cold nuclear matter are
included, the SPS and RHIC data were found to lie on a common suppression 
curve as a function of energy density. However the more precise, final PHENIX 
data, which show that the forward rapidity \jpsi\ yield is significantly more 
suppressed than the midrapidity yield, are inconsistent with such a picture, 
as can be seen on the left-hand side of 
Fig.~\ref{fig:sequential_melting_models} \cite{gunji_2006}. 
The measured $\langle p_T^2 \rangle$ is also inconsistent with this scenario. 

The threshold energy density model applied to RHIC, 
Ref.~\cite{chaudhuri_2006}, is compared to the midrapidity 
PHENIX data on the right-hand side of Fig.~\ref{fig:sequential_melting_models}
for several values of the critical density.  The model behavior is quite 
similar to that of the measured midrapidity \AuAu\ $R_{AA}$ for a critical 
density of 4 fm$^{-2}$. However, since it does not have any rapidity
dependence, it cannot describe the much stronger suppression for \AuAu\ at 
forward/backward rapidity.  It also cannot explain the centrality dependence 
of the \CuCu\ suppression.  

The hydrodynamic model of Ref.~\cite{gunji_hirano_2007} fits 
the critical temperatures to the midrapidity PHENIX \AuAu\ data but disagrees 
with \RAA\ as a function of centrality. 
Since many effects are not included, it is difficult to draw conclusions.

\subsection{Proposed RHIC II quarkonia measurements}

Unlike other probes, quarkonia measurements are guided 
by predictions from lattice QCD calculations.  Color
screening modifies the linear
rise of the QCD potential at large distances.  The quarkonia spectral
functions quantify the temperature dependence of the potential. 
Since quarkonia suppression is determined by the plasma temperature 
and the binding energy (equivalently the
quarkonium size and the Debye screening length), measuring
the sequential disappearance of these states acts as a QCD
thermometer.

Thus the importance of a comprehensive study of \textbf{all} experimentally 
accessible quarkonium states cannot be overstated.  A systematic 
study of heavy quarkonium spectroscopy, with a complete determination of the
suppression pattern of the quarkonium states, remains the \textbf{most direct
probe of deconfinement}. It is also the signature that most closely resembles
a thermometer of the hot initial state which, with future improved
lattice calculations, can be directly compared to QCD.

While $J/\psi$ physics is as compelling as it was in 1986
when first proposed by Matsui and Satz \cite{MS}, the
systematic study of all quarkonia states, and especially
bottomonium, 
feasible at RHIC II, provides a more complete QGP probe than heretofore
possible.  

\begin{figure}[tbh]
\vspace{0.5cm}
 \centering
  \includegraphics[width=0.7\textwidth]{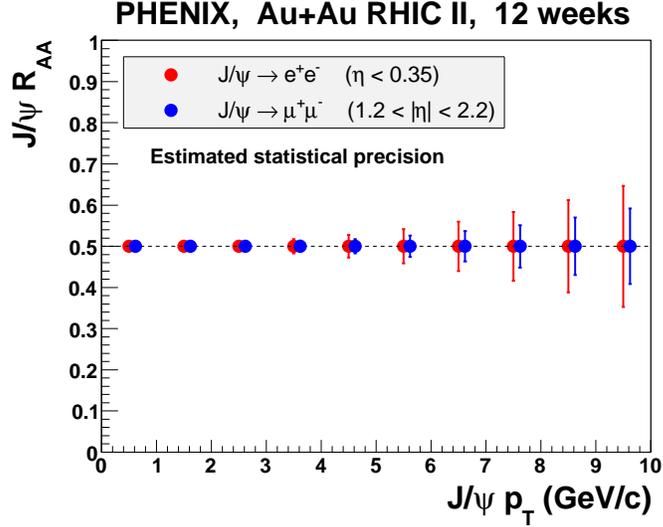}
  \caption{The statistical significance after background subtraction of the 
    \jpsi\ $R_{AA}(p_T)$ in PHENIX at central and forward rapidity from a 12 week
    \AuAu\ run with RHIC II luminosity (22 $nb^{-1}$ delivered). Values 
    above 5 \gevc\ are a $p_T$ extrapolation of current PHENIX data.}
  \label{fig:PHENIX_jpsi_pt_rhic2}
  \vspace{0.5cm}
\end{figure}

\begin{figure}[tbh]
\vspace{0.5cm}
 \centering
  \includegraphics[width=0.7\textwidth]{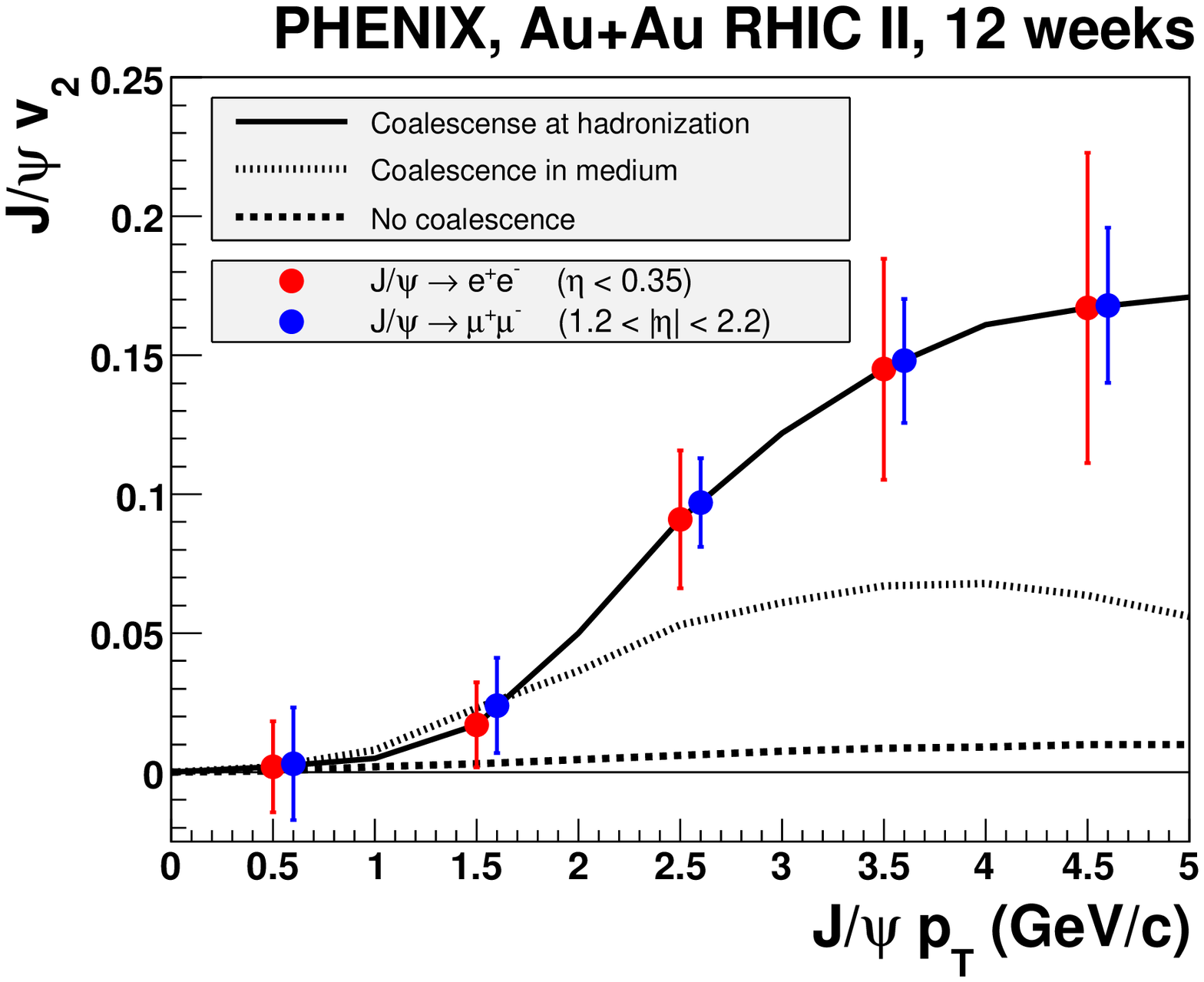}
  \caption{The vertical bars show the estimated absolute precision 
   of the $J/\psi$ $v_2$ measurement in PHENIX 
   at central and forward rapidity in a 12 week \AuAu\ run with RHIC II luminosity 
   (22 $nb^{-1}$ delivered) 
   \protect\cite{wysocki_jpsi_v2}. The precision is 
   compared with several different models
   \protect\cite{Greco:2003vf,yan_zhuang_xu_2006}.}
  \label{fig:PHENIX_jpsi_v2_rhic2}
\vspace{0.5cm}
\end{figure}

Table \ref{tab:quarkonia-req} relates the main physics topics to the
relevant probes and subsequent detector requirements. The ability of a
program at RHIC II to make these measurements can be judged from the
yields given in
Tables~\ref{rhic1_rhic2_yields_phenix},~\ref{rhic1_rhic2_yields_star},
and ~\ref{rhic1_rhic2_yields_500gev}.  The measurements that are
possible at RHIC without the luminosity upgrade are the $J/\psi$
rapidity and $p_T$ distributions at full energy.  The measurements
that are newly possible at RHIC II are those of the excited charmonium
states ($\psi^\prime$ and $\chi_c$) and the bottomonium states
($\Upsilon(1S)$, $\Upsilon(2S)$ and $\Upsilon(3S)$). High $p_T$ \jpsi\ 
measurements, precise measurements of the $J/\psi$ $v_2$ and
polarization, and excitation functions of heavy flavor measurements
will be possible only at RHIC II.  The possible statistical precision 
is illustrated in Fig.~\ref{fig:PHENIX_jpsi_pt_rhic2} 
which shows the statistical
significance of the \jpsi\ nuclear modification factor $R_{AA}$ in
\AuAu\ at high $p_T$, and Fig.~\ref{fig:PHENIX_jpsi_v2_rhic2}, which
presents the expected $J/\psi$ $v_2$ precision in \AuAu\ collisions at
RHIC II \cite{wysocki_jpsi_v2}.  It is evident that a comprehensive
program to use quarkonium as a QCD thermometer to provide direct
evidence of deconfinement is possible only with RHIC II luminosity.

\begin{sidewaystable}
   \begin{tabularx}{\linewidth}{| X | X | X | X |} \hline
        Physics Motivation & Probes & Measurements & Requirements 
\\ \hline  
        Baseline measurements & 
      \jpsi, \psip, $\chi_c$, \upss, \upsss, and \upssss\ decays to dileptons &
        Rapidity and \pT\ spectra 
        in \pA\ and \pp\ as a function of
        \sqrtsNN\
        & High luminosity and acceptance for sufficient statistics, 
        especially for the
        \ups\ family. Good mass resolution to resolve $\psi$ 
        and \ups\ states. \\
        \hline
        Deconfinement and initial temperature &
      \jpsi, \psip, $\chi_c$, \upss, \upsss, and \upssss\ decays to dileptons &
         \AA\ suppression patterns as a function of \sqrtsNN\ and $A$ &
          High luminosity, acceptance and mass resolution for quarkonium, 
        and triggers that 
        work in \AuAu\ collisions.
        \\ \hline 
        Thermalization and transport
        &
        \jpsi 
        &
        $J/\psi$ $v_2$ as function of \sqrtsNN\ and $A$
        &
        High luminosity for good statistics in short runs for \sqrtsNN\ 
        and $A$ scans. \\ \hline
    \end{tabularx}
    \caption{The main physics goals of the RHIC II quarkonium program with
        corresponding probes, studies, and requirements.}
    \label{tab:quarkonia-req}
\end{sidewaystable}

\begin{figure}[tbh]
\vspace{0.5cm}
 \centering
 \includegraphics[width=1.0\textwidth]{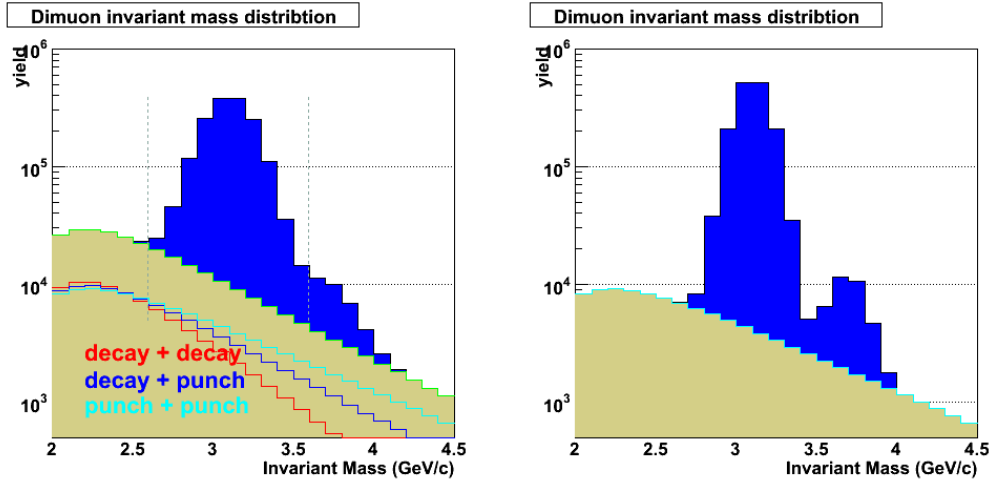}
  \caption{The $J/\psi$ and $\psi^\prime$ invariant mass spectrum in the PHENIX 
    muon arms without (left) and with (right) the improvement in mass resolution from 
    the FVTX detector. }
  \label{fig:fvtx_jpsi_resolution}
  \vspace{0.5cm}
\end{figure}

The measurements needed to study the excited charmonium states,
$\chi_c$ and $\psi^\prime$, have quite different problems. The
$\psi^\prime$ measurement technique is the same as that for the
$J/\psi$, namely reconstruction of dilepton decays, but requires $\sim
100$ times greater integrated luminosity for the same yield. In
addition, the $\psi^\prime$ measurement is more difficult because the 
background under the peak in the invariant mass spectrum is significant, 
increasing the integrated luminosity needed for precision measurements.
The presence of the VTX detector
in PHENIX will lead to significantly better mass resolution in 
the central arms by allowing a measurement of the
full bend angle of the electrons - only the track angle after
the magnetic field is measured at present. The FVTX detector is
crucial to the $\psi^\prime$ measurement in the muon arms because
it both reduces the combinatorial background and improves the mass
resolution, as shown in Fig.~\ref{fig:fvtx_jpsi_resolution}. A
$\psi^\prime$ measurement is certainly feasible at RHIC II.   
The $\chi_c$ measurement can be done with the $\chi_c \rightarrow
J/\psi~\gamma$ channel, where the $J/\psi$ is reconstructed from
dilepton decays and the photon is detected in an electromagnetic
calorimeter.  While the yields are larger than for the
$\psi^\prime$, the need to form the $\chi_c$ invariant mass by
combining each $J/\psi$ candidate with a large number of photons means
that combinatorial backgrounds will be quite large in \AuAu\ 
collisions. Thus the $\chi_c$ measurement will be difficult in central
heavy-ion collisions.  

\begin{figure}[tbh]
  \vspace{0.5cm}
  \centering
  \includegraphics[width=0.85\textwidth]{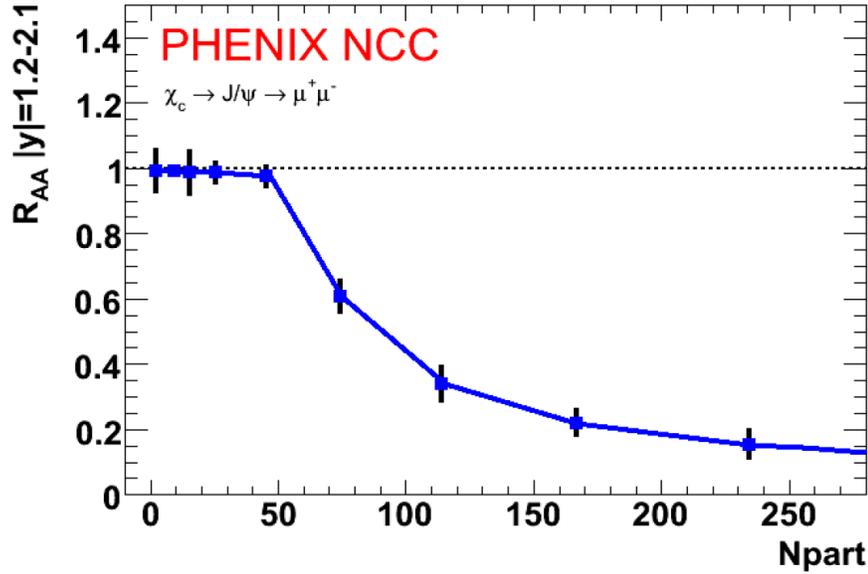}
  \caption{The expected precision of the \RAA\ for the $\chi_c$ from
    PHENIX using $\chi_c \rightarrow J/\psi~\gamma$ decays
    measured in the muon arms and the Nose Cone Calorimeter. The yields
    shown are for a delivered \AuAu\ luminosity of 30 $nb^{-1}$.
  }
  \label{fig:phenix_chic_raa}
  \vspace{0.5cm}
\end{figure}

PHENIX has very recently carried out detailed simulations of the performance
of the PHENIX muon arms in combination with the Nose Cone Calorimeter for
$\chi_c \rightarrow J/\psi~\gamma$ measurements. 
Figure~\ref{fig:phenix_chic_raa} 
shows the expected precision for the \RAA\ of the $\chi_c$ using the suppression
pattern from a simple model in which the $\chi_c$ disappears at $1.16 T_c$.
Note that the maximum \npart\ bin ends at 30\% centrality. Extending the 
measurement beyond 30\% to the most central collisions is still being worked on.

\begin{figure}[tbh]
\vspace{0.5cm}
 \centering
  \includegraphics[width=0.7\textwidth]{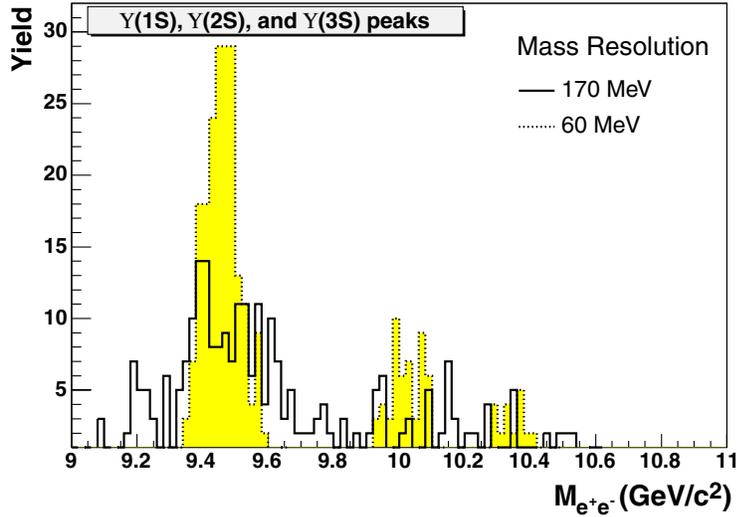}
  \caption{The dielectron mass spectrum for the  $\Upsilon$ family 
from a PHENIX central arm simulation, showing the expected improvement in 
$\Upsilon$ mass resolution provided by the initial direction measurement in 
the SVTX barrel \protect\cite{phenix_vtx_ref}. The number of events 
shown correspond to a delivered luminosity of 30 $nb^{-1}$ for \AuAu.}
  \label{fig:phenix_upsilon_ee_mass}
\vspace{0.5cm}
\end{figure}

\begin{figure}[tbh]
\vspace{0.5cm}
 \centering
  \includegraphics[width=0.7\textwidth]{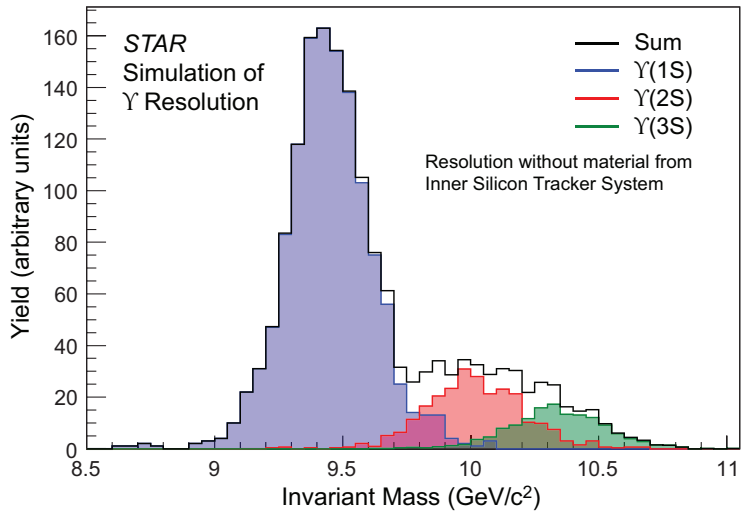}
  \caption{The $\Upsilon$ family dielectron mass spectrum 
simulated by STAR. The three states can be clearly separated in the data given sufficient statistics.}
  \label{fig:star_upsilon}
\vspace{0.5cm}
\end{figure}

As is the case for the $J/\psi$, the bottomonium states are studied
through their dilepton decays.  The bottomonium measurements require
very large integrated luminosity and good invariant mass resolution.
PHENIX expects to be able to resolve the \upss, \upsss\ and \upssss\ 
states, see Fig.~\ref{fig:phenix_upsilon_ee_mass}.  Because of its
larger acceptance, STAR will have substantially greater $\Upsilon$
yields than PHENIX. To clearly resolve the three $S$-wave states, STAR
needs to reduce the amount of material between the interaction point
and the inner field cage of the TPC because energy-loss due to
bremsstrahlung deteriorates the mass resolution.  Figure
\ref{fig:star_upsilon} shows the $\Upsilon$ mass spectrum without the
material from the inner tracking system. Whether the resolution with the
new inner tracking system (HFT) installed is sufficient to resolve all
3 states depends on the final design parameters of the new detectors.

Although the yields are small relative to the $J/\psi$, bottomonium
measurements are quite clean. The states are massive ($\sim 10$
\gevcc) so that their decay leptons have relatively large momenta and
are thus easily distinguished from background leptons.  The
combinatorial background is small and multiple scattering is of less
concern.  While the interpretation of charmonium suppression is made
more difficult by the rather large cross section for nucleon and
comover absorption, the situation for bottomonium is considerably
better. Absorption of directly produced bottomonium by hadronic
comovers was shown to be negligible \cite{lin-ko}.

The \sqrtsNN\ dependence of produced $J/\psi$'s relative to the number
of $c\bar{c}$ pairs, depicted in Fig.~\ref{fig:rapp} \cite{rapp2}, is
striking.  Measurement of the excitation function for this ratio over
$30 < \sqrtsNN \ < 200$ \gev\ could help to disentangle suppression
from enhancement. Such measurements, however, are extremely demanding
statistically since both heavy quarks and quarkonia will need to be
measured with good statistics over a wide energy range.

A measurement of the quarkonium nuclear modification factor at
high~\pT\ can provide a unique experimental probe of energy loss and
color diffusion \cite{Baier1}.  At relatively large transverse
momentum, suppression due to color screening and coalescence are
predicted to be negligible.  Instead, the quarkonium state is a hard
probe that interacts with the medium. In particular, any color octet
can suffer energy loss. The relative abundance of charmonium
resonances can provide an experimental handle on such phenomena as
each resonance may have a different octet contribution. We must
exercise caution, however, as competing charmonium production models
exist.  In parallel with nucleus-nucleus studies, it is therefore
important to investigate and compare production mechanisms in \pp\ and
\pA\ interactions, at both central and forward rapidities
\cite{rvprc,Gavin1,Johnson1}.

In addition to the baseline quarkonium measurements in \pp\ and \pA\ 
collisions listed in Table~\ref{tab:quarkonia-req},
other measurements are required as model input for description of the
$A+A$ measurements. The most prominent of these are
listed in Table~\ref{tab:support-meas}.

\begin{table}[tbh]
    \begin{tabularx}{\linewidth}{| X | X | X |} \hline Topic & Measurements
        & Requirements \\ \hline \hline
        Cold nuclear effects & In \pp\ and \pA\
        collisions:
        \begin{itemize}
        \item $x_{1,2}$, $x_F$ and $y$ 
      dependence of quarkonia production
        \item $A$ dependence
        \end{itemize}
        & Large rapidity acceptance, 
        including forward coverage. 
        \\ \hline
        Suppression vs coalescence &
          In \pp, \pA\ and \AA\ collisions:
        \begin{itemize}
        \item Charm $d\sigma/d\pT dy$
        \item \jpsi\ $v_2$
        \item $R_{AA}(p_T)$
        \end{itemize}
        & High resolution vertex detectors (charm). \\ \hline
        Feed down & $\chi_c$, at least in
        \pp\ and \pA\ & Photon detection over wide rapidity range.
        High rates, good energy and momentum
        resolution to enhance $\chi_c$ signal to background. \\ \hline
        Production mechanism & $\chi_c$, polarization at least in
        \pp\ and \pA\ & Large acceptance for $\cos \theta^*$
        measurement. \\ \hline
    \end{tabularx}
    \caption{Baseline measurements (beyond \AuAu) required in order to
        address the main physics questions.}
    \label{tab:support-meas}
\end{table}

The importance of measuring the underlying charm distributions as input to 
models of $J/\psi$ coalescence is obvious, as is the importance of 
understanding cold nuclear matter effects on quarkonium production.

It is crucial for the interpretation of the \AA\ quarkonia yields 
to understand the feed down contributions from the $\chi_c$ 
states, see Fig.~\ref{fig:feeddown}.  The best feed down measurement will 
be made in 500 \gev\ \pp\ collisions because 
the increased luminosity and increased charmonium production cross sections 
lead to $\sim10$ times larger charmonium yields than in 200 \gev\ \pp\ 
collisions.  Since the $\chi_c$ contribution to the $J/\psi$ yield will not 
change significantly between 200 and 500 \gev, the increased yield at 500 \gev\ 
will provide a definitive
baseline measurement of $\chi_c$ feed down in \pp\ collisions.

Recently, quarkonium polarization measurements were suggested as 
signatures of
QGP formation \cite{Kharzeev1}. The quarkonium yields at RHIC II will 
be large enough to permit a $J/\psi$ polarization measurement at low $p_T$
by both PHENIX and STAR. 


\section{Relationship to the LHC program}

The major differences between quarkonium studies at RHIC II and at the LHC  
will be the temperature and lifetime of the medium, the relative production
cross sections, luminosities and run times.

The initial temperature in $\sqrtsNN = 5.5$ TeV central \PbPb\ 
collisions at the LHC is expected to be $\sim 4~T_c$, while it is $\sim 2~T_c$ 
in \sqrtsNN = 200 \gev\ 
central \AuAu\ collisions at RHIC \cite{rhic_lhc_temperature}. 
The QGP lifetime at the LHC is expected to be two to three times longer 
than at RHIC \cite{rhic_lhc_temperature}.

Heavy flavor production cross sections are much larger at the LHC. 
The open charm and bottom production cross sections are $\sim 15$ and  
$\sim 100$ times higher respectively \cite{Vogt:2001nh}. The
charmonium and bottomonium cross sections are $\sim 13$ and $\sim 55$ times 
higher respectively than at RHIC \cite{hfYR}.
The higher LHC open heavy flavor cross sections increase the number of 
$c\bar{c}$ and $b\bar{b}$ pairs produced in central \AA\ collisions.  There 
are  $\sim 10$ and $\sim 0.05$ pairs, respectively, in \AuAu\ collisions at 
RHIC, while there should be $\sim 115$ and $\sim 5$ pairs
in central \PbPb\ collisions at the LHC \cite{Vogt:2001nh}.

The RHIC II \AuAu\ average luminosity is projected to be 10 times larger 
than the LHC \PbPb\ luminosity ($5 \times 10^{27}$ cm$^{-2}$ s$^{-1}$ 
relative to $5 \times 10^{26}$ cm$^{-2}$ s$^{-1}$).
The yearly heavy-ion runs at RHIC II are also expected to be 
considerably longer than at the LHC. Taking the polarized \pp\ 
program at RHIC II into account, the heavy-ion program is expected 
to get $\sim 12$ week physics runs on average per year while the LHC heavy
ion program will be allocated a one month physics run per year. Thus the annual 
integrated luminosity at RHIC II is expected to be about 30 times larger for 
heavy ions than at LHC.

The larger heavy flavor cross sections at the LHC are approximately 
balanced by the increased luminosity and running times at RHIC II, making
the heavy flavor yields per year similar. Thus the types of measurements 
that can be made at the two facilities will also be similar
as well as of similar quality (see Tables~\ref{rhic1_rhic2_yields_phenix},
~\ref{rhic1_rhic2_yields_star} and ~\ref{LHC_yields_all}). 
However, there will be important differences in the physics environments 
prevailing at the two facilities which will make the two programs 
complementary. 

The higher initial energy density at the LHC means that the QGP will be 
created at a significantly higher temperature with a correspondingly strong 
potential for new physics effects at the LHC. In addition, the factor of ten 
increase in $c\bar{c}$ pairs and the factor of 100 increase in $b\bar{b}$ 
pairs per central collision at the LHC will have a major impact on the 
interpretation of heavy flavor measurements. We will discuss some of those 
differences here.

Lattice calculations suggest that the $J/\psi$ may remain bound at
the highest RHIC temperatures, while the excited charmonium states are 
predicted to be unbound. At the LHC, all of the charmonium states should be 
unbound at the highest temperatures, implying that almost all charmonium 
production in central \PbPb\ collisions at the
LHC will be due to coalescence of $c\bar{c}$ pairs. 
Thus the prompt charmonium yields at the LHC should
reflect only the coalescence mechanism with no contribution from the 
primordial $J/\psi$ production except in very peripheral collisions.
The measurements at RHIC and the LHC will thus
provide very different windows on charmonium suppression in the QGP that 
will help resolve the ambiguities in interpreting data due to the balance 
between destruction and coalescence formation of charmonium at RHIC.

\begin{figure}[tbh]
    \centering\includegraphics[width=0.75\textwidth]{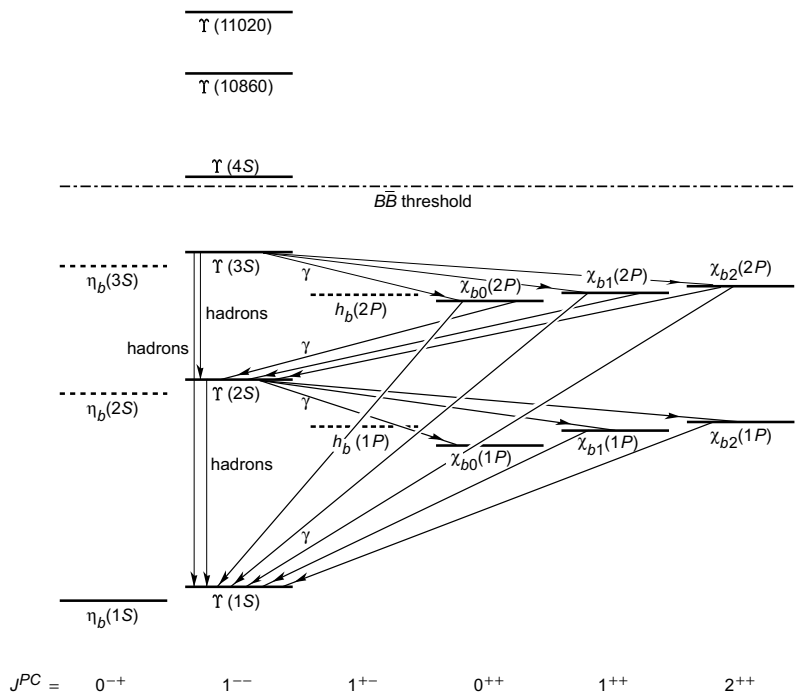}
    \caption{Bottomonium mass levels and spin states.  The common feed down
channels are indicated.}
    \label{fig:bottom_states}
\end{figure}

Because of its higher binding energy, the characteristics of bottomonium 
production at the LHC should be similar to those of charmonium
at RHIC. 
The bottomonium states are shown in Fig.~\ref{fig:bottom_states}.
The $\Upsilon(1S)$ may remain bound at the highest temperatures 
at the LHC while the other bottomonium states will be dissociated. Given 
$\sim 5$ $b\bar{b}$ pairs in central \PbPb\ collisions (relative to 
$\sim 10$ $c\bar{c}$ pairs at RHIC), the 
$\Upsilon$ yield at the LHC is predicted \cite{upsilon_lhc_calc} 
to reflect a balance between dissociation and 
coalescence reminiscent of the RHIC $J/\psi$ production models. 
However, at RHIC, the bottomonium dissociation rates will be significantly 
different. While the $\Upsilon(1S)$ is predicted to be bound, the 
$\Upsilon(2S)$ may also remain bound.  Only the $\Upsilon(3S)$ is likely to
dissociate at RHIC. Also, since the $b\bar{b}$ pair yield at RHIC is 
$\sim 0.05$ per central \AuAu\ collision, no significant bottomonium production
by coalescence is expected. Thus the bottomonium yields at RHIC II 
should reflect only QGP suppression.  Measurements at RHIC II and the LHC 
will thus provide very different windows on bottomonium suppression in the 
QGP that will help to resolve the ambiguities in interpretation due to 
the balance of bottomonium destruction and coalescence at the LHC.

The open heavy flavor programs at RHIC II and the LHC will consist of similar 
measurements with similar goals. 
They will study energy loss, thermalization and flow of heavy quarks in 
systems with very different
energy densities, interaction cross sections and lifetimes.
However, not all challenges in the measurements are similar. 
At $\sqrt{s_{NN}} = 200$ \gev, bottom decays to leptons begin to dominate the
single electron spectrum at $p_T \sim 4$~\gevc.  As the collision energy 
increases, the lepton spectra from $B$ and $D$ decays move closer together
rather than further apart.  Thus,
the large increase in the $b \overline b$ cross section relative to $c 
\overline c$ does not make single leptons from $B$ and $D$ decays
easier to separate.  Preliminary calculations show that the $B \rightarrow e$
decay does become larger than that of $D \rightarrow e$ but only for 
$p_T > 10$~\gevc.
The two lepton sources differ by less than a factor of two to $p_T \sim 
50$~\gevc\ in the range $|y| \leq 1$. Separating single leptons from charm and 
bottom decays will require statistical separation using differences in the
displaced vertex distributions at all $p_T$
at the LHC. 
Thus interpretation of single lepton data
from heavy flavor decays will be more difficult at the LHC. 

ALICE can reconstruct
$D^0$ decays from $p_T \sim 0$ to $p_T \sim 25$~\gevc\ \cite{CJ17}. 
Like STAR, ALICE will be unable to trigger on $D^0$'s and will have to 
obtain these events from the minimum bias sample.  Thus the longer running 
times at RHIC are an advantage since more minimum bias data can be 
taken (see Tables~\ref{rhic1_rhic2_yields_star} and ~\ref{LHC_yields_all}).
While it is not yet clear what CMS and ATLAS will do to reconstruct charm,
they should be able to do $b$ jets well, similar to the Tevatron measurements.
As at RHIC, $B$ mesons can be measured cleanly at the LHC through their 
decays to $J/\psi$, although triggering on low $p_T$ $J/\psi$'s
is difficult at the LHC.  

It has also been suggested that
the $B \overline B$ contribution to the dimuon continuum, the dominant 
contribution above the $\Upsilon$ mass, can be used to measure energy loss
\cite{igor}.  That channel would be fairly clean at the LHC but more difficult
at RHIC. 

\section{Conclusions}

We have shown that, so far, the RHIC heavy flavor physics program has been
very rich and 
stimulating, with many provocative and challenging results. To fully realize
the potential of this compelling program, however, both detector upgrades and
a luminosity upgrade are mandatory. Detector upgrades will improve 
reconstruction of charm hadron decays into hadronic channels and allow 
detection of $B \rightarrow J/\psi X$ decays using secondary vertex 
measurements. Upgrades will also make $\chi_c$ detection possible and, in the 
case of STAR, lead to significant quarkonium yields. However only increased 
luminosities will allow high statistics measurements of all of these yields 
as well as increase the $p_T$ reach of $J/\psi$ and heavy flavor $R_{AA}$ 
and $v_2$.

We have also shown that the RHIC II and LHC heavy flavor physics programs are 
complementary. Both are required for a complete understanding of heavy flavor 
production as a function of energy and temperature. We have also demonstrated 
that, despite lower heavy flavor cross sections at RHIC, the longer running 
times and higher luminosity
of RHIC II make the recorded yields similar at the two facilities.

\ack We thank F. Karsch for the lattice contribution.
The work of R. Vogt was supported in part by the U. S. Department of 
Energy under Contract Nos. DE-AC52-07NA27344 (LLNL) and DE-AC02-05CH11231 
(LBNL) and by the National Science Foundation Grant PHY-0555660.  
The work of T. Ullrich was supported in part by
 the U. S. Department of Energy under Contract No. DE-AC02-98CH10886.  The
work of A. D. Frawley was supported by National Science Foundation 
grant PHY-04-56463.


\end{document}